%% file: IC2431_paper.tex
\documentclass{aastex631}

\usepackage{tablefootnote}

\shorttitle{IC 2431}
\shortauthors{}
\graphicspath{{./}{figures/}}

\begin{document}

\title{
The Complex Multi-Wavelength Morphology of the Peculiar Compact Galaxy Group IC 2431
}

\author[0000-0002-8521-5240]{Beverly J. Smith}
\email{smithbj@etsu.edu}
\affiliation{East Tennessee State University \\
Department of Physics and Astronomy, Box 70652 \\
Johnson City TN  37614, USA}

\author[0000-0002-4622-796X]{Roberto Soria}
\affiliation{INAF-Osservatorio Astrofisico di Torino\\
Strada Osservatorio 20, I-10025 Pino Torinese, Italy}
\affiliation{Sydney Institute for Astronomy,\\
School of Physics A28, The University of Sydney,\\
Sydney NSW 2006, Australia}

\author[0000-0002-2954-4461]{Douglas Swartz}
\affiliation{Science and Technology Institute\\
Universities Space Research Association,\\
Huntsville, AL  35805, USA}

\author[0000-0003-3856-7216]{Mark L. Giroux}
\affiliation{East Tennessee State University \\
Department of Physics and Astronomy, Box 70652 \\
Johnson City TN  37614, USA}

\author[0000-0002-6490-2156]{Curtis Struck}
\affiliation{Iowa State University \\
Department of Physics and Astronomy\\
Ames IA 50011, USA}

\author[0000-0003-1814-8620]{Ryan Urquhart}
\affiliation{Michigan State University \\
Department of Physics and Astronomy\\
East Lansing MI 48824, USA}



\begin{abstract}

We present new Chandra X-ray imaging spectroscopy of the compact
galaxy group IC 2431, and compare 
with archival ultraviolet, optical, infrared, and radio images.
IC 2431 is a starburst system containing
three tidally-distorted disk galaxies. 
All three galaxies may have active nuclei.
One galaxy is classified as an AGN based on its optical spectrum,
a second is identified as a possible X-ray AGN based on the Chandra data,
and the third galaxy may host a radio AGN.
In optical images,
a prominent dust lane crosses the southern galaxy,
while Spitzer infrared images show
a dusty bridge connecting the 
two brightest galaxies.
Chandra maps reveal a massive (2 $\times$ 10$^7$ M$_{\sun}$)
concentration of hot gas between these two galaxies,
as well as several other knots of hot gas and non-thermal emission.
The unabsorbed X-ray luminosity of the hot gas in IC 2431 is 
$\sim$ 1 $\times$
10$^{42}$ erg~s$^{-1}$, 
which is enhanced 
by about a factor of four
relative to the 
star formation rate,
compared to other star-forming
galaxies.
In radio maps, 
a bright jet/ridge of radio continuum emission
extends 4 kpc from one nucleus.
We compare the properties of IC 2431 with those of other interacting
galaxy systems, and 
discuss two different scenarios that may account for
the peculiarities of IC 2431: ram pressure stripping 
of the interstellar medium 
during a head-on collision between two galaxies,
or an AGN-powered radio jet that has been distorted by
an interaction with interstellar gas during a tidal encounter between
galaxies.

\end{abstract}



\section{Introduction} \label{sec:intro}

Galaxies in compact groups evolve quickly.
Tidal distortions and galaxy mergers are frequent
in the dense, low velocity-dispersion environment 
of a compact group
\citep{1994ApJ...427..684M, 1998ApJ...507..691M, 2010AJ....139..545G, 2014MNRAS.442.2188T}, triggering bursts of 
star formation and then
rapid quenching
\citep{2007AJ....134.1522J, 2008ApJ...673..730G,
2010ApJ...716..556T, 2010AJ....140.1254W, 2012AA...546A..48P, 2016MNRAS.459.2948L}.
Tidal stripping of cold interstellar gas 
is common in compact groups, producing
HI deficiencies compared to field galaxies
\citep{2001AA...377..812V, 2019AA...632A..78J, 2023AA...670A..21J}.
Groups that are HI deficient are more likely to be quenched
\citep{2016AJ....151...30W}.
Increased interstellar gas turbulence 
due to shocks driven by interactions and collisions
between galaxies may further suppress
star formation by inhibiting gravitational collapse
of gas clouds
\citep{2015ApJ...812..117A, 2017AA...607A.110L}.
The detection of mid-infrared H$_2$ emission
from some compact groups 
\citep{2010ApJ...710..248C, 2006ApJ...639L..51A, 
2013ApJ...765...93C,
2017ApJ...836...76A}
supports the idea that shock-heating contributes to
quenching in compact groups.
Morphological evolution also occurs quickly in compact groups
\citep{2007AJ....133.2630C, 2023MNRAS.524.5340M}.
Within a few billion years, a compact group of spirals 
may undergo a series of mergers to 
form an elliptical galaxy 
\citep{1989Natur.338..123B,
1993ApJ...416...17B,
1994Natur.369..462P, 2000MNRAS.312..139J}.
So-called `fossil groups', groups containing
a dominant elliptical two magnitudes brighter or more
than the rest of 
the galaxies in the group, tend to have hot gas halos
bright in the X-ray
\citep{1998ApJ...496...73M, 2003MNRAS.343..627J}.

Feedback from active galactic nuclei (AGN) may heat interstellar/intragroup gas
in compact groups and further drive evolution
\citep{2021Univ....7..142E, 2024Galax..12...24E}.
Although compact group galaxies do not show
an excess of optically-selected Seyfert nuclei
(\citealp{1998AA...335..912K, 2013ApJ...771..106S, 2014ApJS..212....9T}, but see
\citealp{2004A&A...418...25K}),
the central dominant elliptical in fossil groups
and other groups
often host
radio-loud AGN \citep{2003ApJ...592..129K,
2012AJ....144...48H, 
2017MNRAS.472.1482O,
2018AA...618A.126O,
2018Kolokythas,
2020MNRAS.497.2163P}.
Outflows from radio AGN may disturb and heat the surrounding medium
\citep{1993MNRAS.264L..25B,
2000ApJ...534L.135M,
2001ApJ...547L.107F,
2003ApJ...592..129K,
2003MNRAS.344L..43F,
2012NJPh...14e5023M,
2020MNRAS.496.1471K},
potentially suppressing star formation
\citep{2006MNRAS.365...11C, 
2010MNRAS.409..985D,
2015MNRAS.449.4105C, 2019SSRv..215....5W, 
2020ApJ...901..159C,
2021A&A...654A...8N,
2023A&A...676A..35D,
2024ApJ...962..196O}.
AGN feedback is routinely included in cosmological simulations
to reproduce the observed properties of galaxies 
\citep{2006MNRAS.365...11C, 
2006MNRAS.370..645B,
2014MNRAS.442..440C,
2018MNRAS.479.4056W,
2025arXiv250116983B}.

Depending upon evolutionary stage and star formation rate (SFR),
X-ray activity may be enhanced in compact groups 
compared to field galaxies 
\citep{1984ApJ...284L..29B, 2013ApJ...763..121D,
2013ApJ...764..175F, 2014ApJS..212....9T, 2016ApJ...817...95T,
2017MNRAS.472.1482O}.
Starburst galaxies tend to have 
large quantities of hot X-ray-emitting gas
\citep{2004ApJS..151..193S,
2012MNRAS.426.1870M,
2018AJ....155...81S,
2019AJ....158..169S}
as well as
enhanced populations of
high mass X-ray binaries (HMXBs)
\citep{2003MNRAS.339..793G,
2012MNRAS.419.2095M}
and ultra-luminous X-ray sources (ULXs) 
\citep{2011ApJ...741...49S,
2012AJ....143..144S}.  
ULXs are extra-nuclear X-ray point sources 
with L$_{\rm X}$ $>$ 
10$^{39}$~erg~sec$^{-1}$ 
which may be HMXBs,
intermediate mass black holes (IMBHs), or pulsars 
\citep{2011ApJ...741...49S,
2014Natur.514..202B,
2017ARA&A..55..303K,
2022MNRAS.509.1587W}.
In addition, in compact groups
the rapid passage of a galaxy through intragroup gas
may lead to shock heating of the gas.
The best-known
example of an X-ray-bright
intergalactic shock front in a compact group
is that in Stephan's Quintet 
\citep{1997AA...322...89P,
2001AJ....122.2993S,
2003AA...401..173T}.
In Stephan's Quintet, a fast-moving intruder galaxy collided
with tidal gas stripped out in earlier interactions 
\citep{2010ApJ...724...80R, 2012MNRAS.419.1780H},
producing a 40 kpc-long shock front
with an X-ray luminosity of 
about 10$^{41}$ erg~s$^{-1}$
\citep{2003AA...401..173T}.
Strong mid-IR H$_2$ lines confirm the presence of shocks
in this region 
\citep{2010ApJ...710..248C, 2006ApJ...639L..51A, 2017ApJ...836...76A}.
The compact group NGC 4410 is another transitional system,
with an extragalactic ridge of X-ray-emitting intragroup hot gas  
coincident with optical emission-line gas
\citep{2003AJ....126.1763S}, an HI tail
\citep{2000ApJ...541..624S},
and a radio-loud AGN
\citep{1986A&A...155..161H, 2000ApJ...541..624S}.

Shocks caused by direct face-on collisions between gas-rich disk galaxies
can also produce large quantities of hot gas outside the main bodies of the
galaxies.
The best-studied example is the Taffy galaxies (UGC 12914/5), a pair of
edge-on side-by-side disk galaxies 
connected by a bridge of radio
continuum emission \citep{1993AJ....105.1730C}.
This bridge produces
5 $\times$ 10$^{39}$~erg$^{-1}$ in diffuse X-ray light
\citep{2015ApJ...812..118A}.  
The bridge is also detected
in the 
mid-IR H$_2$ lines 
\citep{2012ApJ...751...11P},
and has
optical spectral signatures of shocks 
\citep{2019ApJ...878..161J}.
Star formation may be suppressed in the bridge
\citep{2022ApJ...931..121A}, perhaps due to enhanced turbulence
and shock heating \citep{2020MNRAS.492.4892Y, 
2021AA...647A.138V, 2024arXiv240911707Y}.
Head-on collisions between disk galaxies 
may also happen within compact groups,
for example, in the
compact group HCG 57 \citep{2025ApJ...979..240O}.
Shock heating during such collisions
may contribute to the diffuse hot gas seen in compact groups, and may help
suppress star formation.

In a survey of nine compact groups,
seven were found to have hot diffuse X-ray emitting gas 
\citep{2013ApJ...763..121D}.  
The hot gas morphology appears to be correlated with
interaction/merger stage, 
suggesting an evolutionary sequence. 
In early-stage systems, the hot
gas is strictly located within the disks of the galaxies;
in more evolved systems, hot gas is seen in tidal features, 
between the galaxies, or in a group halo.
In the 
\citet{2013ApJ...763..121D} sample,
the diffuse L$_{\rm X}$ from hot gas increases as specific
star formation rate (sSFR) decreases, and the fraction of
galaxies in the group that are elliptical or S0 increases
as the ratio of atomic gas mass to dynamical mass decreases.
In a set of 19 compact groups,
\citet{2014ApJ...790..132D} found that L$_{\rm X}$(gas)
is anti-correlated with the fraction of the baryons that are in interstellar
atomic hydrogen.   These trends suggest that as a compact group ages
and the spirals in the group evolve into early-type galaxies,
cold gas gets used
up in star formation and/or gets heated by shocks,
AGN feedback, or other processes,
producing an X-ray halo.

In this paper, we present new Chandra observations
of the compact group of galaxies 
known as 
IC 2431, and compare with archival data at other wavelengths.
IC 2431, also known as VV 645, UGC 04756, Mrk 1224, V1CG 340, 
IRAS 09018+1447,
MLCG 776, 
and CGCG 090-063, 
was first reported in Javelle's 1908 catalog of nebulae \citep{1908AnNic..11D...1J},
and was occasionally investigated in surveys of highly star-forming
galaxies 
(e.g.,
\citealp{1979Afz....15..201M, 1986ApJS...62..751M, 2000MNRAS.317...55S, 2011PASP..123.1011A,
2019MNRAS.482..560M,
2021MNRAS.506.3079K}).
However, it received
little individual attention until it was `rediscovered' a century
after Javelle by the Galaxy Zoo project \citep{2013MNRAS.435.2835W}
and noted as one of the most dramatic and
oddly-shaped merger systems in the Sloan survey 
\citep{2022AJ....163..150K}.
As discussed below, IC 2431
has some similarities to Stephan's Quintet, NGC 4410, and 
the Taffy galaxies.   
Assuming a Hubble constant of 72 km~s$^{-1}$Mpc$^{-1}$
and correcting for Virgocentric flow,
IC 2431 is at a distance of 207 Mpc.

In Section \ref{sec:IC2431} of this paper, 
we describe the available data,
including archival optical, ultraviolet, infrared, and radio data, as well
as the new Chandra X-ray data, and discuss
the morphology of the system.
A more detailed analysis of these data is provided 
in Section \ref{sec:analysis}, where we investigate the X-ray spectra
and use 
UV/optical/IR photometry
to derive 
star formation rates (SFRs), 
stellar masses, 
stellar population ages, 
and dust extinctions for various portions of the system.
In Section \ref{sec:comparison}, we compare and contrast 
IC 2431 to
other systems.  The evolutionary state of IC 2431 is discussed
in Section \ref{sec:discussion}.
A summary is provided in Section \ref{sec:summary}.

\section{Data} \label{sec:IC2431}

\subsection{Optical Images and Spectra} \label{sec:morphology}

\begin{figure}[ht!]
\plottwo{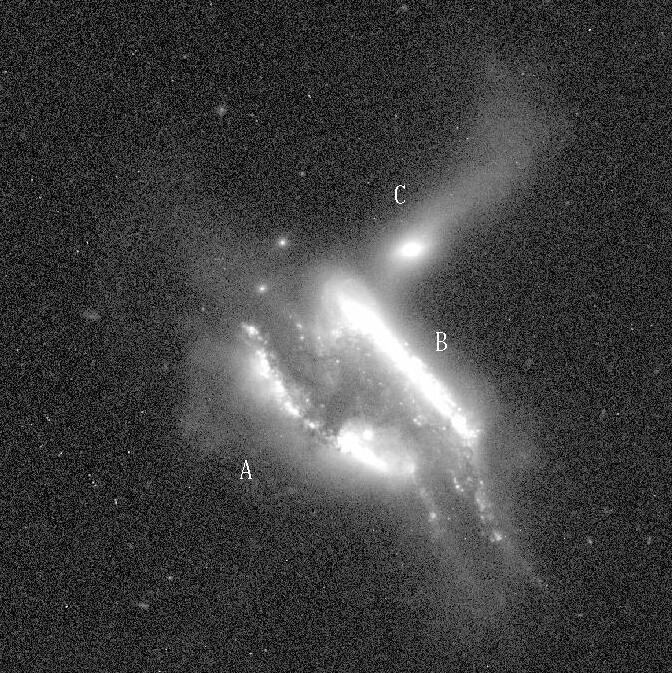}{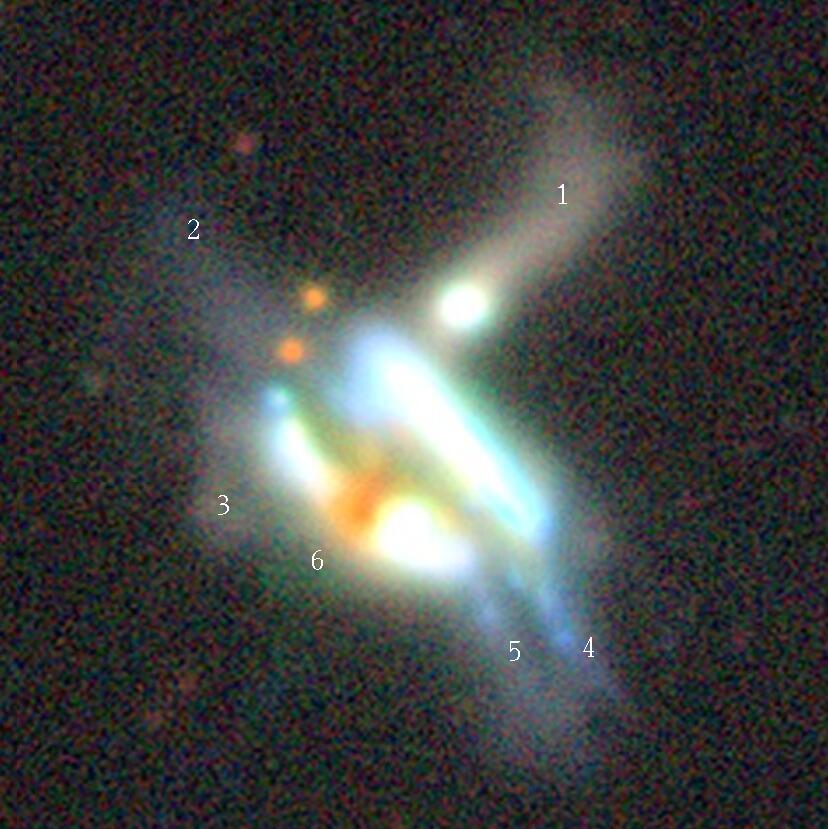}
\caption{
Left: the archival HST F606W image of the IC 2431 group.
Galaxies A, B, and C are marked.
The field of view is 1$'$.
North is up and east to the left.
Right: The DES Legacy Survey RGB image of IC 2431. Note the prominent
red band crossing galaxy A in the south.
The superimposed numbers identify six tidal features in the system.
\label{fig:HST}}
\end{figure}

IC 2431 was observed by the Advanced Camera for Surveys Wide Field Camera on the
Hubble Space Telescope (HST) in the red F606W filter
as part of a large HST snapshot survey (proposal ID 15445; \citealp{2022AJ....163..150K}).
We retrieved the processed HST image from the European
HST Science Archive\footnote{https://hst.esac.esa.int}.
This image has a reported 
registration rms of 0\farcs03 
based on astrometric stars 
from the Gaia DR3 catalog\footnote{https://gea.esac.esa.int/archive/}
\citep{2016A&A...595A...1G}.  This is consistent with our custom registration.
In the left panel of
Figure \ref{fig:HST}, we display this HST
image. 

In the Uppsala General Catalogue of Galaxies (UGC) \citep{1973ugcg.book.....N},
IC 2431
is listed as a quadruple system, and the NASA Extragalactic
Database (NED) lists four components, IC 2431 NED01 $-$ NED04.
Inspection of the HST image and archival
Dark Energy Survey (DES) Legacy grz images\footnote{https://www.legacysurvey.org/} 
\citep{2019AJ....157..168D}
as 
well as near-IR images 
(Section \ref{sec:UV_IR}),
however, suggests
that IC 2431
only contains three main galaxies,
all disk
galaxies.  We label these galaxies A, B, and C,
south to north,
as shown in 
the left panel of Figure \ref{fig:HST}.
Galaxy C is NED01 (SDSS J090434.55+143552.3), 
galaxy B is NED02, and galaxy A is 
NED03 and NED04 together.

The two main galaxies in IC 2431, Galaxy A and Galaxy B,
appear to be two distorted disks both viewed approximately edge-on,
and aligned
face-to-face.  The two galaxies are separated by about 8$''$ on the sky,
which corresponds to 8 kpc at the distance of IC 2431.
The probable reason that earlier studies 
counted more than three galaxies
in this group
is because of a prominent dust lane that bisects galaxy A.
At low spatial resolution
in blue images, galaxy A appears to be two galaxies.
In the higher resolution HST image (left panel, Figure \ref{fig:HST})
and in color DES Legacy images
(right panel, Figure \ref{fig:HST}),
however, the dust lane
is apparent,
projected in front of Galaxy A.
This dust lane may
be a tidal extension of Galaxy C, 
a distorted spiral arm of Galaxy B projected in front of Galaxy A, 
or
interstellar material pushed out of Galaxy B or Galaxy C by a head-on collision 
between galaxies.
This dust absorption is discussed further in the analysis below.

At least five and maybe six tidal features are present in this system,
labeled 1 $-$ 6 in 
the right panel of 
Figure \ref{fig:HST}.
These include:
1)  
a tidal tail extending from C to the northwest,
2) a tail extending from B to the northeast,
3) a curved tail bending to the southeast from the northern
end of A, 
4) a tidal tail extend to the southwest from
the southern end of the disk of B.  
5) another structure that runs parallel to feature 4,
extending southwest from the southern end of disk A.
Feature 5) may connect to disk A, or it may originate
in disk B and cross disk A.
6) The dust lane that crosses disk A may be a sixth tidal
feature, although that is uncertain.

The heliocentric velocities of Galaxies A, B, and C,
respectively, 
are 
14840 $\pm$ 36 km~s$^{-1}$,
14926 $\pm$ 39 km~s$^{-1}$, 
and 14870 $\pm$ 3 km~s$^{-1}$ 
\citep{2016ApJS..225...23S},
giving
a radial velocity dispersion for the group of only 31 km~s$^{-1}$.
Given the prominent tidal features visible in this system, this small
observed velocity dispersion
suggests that most of the relative motion of the galaxies
in the group are in the plane of the sky.
Galaxy C is classified as an optical AGN by \citet{2017ApJ...835..280L}.
Galaxy B is classified as an H~II region galaxy by 
\citet{1998ApJ...494..150S} based on optical spectroscopy.
A rotation curve for Galaxy B
was derived 
by \citet{1996ApJS..104..217S}
from a long slit optical
spectrum.   
This rotation curve rises steeply in the inner 3$''$ radius
then flattens,
with a total velocity width of 434 km~s$^{-1}$ and a derived
dynamical mass of 1.0 $\times$ 10$^{11}$~M$_{\sun}$.


\subsection{Atomic Hydrogen Gas}

IC 2431 was detected in the 21 cm HI line as part of the 
ALFALFA Arecibo HI Survey (source AGC 190759)
\citep{2018ApJ...861...49H}.
The IC 2431 group is unresolved with the Arecibo beam of 3\farcm8 $\times$ 3\farcm2.
The HI centroid is offset about 18$''$ north of the group center,
close to Galaxy C.
The HI flux of 0.57 $\pm$ 0.07 Jy~km~s$^{-1}$ corresponds to 
an HI mass of 6 $\times$ 10$^9$~M$_{\sun}$.
The HI heliocentric velocity is 14,871 km~s$^{-1}$
and the 
FWHM line width is 124 km~$s^{-1}$
\citep{2018ApJ...861...49H}.

\subsection{UV and IR Images} \label{sec:UV_IR}

\begin{figure}[ht!]
\plotone{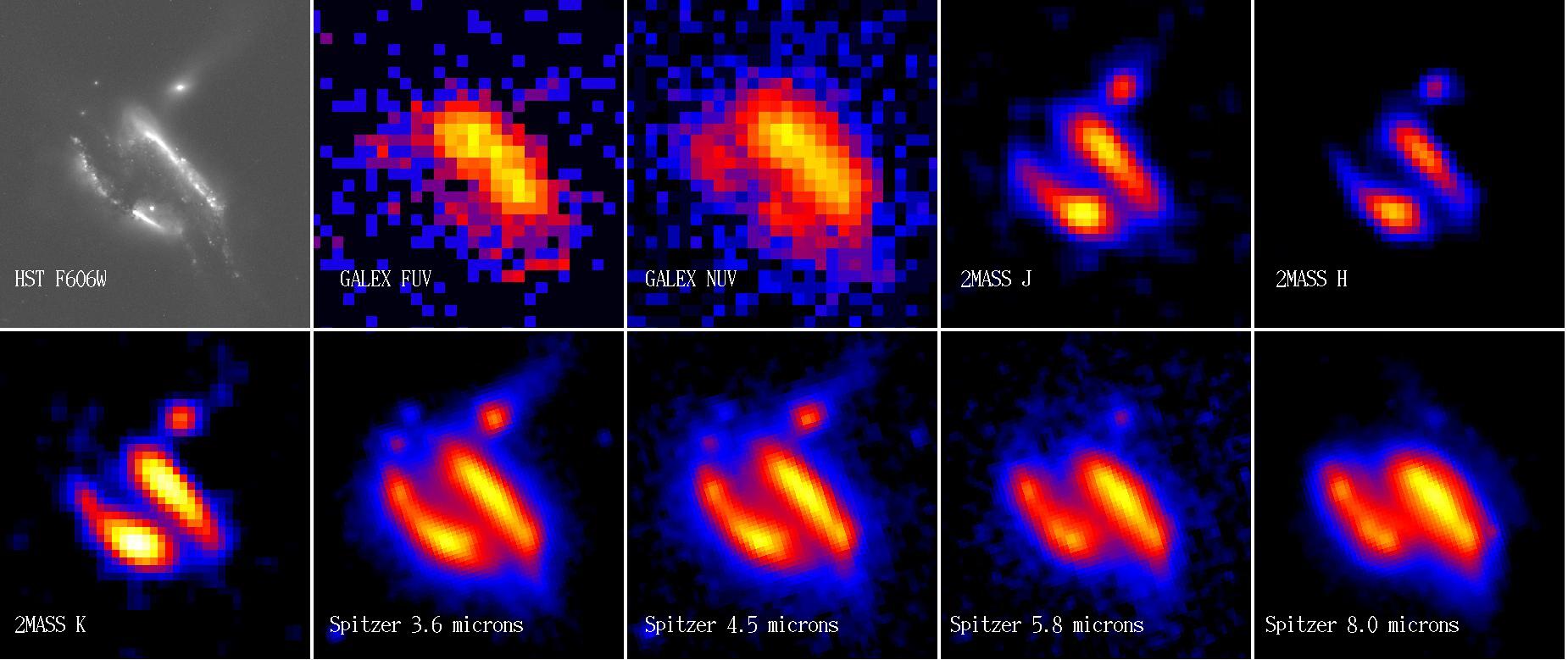}
\caption{
Images of IC 2431 at various wavelengths.
From left to right, top to bottom:
the images are:
HST F606W, GALEX FUV, GALEX NUV, 2MASS J, 2MASS H,
2MASS K, Spitzer 3.6 $\mu$m, Spitzer 4.5 $\mu$m,
Spitzer 5.8 $\mu$m, and Spitzer 8.0 $\mu$m.
The field of view is 0\farcm7, with north up and east to the left.
\label{fig:multi}}
\end{figure}

In Figure \ref{fig:multi}, we compare the optical HST image
of IC 2431 with far-UV and near-UV images from the GALEX 
All-Sky Survey \citep{2005ApJ...619L...1M},
near-infrared J, H, and K images from the 2MASS 
Atlas\footnote{https://irsa.ipac.caltech.edu/data/2MASS/docs/releases/allsky/doc/explsup.html}, 
and 
archival Spitzer Infrared Array Camera (IRAC) \citep{2004ApJS..154...10F}
images 
at 3.6 $\mu$m, 4.5 $\mu$m, 5.8 $\mu$m, and 8.0 $\mu$m.
The GALEX FUV and NUV bands have approximate FWHM
resolutions of 4\farcs0 and 5\farcs6, 
respectively\footnote{https://asd.gsfc.nasa.gov/archive/galex/Documents/MissionOverview.html},
the effective resolution of 2MASS is about 
4\farcs0 in the Atlas images\footnote{https://www.ipac.caltech.edu/2mass/overview/about2mass.html}, and 
Spitzer provided effective resolution of 1\farcs66, 1\farcs72, 1\farcs88, and 1\farcs98 at
3.6, 4.5, 5.8, and 8.0 $\mu$m, respectively\footnote{https://irsa.ipac.caltech.edu/data/SPITZER/docs/irac/}.
We also utilize 
co-added
images 
from the 
WISE Atlas\footnote{https://wise2.ipac.caltech.edu/docs/release/allsky/expsup/}
in the mid-infrared W1 (3.4 $\mu$m), W2 (4.6 $\mu$m),
W3 (12 $\mu$m), and W4 (22 $\mu$m) filters.
These WISE images have approximate FWHM point spread functions
of 8\farcs3, 9\farcs1, 9\farcs5, and 16\farcs8 in W1, W2, W3, and W4,
respectively\footnote{https://wise2.ipac.caltech.edu/docs/release/allsky/expsup/sec1\_4c.html}, thus the galaxies are not well-resolved with WISE.

Figure 
\ref{fig:multi} shows that 
Galaxy B contributes a large majority of the observed
light in the two UV bands.  
In contrast, in the 2MASS NIR images, 
Galaxy A is comparable to Galaxy B in brightness.
In the 2MASS images the southern portion of
Galaxy A is considerably brighter
than the northern portion.  
In the Spitzer 5.8 $\mu$m and 8.0 $\mu$m images, Galaxy B
is clearly brighter than Galaxy A; the two galaxies
are more comparable in the 3.6 $\mu$m and 4.5 $\mu$m images.
In the Spitzer images, the relative brightness of Galaxy C decreases with increasing
wavelength from 3.6 $\mu$m to 8.0 $\mu$m.

\begin{figure}[ht!]
\plotone{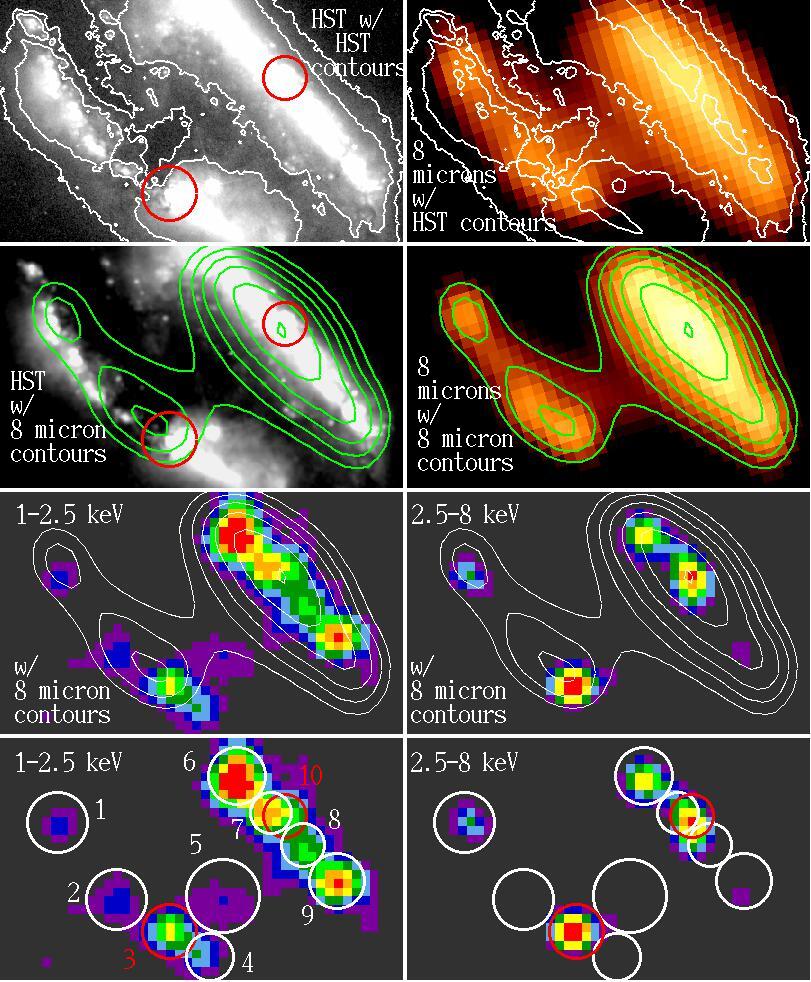}
\caption{
Zoomed-in views of Galaxy A and Galaxy B in various bands.
The field of view is 22\farcs2 $\times$ 13\farcs3, with north up and 
east to the left.
Top left: 
HST F606W image of IC 2431 with F606W contours 
of 
4, 5, and 20 $\times$ 10$^{-20}$~erg~s$^{-1}$~cm$^{-2}$~\AA$^{-1}$~
displayed.
These contours were selected to outline the dust 
features and mark the optically-bright areas.
The presumed galactic nuclei 
are marked in red.
Top right: Spitzer 8 $\mu$m image with HST contours overlaid.
Second row left: HST image with contours showing Spitzer
8 $\mu$m isophotes 
at levels of 20, 27, 36, 49, 69, 102, and 155 MJy~SR$^{-1}$.
Second row right: Spitzer 8 $\mu$m map with 8 $\mu$m contours.
Third row left: 
medium energy (1.0 $-$ 2.5 keV) Gaussian-smoothed Chandra map
with Spitzer 8 $\mu$m contours.
Third row right:
high energy (2.5 $-$ 8 keV) Gaussian-smoothed Chandra map
with Spitzer 8 $\mu$m contours.
Bottom left:
the smoothed Chandra medium-energy map,
with our ten targeted X-ray sources marked and labeled.
The assumed galactic nuclei for Galaxy A and Galaxy B, sources
\#3 and \#10, are circled and labeled in red.
Bottom right: the smoothed Chandra high energy 
map with our targeted X-ray sources marked.
\label{fig:spitzer_xray}}
\end{figure}

A zoomed-in comparison of the HST F606W image and the Spitzer 8 $\mu$m 
image is provided in the top two rows of Figure 
\ref{fig:spitzer_xray}.
The top two left panels display the HST image, while the top two right panels
show the Spitzer image. 
In the top row, 
contours from the 
HST image are superimposed, while
in the second row, contours from the Spitzer 8 $\mu$m image are 
overlaid.
Galaxy A looks very thin in the red HST 
image, suggesting an edge-on disk without a large bulge
component.
The 
position of the mid-infrared maximum in Galaxy A is coincident within the
uncertainties with the hard X-ray peak (see Section \ref{sec:chandra}), which
we assume is the galactic nucleus. 
The mid-infrared peak in Galaxy A lies just to the
south of the dust lane.   From this peak, a ridge of mid-IR emission
extends along the disk of Galaxy A to the northeast, to a secondary peak about 9$''$ (9 kpc) away
from the nucleus.
This ridge of mid-IR emission in the north
is slightly offset (about 0\farcs6 = 0.6 kpc) 
to the northwest of star forming 
regions visible in the optical.  
Along the ridge,  the mid-IR light is particularly bright where the dust lane
crosses the disk of Galaxy A,
implying strong star formation in that region.
Fainter diffuse 8.0 $\mu$m emission 
is seen extending 
towards Galaxy B, starting at the location 
where the dust lane crosses
Galaxy A.  This morphology
suggests a bridge of gas and dust connecting the two
galaxies. 
This mid-IR bridge 
is especially apparent in 
the 8 $\mu$m
contours
in Figure \ref{fig:spitzer_xray}.

South of the mid-IR peak in Galaxy A,
little mid-infrared emission is detected.
The mid-IR/optical spatial 
offsets in Galaxy A and the apparent mid-infrared bridge between the two galaxies
suggest star formation triggering caused by ram pressure stripping during a collision
between two galactic disks.   This possibility is discussed further
in Section \ref{sec:discussion}.

In Section \ref{sec:photometry}
of this paper, we extract UV/optical/IR photometry from these images.
In Sections \ref{sec:SFR} and \ref{sec:pop_syn}, we use this photometry to derive
SFRs, stellar masses, population ages, and extinctions for
various regions within IC 2431.

\subsection{Chandra Observations and Images} \label{sec:chandra}

A total of eight observations of IC 2431 were made with Chandra 
(Table \ref{tab:observations}).  All of these data were
obtained with the Advanced CCD Imaging Spectrometer (ACIS) S-array.
In the following analysis, we only use the data from the S3 chip on ACIS, since
IC 2431 was centered on S3 and the entire
group of galaxies fits within the 8\farcm3 $\times$ 8\farcm3
field of view of the 
ACIS-S3 chip.  
Basic data reduction was accomplished using
the Chandra Interactive Analysis of Observations (CIAO) software.
The data were reprocessed
using the chandra$\_$repro script, and the data were filtered
to
exclude all but grades 0, 2, 3, 4, and 6. 
Background regions on the S3 chip were selected to
exclude the IC 2431 system and bright background point sources.
The data were searched for background flares using the 
CIAO routine deflare with the sigma clipping routine.  
Only one of the eight datasets showed flare activity,
eliminating only 0.07 ksec of data. The final exposure times
are provided in Table  
\ref{tab:observations}.

\begin{deluxetable*}{crc}
\tablecaption{Chandra Observations \\ of IC 2431\label{tab:observations}}
\tablewidth{0pt}
\tablehead{
\colhead{Obs ID} & \colhead{Exposure} & \colhead{Date}\\
\colhead{} & \colhead{(ksec)} & \nocolhead{}
}
\startdata
26932 	& 	11.92 &  2024-01-29 \\
27202 	&  	11.85 & 2024-02-13 \\
27203 	& 	15.46 & 2023-01-16 \\
27673 	&  	15.73 & 2023-01-18 \\
29230 	&  	11.92 & 2024-01-29 \\
29231 	&  	18.29 & 2024-01-29 \\
29273 	&  	10.93 & 2024-02-13 \\
29274 	&  	9.70 & 	2024-02-16 \\
\enddata
\end{deluxetable*}

Point sources in the field were used to register each individual
dataset using Gaia and Dark Energy Survey positions.
For display purposes, after registration we combined the datasets 
using the CIAO routine reproject$\_$obs.
After Gaussian smoothing 
using the ds9 
software\footnote{https://sites.google.com/cfa.harvard.edu/saoimageds9},
the X-ray images are 
shown in the bottom two rows of Figure 
\ref{fig:spitzer_xray}.
The left panels give the
medium energy map (1.0 $-$ 2.5 keV), while the right panels
display the high energy map (2.5 $-$ 8 keV).
Contours from the Spitzer 8 $\mu$m image are overlaid on the X-ray maps in
the second-to-the-bottom row of 
Figure
\ref{fig:spitzer_xray}.

Because of the considerable amount of diffuse X-ray emission in IC 2431,
the CIAO point source detection routine {\it wavedetect} is not able to reliably
identify point sources in the system.  We therefore
selected the brightest X-ray sources by eye for further analysis,
using extraction apertures that range from 1\farcs2 $-$ 2\farcs0 based on the
appearance of the source.
These ten sources are marked 
in the bottom panels of 
Figure
\ref{fig:spitzer_xray}.
Source \#1 $-$ \#9 were selected as local maxima in the 1.0 $-$ 2.5 keV map,
while 
source \#10 was chosen based on a local maximum
in the high energy 2.5 $-$ 8.0 keV map.
Source \#10
does not have a discrete counterpart in the middle-energy map.
Note that our region for source \#10 overlaps with those of sources \#7 and \#8.

We assume that 
the X-ray source labeled \#3 
in Figure 
\ref{fig:spitzer_xray}
at 9h 4m 34.98s, 14$^\circ$ 35$'$ 35\farcs8
is the 
galactic nucleus of Galaxy A, as its position agrees with 
that of the 4.86 GHz radio
continuum maximum (see Section \ref{sec:radio}),
and it is
coincident within the positional uncertainties 
with the Spitzer mid-IR maximum and the 2MASS near-IR  
peak in Galaxy A.  
The hard X-ray source
labeled \#10 in
Figure 
\ref{fig:spitzer_xray},
at 9h 4m 34.54s, 14$^\circ$ 35$'$ 42\farcs32, 
is 
coincident with the mid-IR maxium in Galaxy B, 
thus we assume it is the nucleus of Galaxy B.
In addition to the two galactic nuclei and the overlapping source \#7,
two other sources, source \#1 in the north of Galaxy A and source \#6 in
the north of 
Galaxy B, also have $>$10 net counts in the high energy map.

In the medium-energy Chandra map,
four knots of emission are seen along the disk of Galaxy B
(regions \#6, \#7, \#8, and \#9).
As noted earlier, the apparent nucleus of Galaxy B, the hard source \#10, lies between
two sources visible in the medium-energy maps (sources \#7 and \#8),
but is not detected as a discrete source in the medium-energy
map.  No photons with energies less than 4 keV are detected from this
apparent
nuclear source, implying a high extinction.  
The X-ray spectra of 
these ten sources 
are
analyzed 
in Section \ref{sec:xray_spectra}.
Since the extraction aperture of source \#10 overlaps with that
of sources \#7 and \#8, the light from these sources will be blended
in the spectral analysis.

Over the lifetime of its mission, 
Chandra has become less sensitive to low energy X-ray photons
\citep{2022SPIE12181E..6XP}.
In the low energy range (0.5 $-$ 1 keV), only a few dozen 
photons are detected from IC 2431, mostly from X-ray sources 
\#5, \#6, and \#9, thus we do not display a low energy map.  
This drop in sensitivity is incorporated into our
calibration of the 
spectral fits (Section \ref{sec:xray_spectra}).

\begin{figure}[ht!]
\gridline{\fig{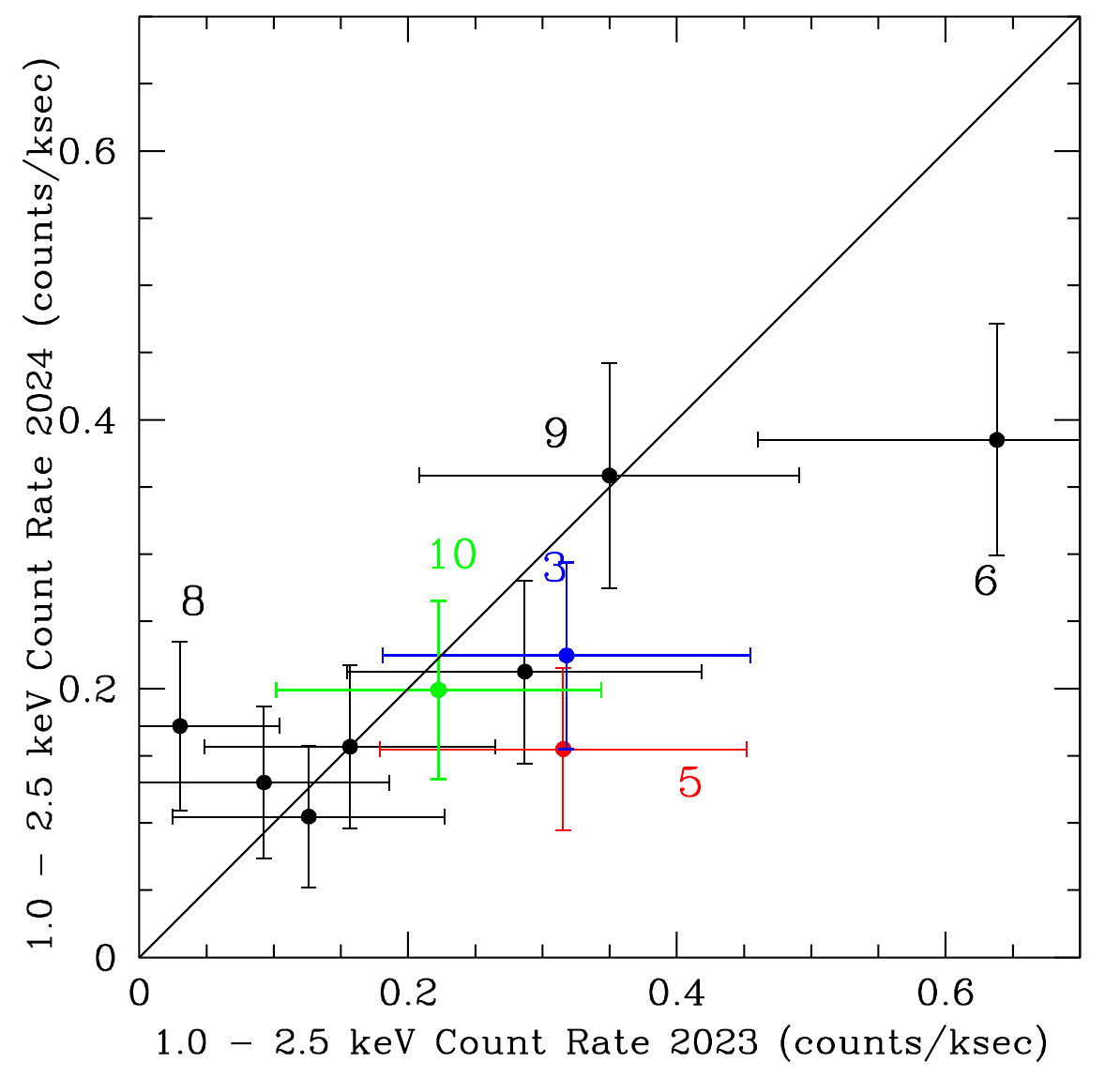}{0.35\textwidth}{(a)}
\fig{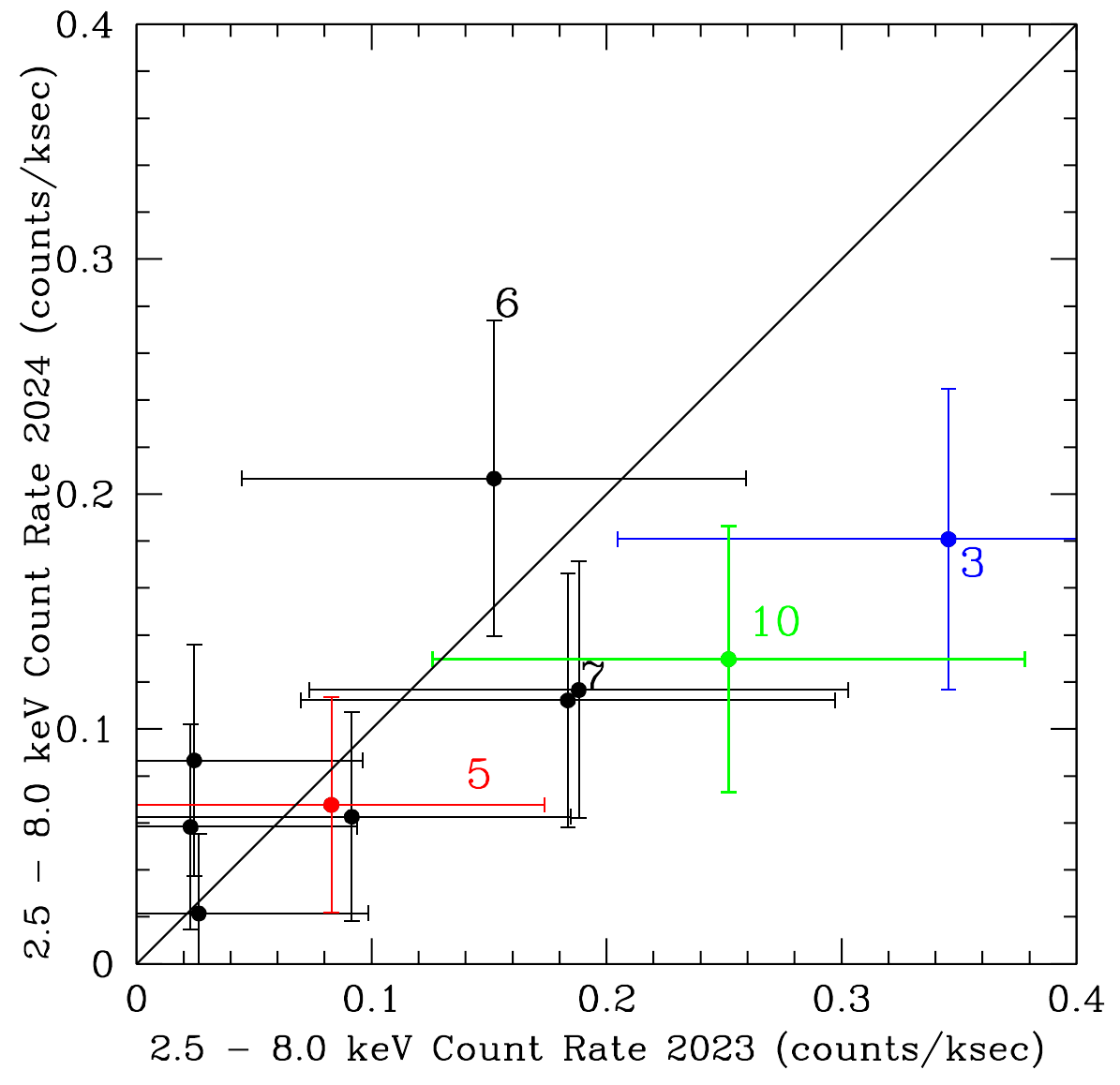}{0.35\textwidth}{(b)}
\fig{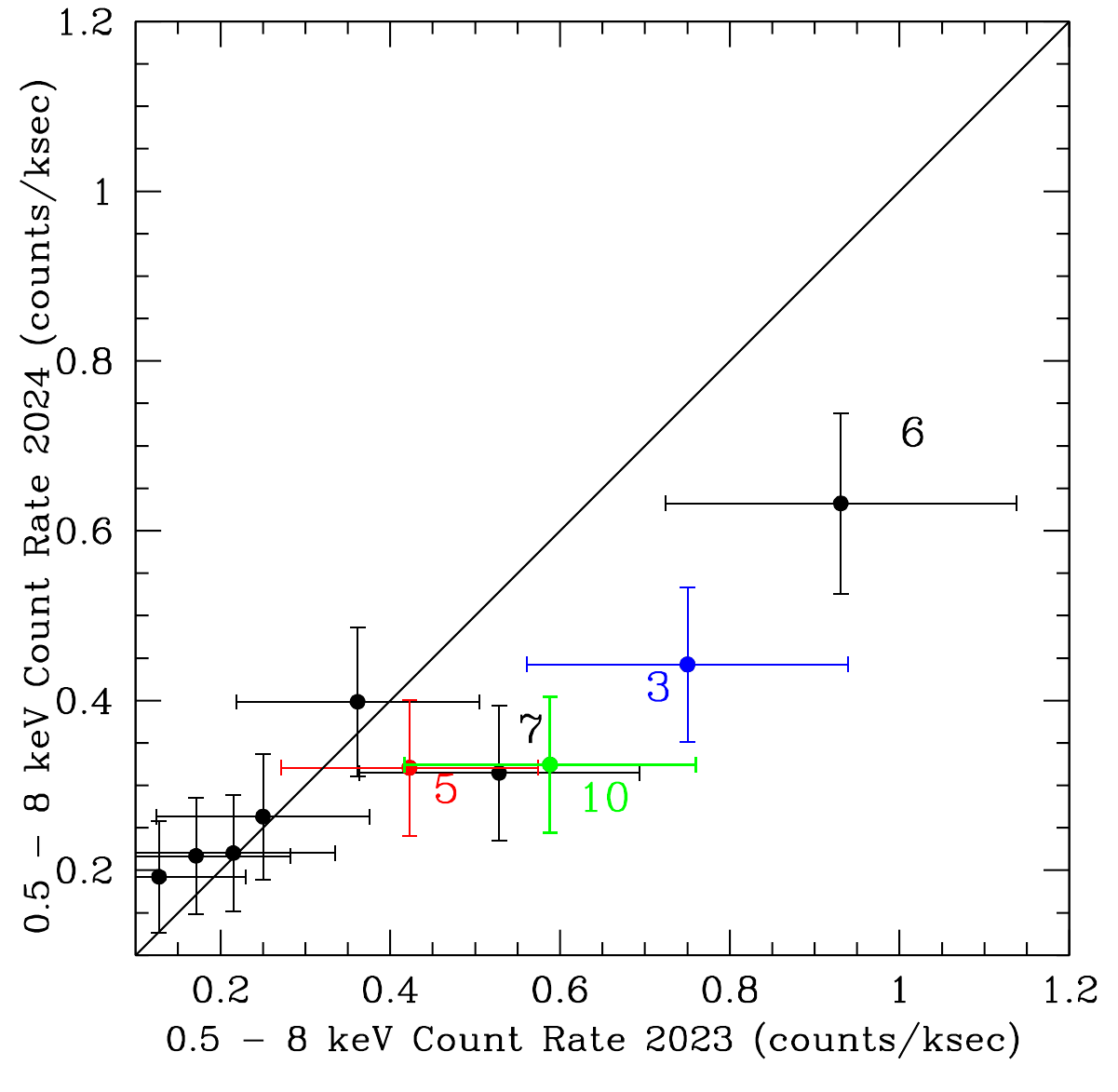}{0.35\textwidth}{(c)}
}
\caption{
The Chandra count rates for the 2024 observations vs.\ the
2023 dataset, for the ten X-ray sources in IC 2431. 
These were extracted using the CIAO routine {\it dmextract}.
Left: middle energy range (1.0 $-$ 2.5 keV).
Middle: high energy range (2.5 $-$ 8.0 keV).
Right: full range (0.5 $-$ 8.0 keV).
Source \#3 (the apparent nucleus of Galaxy A) is plotted in blue;
Source \#10 (the apparent nucleus of Galaxy B) is plotted in green.
Source \#5 (between the two galaxies) is plotted in red.
Some additional sources are labeled.
The uncertainties on these rates were calculated using 
\citet{1986ApJ...303..336G} statistics, i.e.,
$\sigma$ = 1 + $\sqrt{N + 0.75}$.
\label{fig:rates}}
\end{figure}

None of the X-ray sources in IC 2431 have sufficient counts for a reliable light curve.
However, we can test whether these sources are variable by splitting the
registered data into two sets and searching for differences.  
In Figure \ref{fig:rates}
we plot the count 
rates in 2024 vs.\ the count rates in 2023,
in the medium energy range 1 $-$ 2.5 keV (left panel), the high 
energy range 2.5 $-$ 8 keV (middle panel), and the full energy range 0.5 $-$ 8 keV (right panel).
Between 2023 and 2024,
the count rates of 
the apparent nuclei of both Galaxy A (source \#3) and Galaxy B (source \#10) dropped
1$\sigma$ or more
in both the high energy range
and the full 0.5 $-$ 8 keV band.
Source \#6 in the northern part of Galaxy B also shows
some evidence of variability 
at medium energies and in the full
0.5 $-$ 8 keV band.
In the 1.0 $-$ 2.5 keV range, 
source \#8 shows a 1$\sigma$ increase,
and source \#5 a 1$\sigma$ decrease.

\subsection{Radio Continuum Maps} \label{sec:radio}

IC 2431 was mapped at 4.86 GHz (C band; 6 cm) and 1.49 GHz (L band; 20 cm) 
by the Karl G. Jansky Very Large Array (VLA)
\citep{1992ApJS...81...49S,
1996ApJ...460..225C}. 
Two resolved sources are seen in the published VLA maps, associated
with Galaxy A and Galaxy B.
The quoted positions of the peak radio emission
for the two galaxies agree within the uncertainties with the nuclear
X-ray positions.
According to 
\citet{1996ApJ...460..225C},
the nuclear radio source in Galaxy A may be variable, making it
a possible AGN.

We have retrieved these VLA data from the archives, and re-reduced the data using the Common Astronomy Software Application (CASA; \citealt{2007ASPC..376..127M}) version 6.5.4.9. Standard flagging and calibration techniques were followed. Imaging was performed with the \verb|tclean| task, selecting a Briggs weighting at a robust value of one. Additionally, self-calibration was applied to the 4.86\,GHz image. The final 4.86\,GHz image has a local rms of 
0.06\,mJy~beam$^{-1}$, with a restoring beam of 0\farcs61 $\times$ 0\farcs41. 
There are 
two 1.49\,GHz datasets, from 1987 and 1991.  These have local rms noise levels of 0.4 and 0.3\,mJy~beam$^{-1}$, and 
restoring beams of 1\farcs71 $\times$ 1\farcs45 and 2\farcs19 $\times$ 1\farcs40, respectively. 
In the following analysis
we use the 1991 dataset since it is deeper.

\begin{figure}[ht!]
\plotone{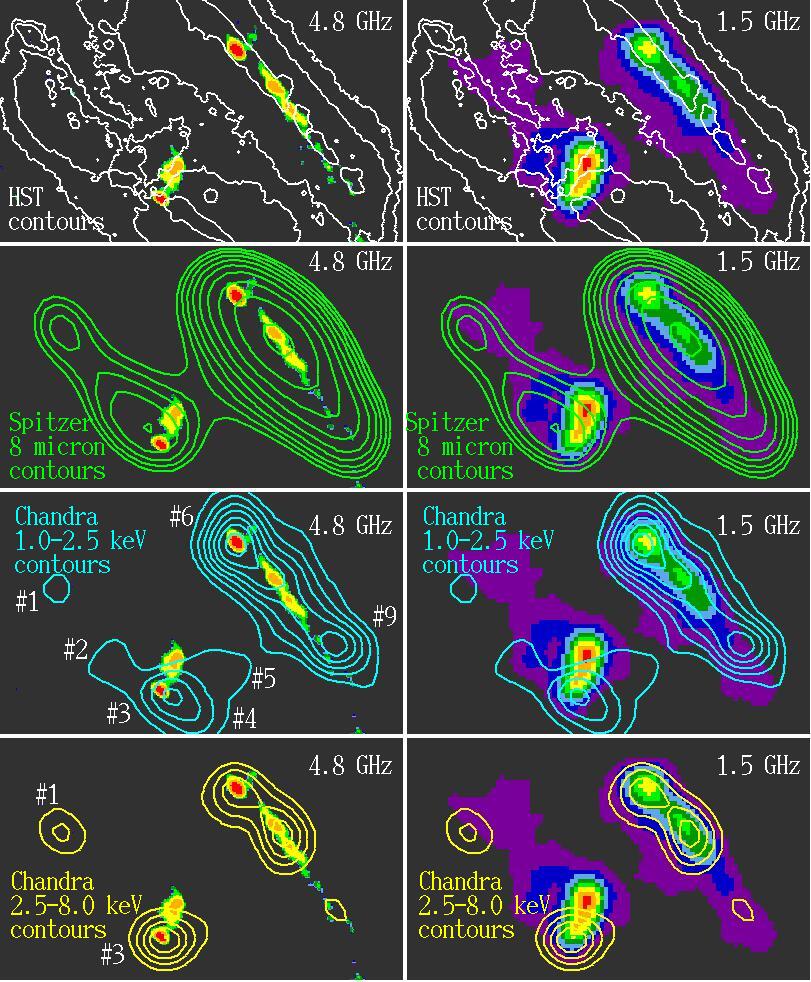} 
\caption{
Comparison of the radio continuum maps with data at other
wavelengths. 
Left panels: the 4.86 GHz map.  Right panels: the 1.49 GHz map.
The contours in the first row are from the HST F606W map,
selected to delineate the
dust feature crossing Galaxy A.
The contours in the second row are from the Spitzer 8 $\mu$m map,
chosen to emphasize the mid-infrared bridge and the star-forming regions
in Galaxy A.
The contours in the third row are from the Chandra 1.0 $-$ 2.5 keV map,
selected to outline 
some of the 
X-ray sources mentioned in the text.  
Sources \#1, \#2, \#3, \#4, \#5, \#6, and \#9 are labeled 
in the left panel of the third row.
The contours in the fourth row are from the Chandra 2.5 $-$ 8.0 keV map.
X-ray sources \#1 and \#3 are labeled in the left panel.
The spatial resolution for the 1.49 GHz map is 2\farcs19 $\times$ 1\farcs40,
and for the 4.86 GHz map is 0\farcs61 $\times$ 0\farcs41.
The field of view is 
22\farcs2 $\times$ 13\farcs3, with north up and east to the left.
\label{fig:radio_multi}}
\end{figure}

We compare 
the VLA maps of Galaxy A and Galaxy B with the 
HST, Spitzer, and Chandra maps 
in 
Figure \ref{fig:radio_multi}
(Galaxy C is undetected in the VLA maps).
The 4.86 GHz and 1.49 GHz images are presented in the left and right columns of panels,
respectively, of 
Figure \ref{fig:radio_multi}.
The superimposed contours are from the HST F606W image (top row),
the Spitzer 8 $\mu$m map (second row), the 1.0 $-$ 2.5 keV Chandra image
(third row), and the 2.5 $-$ 8.0 keV Chandra map (bottom row).
The brightest source in Galaxy A at 4.86 GHz is coincident with the hard X-ray
peak, the presumed galactic nucleus (bottom left panel).
An intriguing ridge-like/jet-like feature extends 
about 4$''$ to the northwest of this nucleus at both radio 
frequencies.  This feature is discussed
further below. 

In the 1.49 GHz map, a larger but fainter 
filament 
of diffuse emission extends about 
8$''$ to the northeast of this
radio ridge/jet.  
This radio filament is offset from the disk
of Galaxy A as seen in the optical (top right panel), and is slightly
northwest of the star-forming region in the north
of Galaxy A seen in the Spitzer map (right panel, second from top).
This radio structure is roughly coincident
with a dust feature seen in the optical map
(top right panel).

Galaxy B has diffuse 1.49 GHz 
emission spread out over
about 10$''$ 
along the galactic disk.   
The extent of this diffuse
emission
approximately 
matches that of the 8 $\mu$m emission.
The brightest region in Galaxy B at both radio frequencies 
is not the apparent nucleus, but instead is close to X-ray source \#6
in the north.
Another radio knot is visible in the southwest of Galaxy B,
approximately coincident with X-ray source \#9 
in the medium energy Chandra map.
Sources \#6 and \#9 may be intense star-forming regions.  

\begin{figure}[ht!]
\plotone{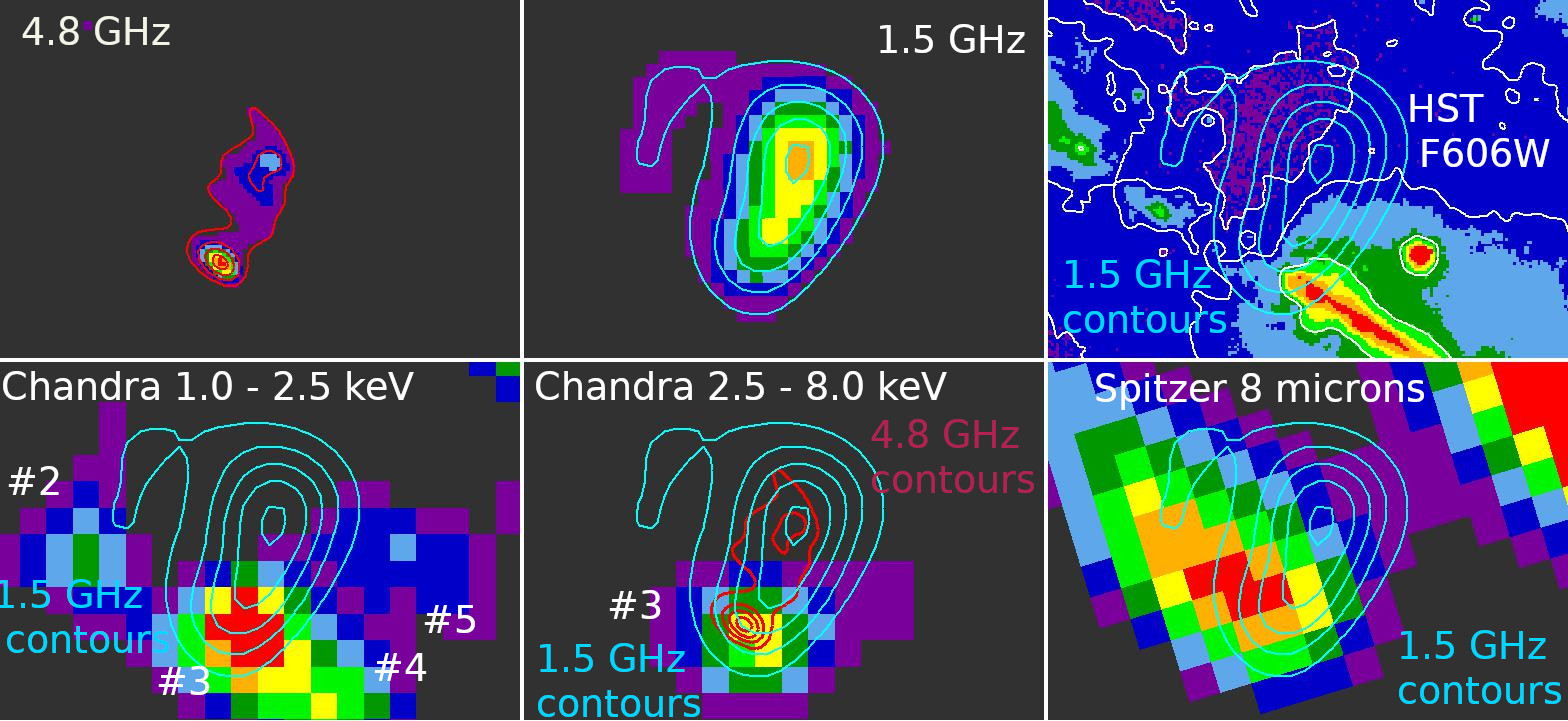}
\caption{
A multi-wavelength comparison of the region around the nucleus of Galaxy A.
Top left: the 4.86 GHz map.  Top middle: the 1.49 GHz map.  Top right: the HST F606W image, with 
1.49 GHz contours (cyan), and contours from the HST image itself (white contours).
The contours on the HST image are selected to outline dust features; the dust lane
is visible in the upper middle of this picture, color-coded in purple.
Bottom left: the medium-energy Chandra map, with 1.49 GHz contours.
X-ray sources \#2, \#3, \#4, and \#5 are marked.
Bottom middle: the high-energy Chandra map, with contours from both 1.49 GHz (cyan) and 
4.86 GHz (red).
X-ray source \#3 is marked.
Bottom right: The Spitzer 8 $\mu$m image, with contours from the 1.49 GHz map.
The field of view is 9\farcs5 $\times$ 6\farcs6, with north
up and east to the left.
\label{fig:nucA}}
\end{figure}

In Figure \ref{fig:nucA}, we zoom in on the nuclear region of Galaxy A,
and compare the two radio maps with maps at other wavelengths.
In the 4.86 GHz map (top left panel), the brightest source in Galaxy A
is a point source coincident with the apparent nucleus as 
seen in the X-ray, source \#3 (bottom middle panel).
Extending about 3$''$ $-$ 4$''$ to the northwest of this nucleus in the radio maps
is the 
ridge/jet, which has a maximum about 2$''$ 
away from
the nucleus.   
This feature is approximately 
perpendicular to the galactic disk as seen in the HST data
(top right panel).
The peak brightness of this ridge lies about 3$''$ east of 
X-ray source \#5 (bottom right panel).
The radio ridge/jet is approximately aligned along the 8 $\mu$m bridge
(bottom right panel).
In the 1.49 GHz maps,
the ridge is brighter
than the nucleus, while in the 4.86 GHz map,
the nucleus is brighter. 
This indicates different radio spectral indices 
for the nucleus and the ridge, as discussed below.
With the exception of the nucleus itself, the radio ridge/jet is
anti-coincident with the X-ray emission in the region (bottom left panel).
In the 1.49 GHz map (top middle panel),
fainter diffuse
emission is seen 
extending about 4$''$
to the northeast of the radio ridge.
At 4.86 GHz, the integrated flux for the ridge plus nucleus is 
5.3 $\pm$ 0.2 mJy (spread over 13.4 beams) and the point source
at the nucleus is 1.04 $\pm$ 0.06 mJy~beam$^{-1}$. 
At 1.49 GHz, the integrated flux of this region is 19.9 $\pm$ 0.6 mJy
(spread over 3.2 beams).  Our approximate estimate
for the nucleus at 1.49 GHz is 3.4 $\pm$ 0.3 mJy~beam$^{-1}$,
but this is uncertain because of blending with the extended emission
from the ridge.  At 1.49 GHz, the source is not centrally-peaked;
there is enhanced flux near the nucleus but also at the end of the ridge.
This morphology is what is expected if there is an active AGN
and a jet interaction with 
interstellar gas downstream.   This possibility is 
discussed further in Section 
\ref{sec:radioAGN}.

\begin{figure}[ht!]
\plotone{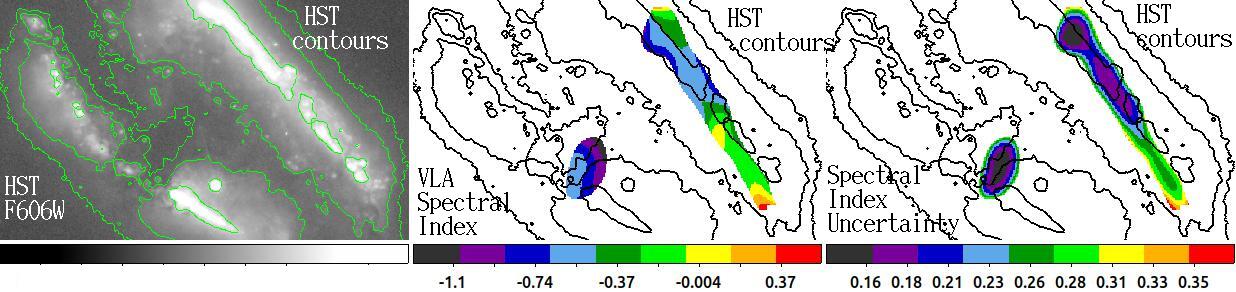}
\caption{
Left: the HST F606W map.  Middle: the VLA spectral index map,
defined by flux density S$_{\nu}$ $\propto$ $\nu$$^{\alpha}$.
Right: the uncertainty in the radio spectral index.
Contours from the HST map are overlaid on all of the images.
The field of view is 
22\farcs2 $\times$ 13\farcs3 with north up and east to the left.
\label{fig:spectral_index}}
\end{figure}

We also constructed a radio spectral index map 
(Figure \ref{fig:spectral_index}). 
The 4.86\,GHz image was convolved with the beam of the 1991 1.49\,GHz image. Next, using the CASA task \verb|regrid|, we re-gridded the 1.49\,GHz image to match the 4.86\,GHz image. Each image is then masked to their respective $3\sigma$ thresholds and, finally, we created a two-point spectral image map.
We 
define the radio spectral $\alpha$
by flux density S$_{\nu}$ $\propto$ $\nu$$^{\alpha}$. 
In Figure \ref{fig:spectral_index}, we compare the spectral index (middle
panel) with the 
HST F606W image (left panel), and a map of the uncertainty in $\alpha$
(right panel).  

\begin{figure}[ht!]
\plotone{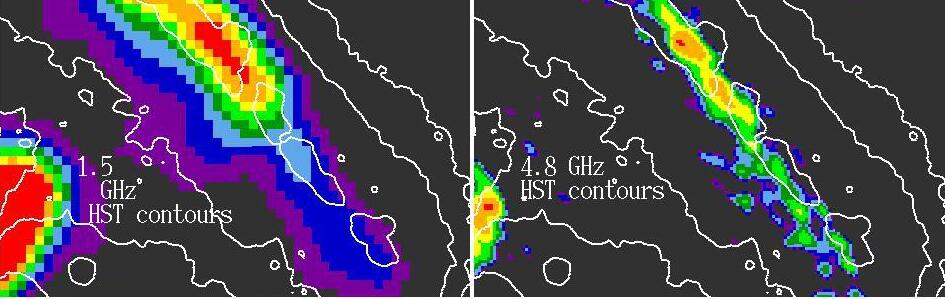}
\caption{
The southern portion of Galaxy B.
Left: 
The VLA 1.49 GHz map, with HST contours superimposed.
Right: The VLA 4.86 GHz map, with HST contours superimposed.
The field of view is 
about 18$''$ 
$\times$ 8$''$, with north up and east to the left.
\label{fig:south_of_B}}
\end{figure}

The nuclear region of Galaxy A has a spectral index of about $-$0.6, with
$\alpha$ decreasing (steepening) 
along the ridge/jet to a minimum of around $-$1.1 near the NW tip.
The northern part of Galaxy B has a spectral index of approximately $-$0.6; this
increases (flattens) to $\alpha$ = $-$0.1 to $-$0.2 in the southern part of Galaxy B.
In the middle panel of Figure 
\ref{fig:spectral_index}, a striking offset is seen in the southern
part of Galaxy B between the radio continuum and the optical,
with the sources seen in the optical shifted to the north of the 
radio emission. This offset is seen more clearly in Figure 
\ref{fig:south_of_B},
a zoomed-in picture of the southern portion of Galaxy B.

\section{Analysis} \label{sec:analysis}

\subsection{UV/Optical/IR Photometry} \label{sec:photometry}

\begin{figure}[ht!]
\plotone{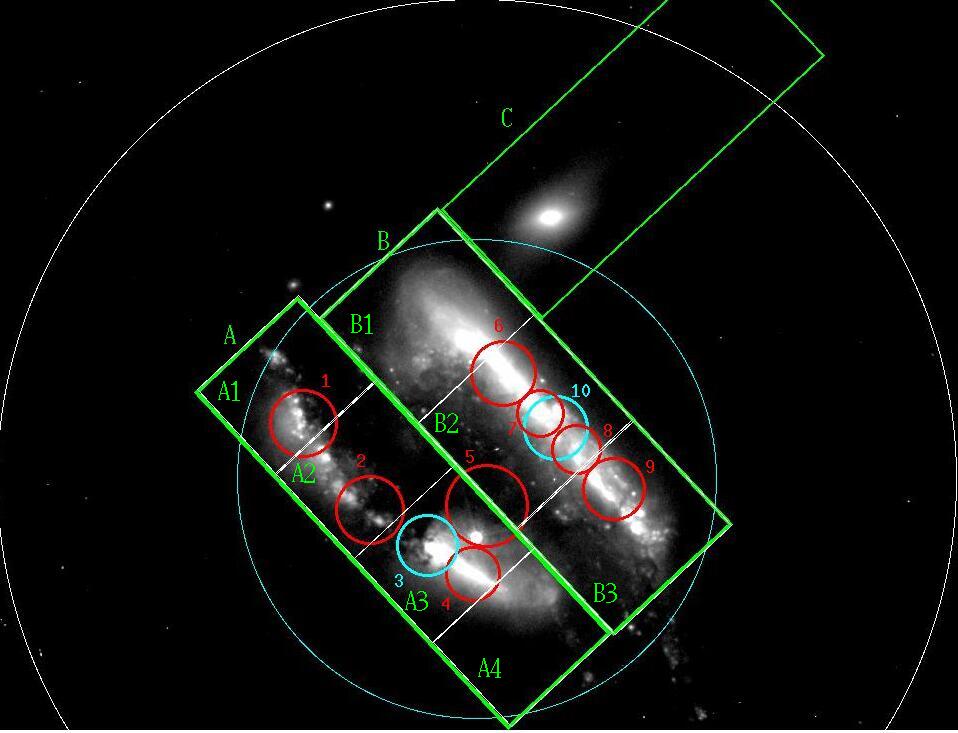}
\caption{
The HST F606W image with various regions overlaid.
The large rectangular regions in green are the regions in 
which the fluxes for Galaxies A, B, and C were extracted.
The regions for Galaxy A and B are divided into smaller 
rectangular regions as shown by the white lines (four regions for Galaxy A and three
for Galaxy B).  These smaller regions are labeled 
A1, A2, A3, A4, B1, B2, and B3.
The 12$''$ radius circle
that includes most of the Chandra X-ray flux is shown in cyan.   
A larger circle
with a 24$''$ radius (in white) includes most of the tidal tails as seen
in the optical.
The ten smaller regions marking the X-ray sources are shown in red
or (for the galactic nuclei) in cyan.
The field of view is approximately 47$''$ $\times$ 36$''$, with north
up and east to the left.
\label{fig:boxes}}
\end{figure}

We extracted fluxes 
in the GALEX, SDSS, DES, HST, 2MASS, and Spitzer bands
in three large rectangular regions 
enclosing
Galaxy A, Galaxy B, and Galaxy C
(see Figure 
\ref{fig:boxes}).
For sky determination, we selected large regions 
off the galaxies away from foreground stars. 
For all of the available filters plus the four WISE bands,
we also extracted fluxes for 
the full IC 2431 system within a 24$''$
radius aperture, and for
the combined A+B system within a 12$''$ radius
system.  
The 
photometry 
was corrected for Galactic absorption 
as in Schlafly and Finkbeiner (2011) and Yuan et al. (2013).
The final fluxes for each region
are given in 
Appendix A,
along with the central
coordinates of the region, the major and minor axes of the
rectangular region, and its position angle on the sky.

As shown in Figure \ref{fig:boxes},
we sub-divided the Galaxy A and Galaxy
B rectangles into four and three slices, respectively.
These regions are large enough for reliable photometry
of extended sources with GALEX.
Photometry for these smaller rectangular 
regions are also given in Appendix A.

In Sections \ref{sec:SFR} and 
\ref{sec:pop_syn}, we use these fluxes to derive SFRs, stellar masses, and 
extinctions.   

\subsection{Star Formation Rates and Dust Absorption from UV and IR Fluxes, vs.\ Radio Continuum Fluxes} \label{sec:SFR}

From our 22 $\mu$m and FUV fluxes in the 24$''$ radius aperture, 
we derive a total SFR for IC 2431
of 
37 M$_{\sun}$~yr$^{-1}$ 
using the \citet{2011ApJ...741..124H} formula assuming a Kroupa initial
mass function (IMF; \citealp{2003ApJ...598.1076K}).
We also derive an FUV extinction A$_{\rm FUV}$ = 2.3
from the \citet{2011ApJ...741..124H} relation.
This implies
A$_{\rm V}$ = 0.92, using the \citet{2000ApJ...533..682C}
extinction law,
and E(B$-$V) = 0.30, using A$_{\rm V}$/E(B$-$V) = 3.1.  
From this, we infer a hydrogen column density of 
N$_{\rm H}$ = 1.7 $\times$ 10$^{21}$~cm$^{-2}$ using the 
\citet{1978ApJ...224..132B}
relation of N$_{\rm H}$/E(B$-$V) = 5.8 $\times$ 10$^{21}$ cm$^{-2}$~mag$^{-1}$.

The individual galaxies that make up IC 2431 are not resolved
in the WISE 22 $\mu$m image, so we cannot derive SFRs
for the three galaxies separately using WISE data.
However, the galaxies are resolved in the Spitzer 8 $\mu$m image.
The 8 $\mu$m luminosity of star-forming galaxies is also
correlated with SFR \citep{2005ApJ...632L..79W,
2008ApJ...686..155Z}, 
so can be used as a rough measure
of SFR, however, there are some complications.
First, one needs to correct for contributions from the
underlying stellar continuum in the Spitzer 8 $\mu$m band 
\citep{2004ApJS..154..253H}.
Second, in low metallicity systems such as dwarf galaxies,
the polycyclic aromatic hydrocarbon (PAH) features which dominate
the broadband Spitzer 8 $\mu$m filter tend to be weaker for the
same SFR \citep{2000NewAR..44..249M,
2005ApJ...628L..29E, 
2008ApJ...678..804E,
2006ApJ...639..157W,
2007ApJ...668...87W}.
This should not be a problem for IC 2431, however, since the 
stellar masses
of the two main galaxies are relatively high (see below),
and therefore they are not dwarf galaxies.
Third, PAH features can be destroyed in star forming regions
with intense UV radiation fields 
\citep{1999AJ....118.2055D, 
2018ApJ...864..136B, 
2023A&A...678A.129B}.
A fourth issue is that, in regions with strong interstellar shocks,
the mid-infrared H$_2$ S(5) and S(4) emission lines at 6.91 $\mu$m and
8.03 $\mu$m, respectively, may contribute to the 
broadband Spitzer flux in the 8 $\mu$m filter \citep{2010ApJ...710..248C}.
A fifth issue is that the 7.7 $\mu$m PAH feature shifts with
redshift relative to the broadband Spitzer 8.0 $\mu$m bandpass,
thus ideally a K-correction should be applied to the data to 
account for this shift.   
The latter three issues are discussed further
in Section \ref{sec:spitzer_colors}.

To estimate the SFR from the Spitzer 8 $\mu$m luminosity,
SFR$_{8~{\mu}m}$, 
we use the relation from 
\citet{2008ApJ...686..155Z} 
after converting to a Kroupa IMF:
SFR$_{8~{\mu}m}$
= 1.04 $\times$ 10$^{-43}$ L$_{8~{\mu}m}$,
where 
the SFR is in M$_{\sun}$~yr$^{-1}$, 
L$_{8~{\mu}m}$ = $\nu$L$_{\nu}$ in erg~s$^{-1}$,
$\nu$ is the frequency in Hertz, 
and L$_{\nu}$ is the monochromatic luminosity per frequency in units
of erg~s$^{-1}$~Hz$^{-1}$ after correction for
the stellar continuum.  We assume that the flux density
of the stellar
continuum contribution to the 8 $\mu$m band, F$_{8~{\mu}m}$(stars),
is equal to 
0.232 F$_{3.6~{\mu}m}$
\citep{2004ApJS..154..253H},
where F$_{3.6~{\mu}m}$ is the flux
density in the Spitzer 3.6 $\mu$m filter.
We combine SFR$_{8~{\mu}m}$ with the unobscured
SFR seen in the GALEX FUV band, SFR$_{\rm FUV}$,
defined as SFR$_{FUV}$ = 4.47 $\times$ 10$^{-44}$L$_{FUV}$
\citep{2011ApJ...737...67M, 2011ApJ...741..124H}, where L$_{FUV}$
is 
$\nu$L$_{\nu}$ in the FUV.
The total
composite SFR formula from the FUV and 8 $\mu$m together
is thus
SFR$_{\rm total}$ = 
4.47 $\times$ 10$^{-44}$ (L$_{\rm FUV}$ + 2.33 L$_{8~{\mu}m}$).
We use these formulae to calculate 
SFR$_{\rm total}$, as well as the implied extinction
E(B$-$V) and estimates of N$_{\rm H}$.  
The total SFRs and absorption estimates for the individual galaxies as well
as the smaller rectangular areas in IC 2431 are given 
in columns 2 $-$ 4 of 
Table \ref{SFR_table}.
In Section 
\ref{sec:pop_syn}
we compare the above estimates
of N$_{\rm H}$ with values obtained from population synthesis
modeling.

For all of our targeted regions in IC 2431,
SFR$_{8~{\mu}m}$ is considerably larger than 
SFR$_{\rm FUV}$, implying relatively high absorption.
The lowest absorption is seen
towards Galaxy C, while 
the three slices in the northern part of Galaxy A (regions A1, A2, and A3)
have the highest inferred absorption, N$_{\rm H}$
= 2.4 $-$ 3.1 $\times$ 10$^{21}$ cm$^{-2}$.  
According to these calculations, the SFR of Galaxy A
is about 38\% of that of Galaxy B.
The total SFR for the full 24$''$ radius region of
39 M$_{\sun}$~yr$^{-1}$ obtained
from the 8 $\mu$m + FUV measurements agrees well with 
that derived from 
22 $\mu$m + FUV.    
This agreement argues that the 8 $\mu$m is a reasonably
good measure of the SFR in this system.

\citet{1991MNRAS.253..485R}
report 60 $\mu$m and 100 $\mu$m
flux densities for IC 2431 from the Infrared Astronomical Satellite (IRAS)
of 4.43 Jy and 7.63 Jy, respectively, extracted from the IRAS
Point Source Catalog \citep{1988iras....7.....H}.
Refining these values using the xscanpi 
software\footnote{https://irsa.ipac.caltech.edu/applications/Scanpi/},
we obtain fluxes of 4.4 Jy and 5.43 Jy at 60 $\mu$m and 100 $\mu$m,
respectively.
This implies a far-infrared luminosity (42.5 $-$ 112.5 $\mu$m)
of 2.8 $\times$ 10$^{11}$ L$_{\sun}$, using the 
\citet{1985ApJ...298L...7H} relation.  Converting to
total infrared (3 $-$ 1000 $\mu$m) luminosity
using the method of 
\citet{2002ApJ...576..159D}
gives L(IR) 
of 5.3 $\times$ 10$^{11}$ L$_{\sun}$. 
Converting this to a
SFR using the relation of 
\citet{2011ApJ...737...67M}
yields a SFR of 80 M$_{\sun}$~yr$^{-1}$.
This is 2.2 times larger than our global estimate of the SFR from the
UV plus mid-IR.

\citet{1992ApJS...81...49S} compared the radio and far-infrared 
luminosities of
30 starburst galaxies, including IC 2431.  
Of these 30 galaxies, IC 2431 has the lowest
far-infrared to radio ratio, a factor of $\sim$ 5 times less
at both 6 cm and 20 cm
than the mean for the other galaxies. 
Compared to a sample of other luminous infrared galaxies,
\citet{1996ApJ...460..225C} find a FIR to 20 cm ratio
for IC 2431 that is low by about a factor of five.
This apparent radio excess from IC 2431 is especially intriguing given
that the radio continuum from this system is resolved;
for both Galaxy A and Galaxy B, 
the integrated light at 6 cm is about twice that
of the peak emission 
\citep{1996ApJ...460..225C}.
A majority of the excess radio emission from IC 2431 is apparently coming
from Galaxy A. 
Comparing with our SFRs for the two
galaxies obtained from the UV plus 8 $\mu$m fluxes,
the radio to SFR ratios for Galaxy A are about twice that of Galaxy B.
This supports the idea that Galaxy A hosts a radio AGN, or some
other process has enhanced the radio continuum emission
relative to that expected from star formation alone.

\input SFR_table.tex

\subsection{UV/Optical/IR Spectral Energy Distributions}

\begin{figure}[ht!]
\gridline{\fig{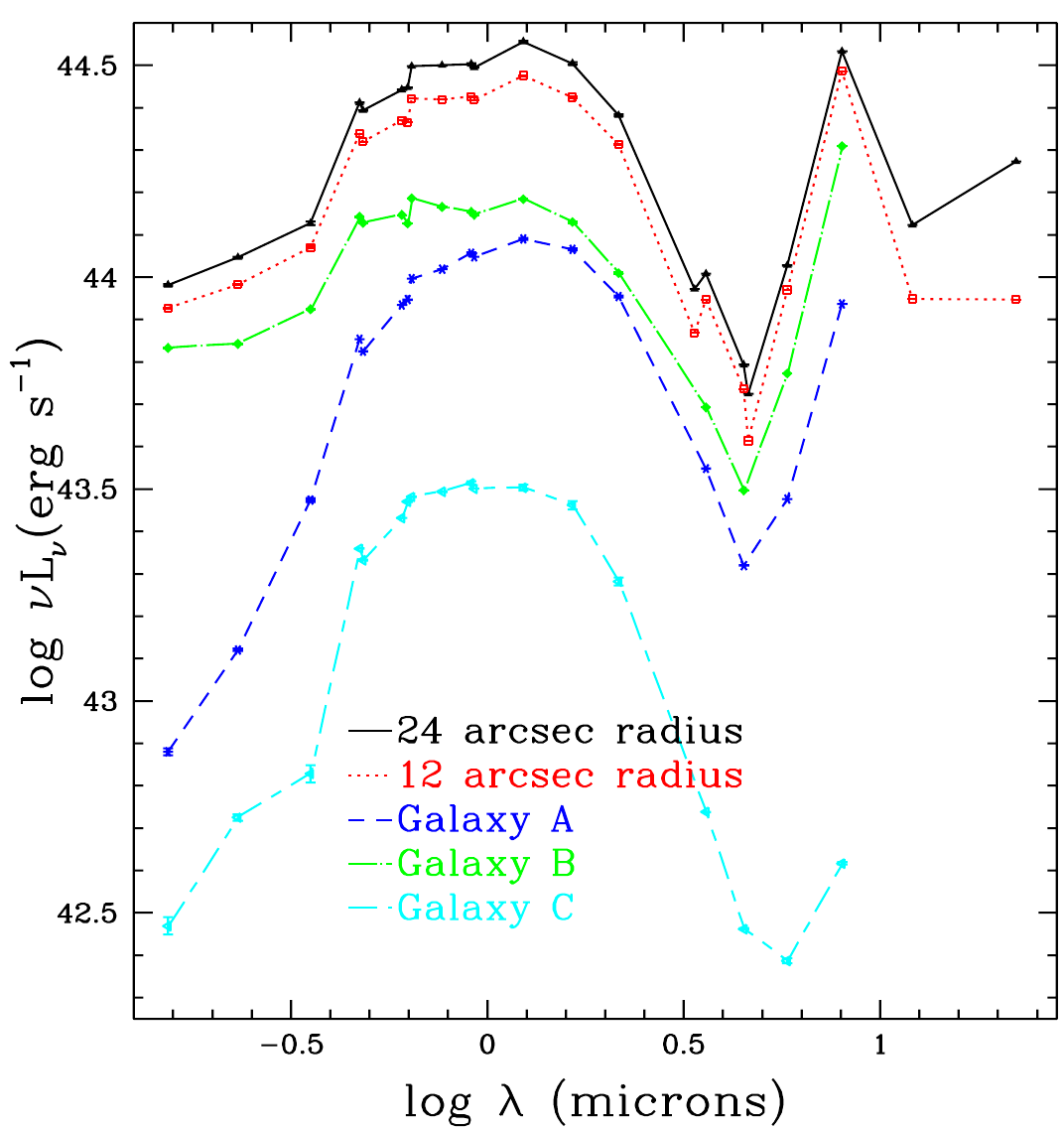}{0.35\textwidth}{(a)}
\fig{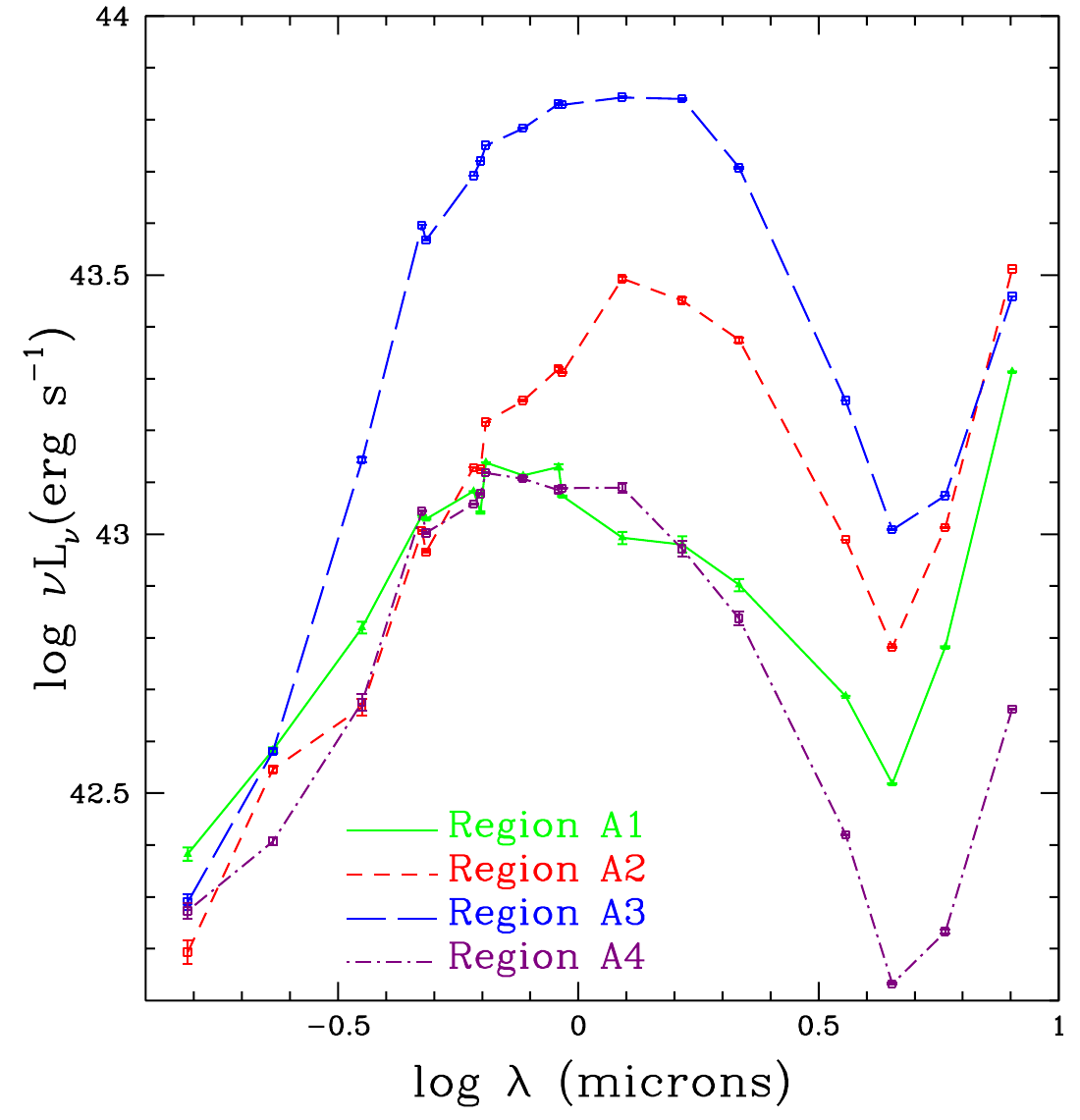}{0.35\textwidth}{(b)}
\fig{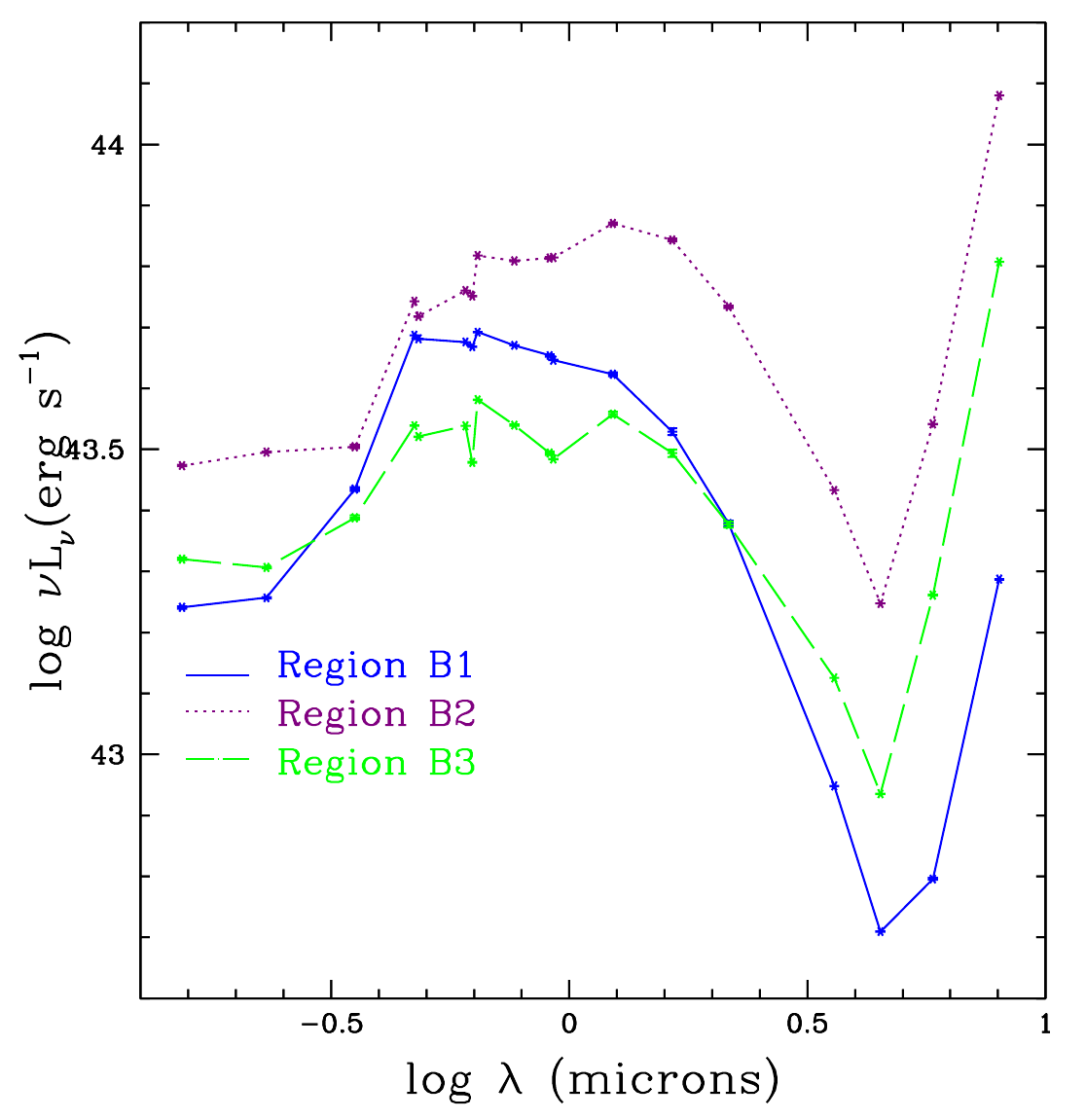}{0.35\textwidth}{(c)}
}
\caption{
Left: Spectral energy distributions for the full IC 2431 system,
compared to the SEDs of the individual galaxies
(see Table
\ref{Photometrytab}).   
Middle: Spectral energy distributions of the four rectangular regions
within 
Galaxy A, as identified in the caption.  From north to south in the galaxy, they are regions A1, A2,
A3, and A4.
Right: Spectral energy distributions of regions B1, B2, and B3
in IC 2431
(see the Appendix for the photometry).
\label{fig:SEDs}}
\end{figure}

In the left panel of Figure \ref{fig:SEDs}, we provide spectral energy distribution (SED) plots 
for the three galaxies in IC 2431. The SEDs of Galaxy A and Galaxy B differ
dramatically; Galaxy B has significantly more UV and blue light
than Galaxy A, although the near-IR fluxes of the two galaxies are similar.  
Galaxy B is also much brighter in the mid-IR than Galaxy A.
The SED of Galaxy C suggests a quiescent but relatively unobscured
stellar population, with proportionally little mid-infrared and UV compared to
the visible.  

The middle panel of Figure 
\ref{fig:SEDs} gives SED plots for the four
rectangular regions in Galaxy A.
Region A3, which contains
the nuclear region, 
is by far the brightest of the four Galaxy A regions 
in the optical and near-infrared,
and is proportionally fainter
in the UV and mid-IR. This implies a large older stellar population
and proportionally fewer young stars.
In contrast,
the most northern region, region A1, is the brightest of the four
regions in the UV and is the bluest
in the optical, and is also quite bright at 8 $\mu$m, but is
faint in the near-IR.
This suggests very recent star formation, but not a large underlying
older stellar population.
Region A2 (which contains part of the dust lane)
is quite bright in the near-IR and mid-IR, but fainter in the UV and short-wavelength
optical, implying a lot of dust extinction.  
Region A4, the most southern region, has a SED that peaks in the optical,
but has proportionally far less mid-IR than the other regions.   This
indicates that Region A4
is more quiescent.

The right panel of Figure \ref{fig:SEDs} compares the SEDs for the three 
rectangular 
regions
in Galaxy B.   
The northern region B1 is brightest in the optical, drops with increasing wavelength to 4.5 $\mu$m,
then increases somewhat at 8 $\mu$m. 
The SEDs of 
both the middle region B2 and the southern region B3 peak at 8 $\mu$m
and are faintest at 4.5 $\mu$m, but B2 is much brighter in the 2MASS bands.
This suggests that B2 includes the galactic bulge, and B3 has a younger stellar
population than B1.  The northern part of Galaxy B is more quiescent than the south.
These differences are discussed further below.

\subsection{Population Synthesis Modeling} \label{sec:pop_syn}

We fit the SEDs for the various regions 
in IC 2431
to population synthesis models using the 
Code Investigating GALaxy Emission
(CIGALE) software \citep{2019A&A...622A.103B}.   
The framework we use to establish
these parameters follows
that of 
 \citet{2023AJ....165..260D}.
We used single stellar populations from \citet{2003MNRAS.344.1000B},
dust emission models from 
\citet{2014ApJ...780..172D},
and the 
\citet{2000ApJ...533..682C}
modified dust law 
with the UV bump amplitude set to zero.
We assumed solar metallicity.  We used
the \citet{2003ApJ...586L.133C} IMF, which produces almost identical
SFRs as the Kroupa IMF \citep{2011AJ....142..197C, 2012ARA&A..50..531K}.
For the star formation history,
we used the `delayed star formation history with a
late burst' module in CIGALE, which can include
either a recent burst or a recent quenching episode.

CIGALE calculates a 
Bayesian-like estimate of various parameters and their uncertainties
based on a likelihood-weighting scheme involving the $\chi$$^2$ value for each
grid point.  
Since the nominal statistical uncertainties in our photometry
are likely underestimates of the actual uncertainties,
we use CIGALE's default across-the-board flux uncertainty of 10\%,
which is combined in quadrature with the 
listed uncertainties.   
In columns 5 $-$ 13 of 
Table \ref{SFR_table}, we 
provide the CIGALE fits for several parameters,
including 
the SFR, the stellar mass,
the age of the late burst/quench, 
the ratio of the SFR after vs.\ before the burst
(r$_{\rm SFR}$, a measure of the relative intensity of the starburst), 
the B $-$ V color excess E(B$-$V) for the stars and, separately, E(B$-$V) for the emission lines.
We use the GALEX, SDSS, 2MASS, and Spitzer photometry in these fits. 
The SFRs obtained from CIGALE are averaged over the last 100 Myrs, thus, as expected,
they agree fairly well with 
the SFRs derived from the FUV plus 8 $\mu$m fluxes, which average over
similar timescales.
More details on the CIGALE fitting and additional parameters are
provided in Appendix B.

Galaxy B has the highest sSFR of the galaxies in the group, 
and Galaxy C the lowest.   
Among the smaller rectangular regions measured, the southernmost region in Galaxy B, region B3,
has the highest sSFR, followed closely by A1 in the north of Galaxy A.
Region A3, which includes the galactic 
bulge of Galaxy A, has the lowest sSFR of all the smaller regions measured.  
Region A3 also has the lowest pre/post-starburst SFR ratio of all of the regions
measured, r$_{\rm SFR}$ = 0.28 $\pm$ 0.24, implying recent quenching
rather than a starburst.  
In contrast, 
regions A1, A2, B2, and B3 all have high 
r$_{\rm SFR}$ of 6 $-$ 8 $\pm$ 2, 
indicating a starburst,
with 
region B3
having the highest ratio.
The northern region in Galaxy B, B1, and the southern region
in Galaxy A, A4, have r$_{\rm SFR}$ $\sim$ 1, meaning little
change in the SFR over the last 100 Myrs.
All of the regions identified as starbursts (A1, A2, B2, and B3)
have similar burst ages of 30$-$55 $\pm$ 35 Myrs,
with A2 being the youngest and A1 the oldest.

In Table 
\ref{SFR_table}, we convert the CIGALE E(B$-$V) values to 
N$_{\rm H}$ using the 
\citet{1978ApJ...224..132B} relation.
The extinctions found by CIGALE are reasonably consistent with those implied
by the 8 $\mu$m to FUV flux ratios.  CIGALE found the lowest dust 
absorption for the northern
region of Galaxy B, region B1, and the highest extinction for region A2, 
which covers part of the dust lane.
In Section \ref{sec:xray_spectra}, we compare
these estimates of N$_{\rm H}$ to the X-ray absorptions inferred from
the Chandra spectra.

\subsection{Chandra X-Ray Spectral Analysis} \label{sec:xray_spectra}

\subsubsection{Global Galaxy Fits}

\begin{figure}[ht!]
\gridline{\fig{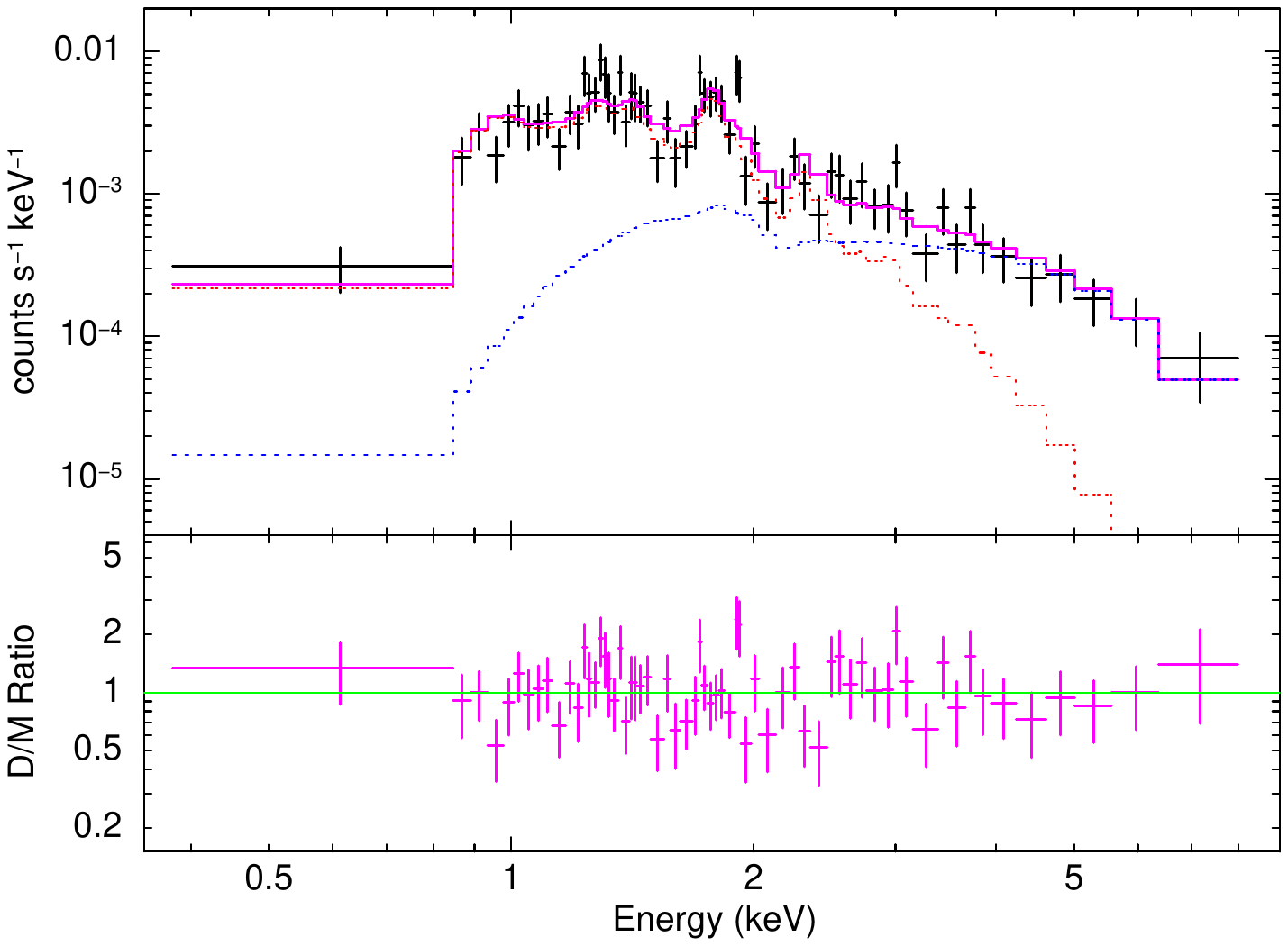}{0.35\textwidth}{(a)}
\fig{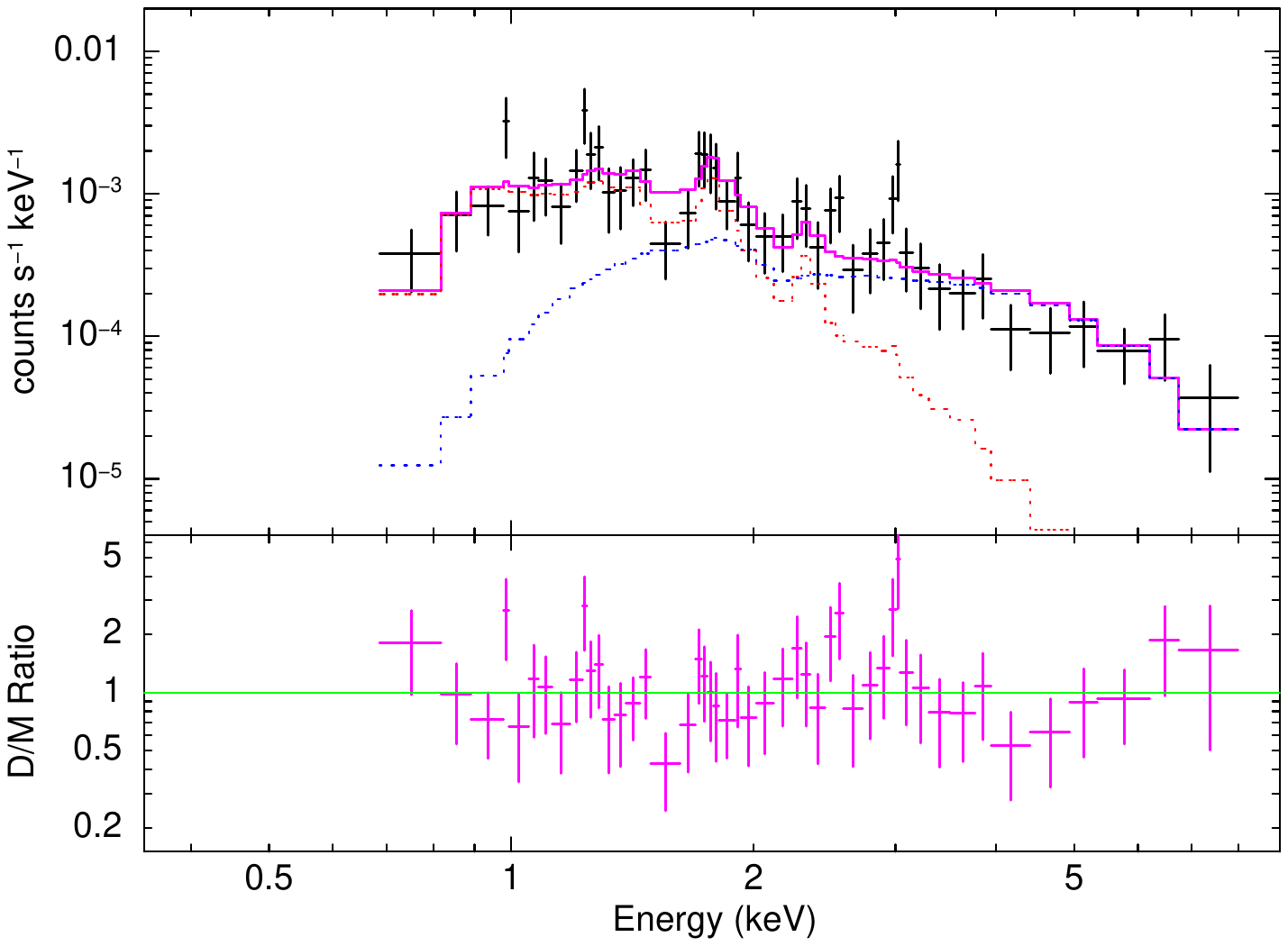}{0.35\textwidth}{(b)}
\fig{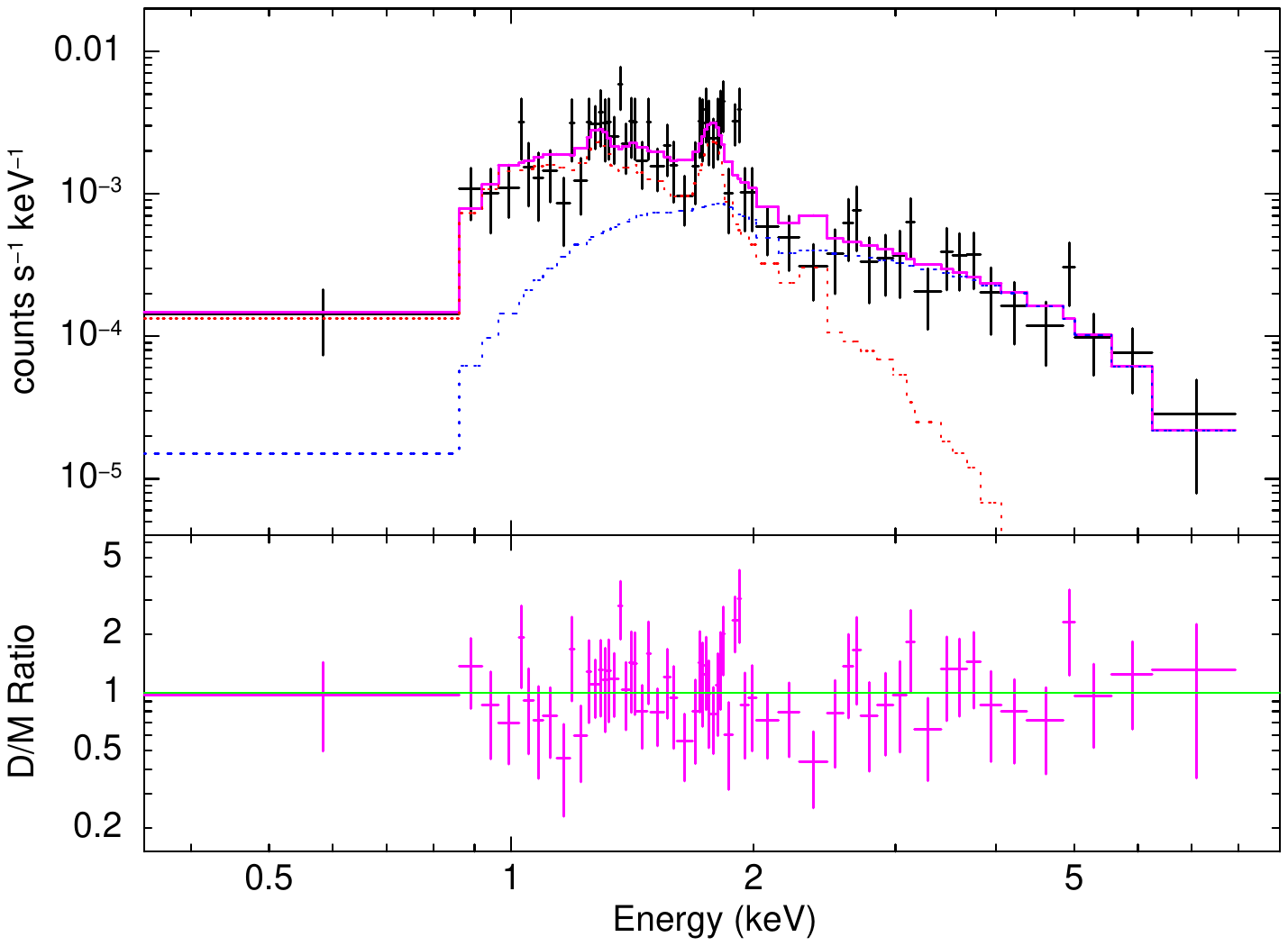}{0.35\textwidth}{(c)}
}
\caption{
Top panels:
The Chandra spectra for the diffuse X-ray emission
in the IC 2431 group within
a 12$''$ radius (left), 
Galaxy A (middle) and Galaxy B (right).
The best-fit 2-component model is overlaid as a solid magenta curve.
The red
dotted
component dominating at lower energies is the APEC component;
the blue dotted dashed component dominating at high energies is
the power law component.  
Bottom panels: the data/model ratio.
For plotting purposes only, the spectra for Galaxy A and Galaxy B
are rebinned to a minimum of S/N = 2 per bin, and the spectrum
for the 12$''$ radius is rebinned to $>$2.7 per bin.
\label{fig:xspec_12arcsec}}
\end{figure}

We used the CIAO routine specextract to extract and 
combine the Chandra X-ray spectra within 
various
regions in IC 2431.
Figure \ref{fig:xspec_12arcsec} displays 
the Chandra spectrum for the full system within a 12$''$
radius region (left panel), the rectangular region around Galaxy A (middle panel),
and the rectangular region enclosing Galaxy B (right panel).
We used the xspec software
to fit these spectra to two components, a power
law and a thermal APEC (Astrophysical Plasma Emission Code) function.
In these fits,
we fixed the redshift to 0.0497, the metallicity to solar, 
and the Galactic absorption to
4 $\times$ 10$^{20}$~cm$^{-2}$
(Dickey and 
Lockman 1990)$\footnote{from COLDEN at https://cxc.harvard.edu/toolkit/colden.jsp}$.
We used the \citet{2000ApJ...542..914W} abundance table.
We limited the fit to the energy range 0.5 $-$ 7 keV.
In these fits, 
we fit for a single value of the internal absorption, 
the photon index $\Gamma$ of the power law component, 
and the temperature of the thermal gas.

Results of the fitting are given 
in Table \ref{xspectab}.
Because of the relatively low number of counts from this system,  
we report Cash statistics (C-statistics; 
\citealp{1979ApJ...228..939C}) for these fits, along with the degrees of freedom
(DOF).
Our best-fit models
are overlaid on the data in Figure 
\ref{fig:xspec_12arcsec}, 
along with the best-fit thermal and power law components.
The total observed 
Chandra 
flux in the 12$''$ radius region 
is 9.9 $\pm$ $^{0.1}_{0.4}$ $\times$ 
10$^{-14}$ erg~s$^{-1}$~cm$^{-2}$.  
In the global spectrum and the spectrum
for Galaxy B, the Si XIII 1.87 keV feature is seen,
as is the Mg XI 1.343/1.352 keV feature.
The 0.8 keV Fe L complex is also visible.
In the global spectrum and in Galaxy B,
the thermal component dominates at energies below 2.5 keV, while
the power law component dominates at higher energies.
In Galaxy A, the power law component dominates down to about 1.5 keV.

For the system as a whole,
the best-fit temperature of the APEC component, kT = 0.92 
$\pm$ $^{0.13}_{0.58}$ keV,
is relatively well-constrained, but the photon index $\Gamma$
of the power law component is not (1.06 $\pm$ $^{0.78}_{0.91}$).
The best-fit model for the system as a whole
gives a high internal N$_{\rm H}$ of
7.6 $\pm$ $^{2.1}_{2.6}$ $\times$ 10$^{21}$ cm$^{-2}$.
For Galaxy A alone, 
the best-fit N$_{\rm H}$ = 5.8 $\pm$ $^{3.6}_{4.6}$ 
$\times$ 10$^{21}$ cm$^{-2}$ agrees well 
with the CIGALE results for line emission of 
N$_{\rm H}$ = 4.7 $\pm$ 0.16
$\times$ 10$^{21}$ cm$^{-2}$.  The best-fit temperature for
Galaxy A, 0.85 $\pm$ $^{0.16}_{0.48}$ keV, is moderately-well constrained,
though $\Gamma$ = 1.05 $\pm$ $^{0.64}_{0.85}$ is not.
This $\Gamma$, and the $\Gamma$ for the system as a whole,
are significantly smaller than typical for
AGNs and HMXBs.  AGNs usually have $\Gamma$ in the range 1.6 $-$ 2.0
\citep{1994MNRAS.268..405N,
2000ApJ...531...52G,
2003A&A...412..689P, 2006A&A...451..457T}, while
luminous HMXBs typically have $\Gamma$ = 1.8 $\pm$ 0.1
\citep{2002ApJ...577..738K, 2004ApJS..154..519S, 2022ApJ...930..135L}.
Variations
in absorption and X-ray spectral shape
across Galaxy A may be responsible
for these unusually low values of $\Gamma$.

\input roberto_global.tex

For Galaxy B alone, 
the best-fit 
N$_{\rm H}$ 
of 9.0 $\pm$ $^{3.8}_{5.0}$ $\times$ 
10$^{21}$ cm$^{-2}$ is 2.3 $\times$ the 
value from CIGALE for line emission.
This suggests that the sources of X-ray emission are more deeply
embedded on average than typical star-forming regions dominating
the UV/optical/IR emission.
The temperature for Galaxy B, 0.58 $\pm$ $^{0.35}_{0.21}$ keV,
is moderately-well constrained, and is lower than for Galaxy A.
The best-fit $\Gamma$ for Galaxy B, $\Gamma$ = 1.80 $\pm$ $^{0.71}_{0.60}$,
is consistent with values expected for HMXBs and AGN.

For each of these fits,
we provide the unabsorbed thermal and power law 0.3 $-$ 8 keV luminosities
in Table \ref{xspectab}.
In the best fit model for the total spectrum,
the APEC component dominates the X-ray luminosity,
with L$_{\rm X}$(0.3 $-$ 8 keV) for the APEC component being
about 4.6 $\times$ that from the power law component.
For Galaxy A, the intrinsic luminosity
for the APEC component is 2.3 $\times$ that of the power law component.
For Galaxy B, 
the absorption-corrected L$_{\rm X}$(0.3$-$8 keV)
for the APEC component is about 4.6 $\times$ larger than that of the 
power law component.

\subsubsection{X-ray Spectra of the Smaller X-ray Sources} \label{sec:smaller_xray}

\begin{figure}[ht!]
\gridline{\fig{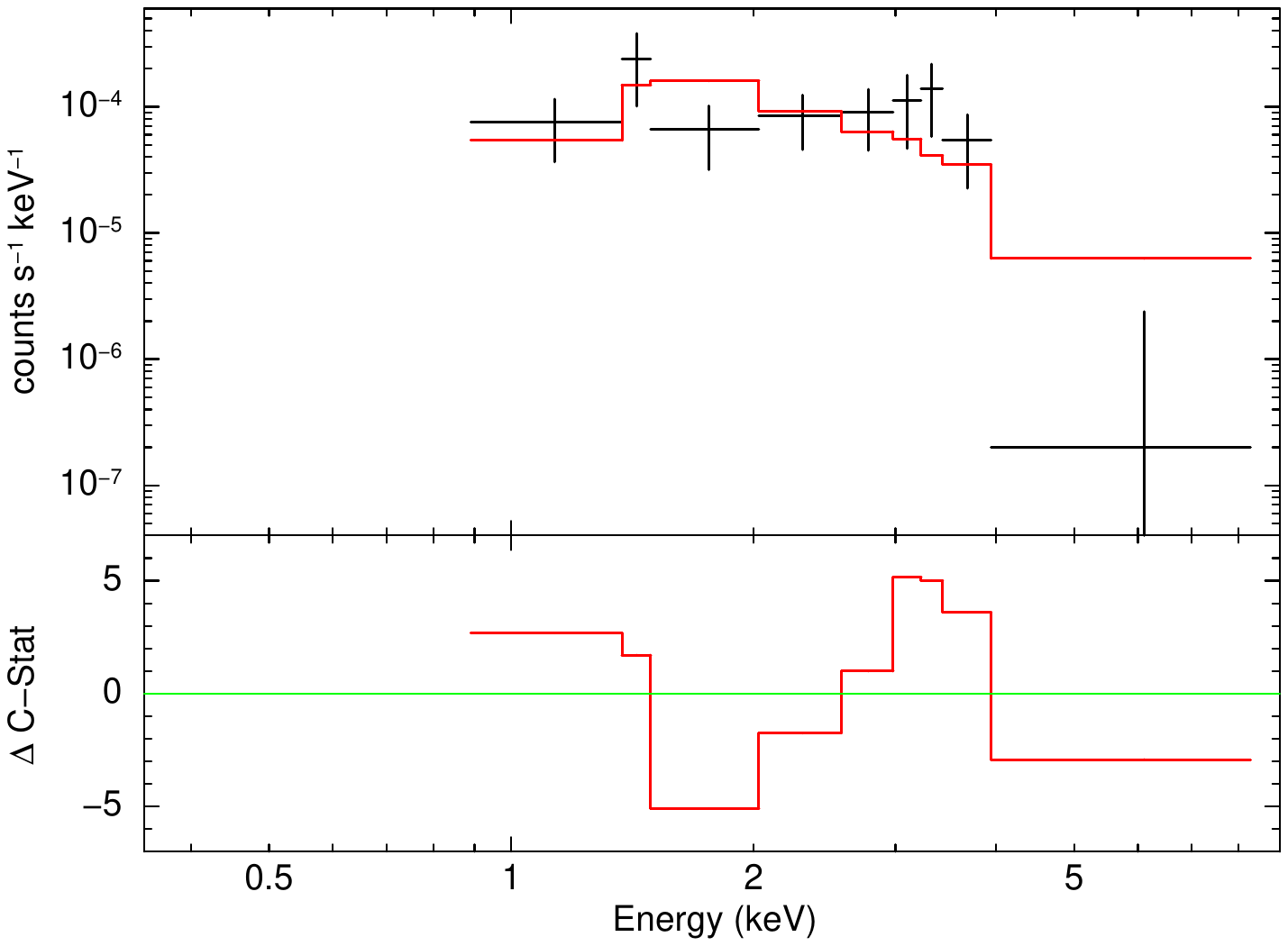}{0.35\textwidth}{(a)}
\fig{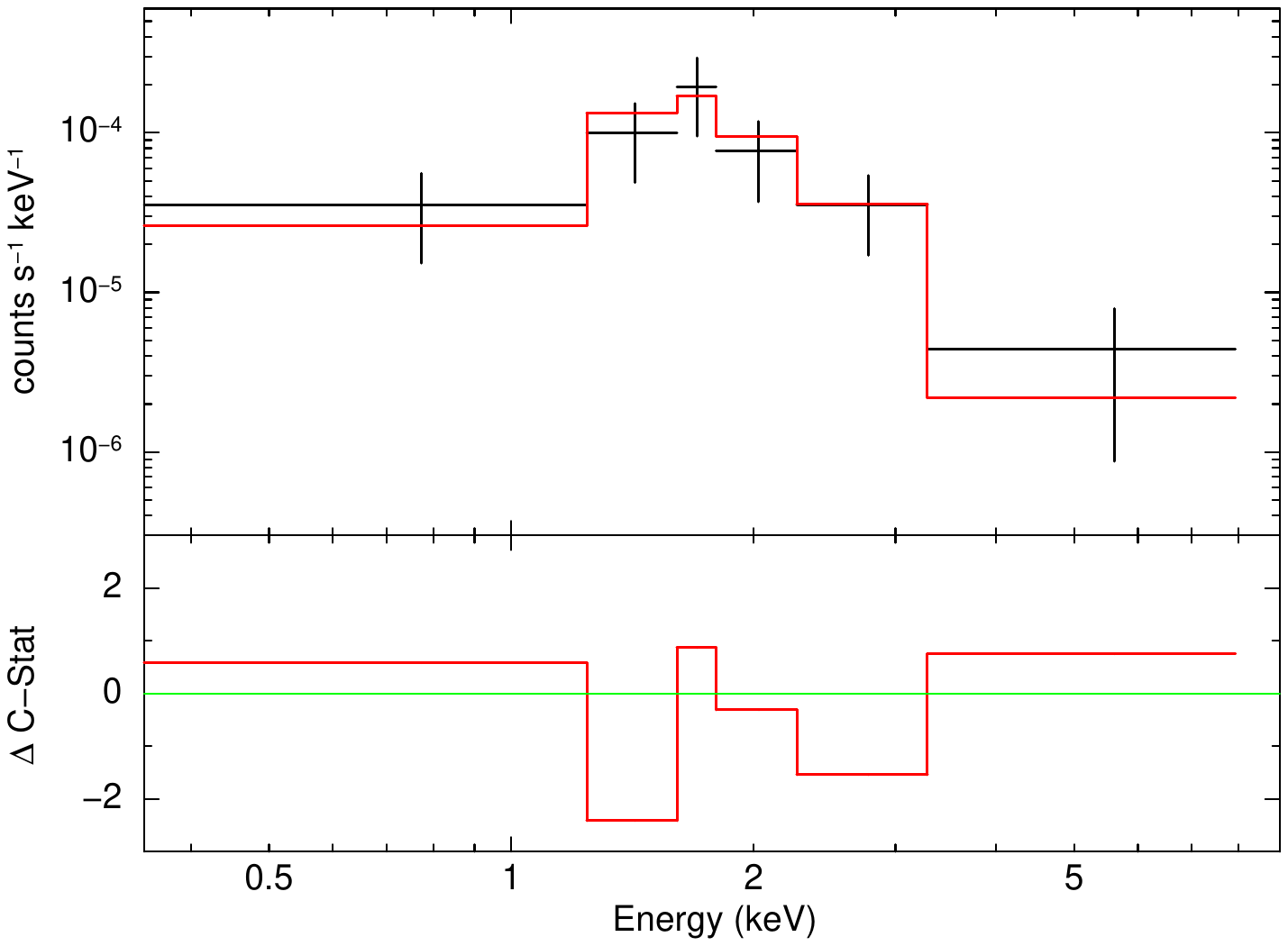}{0.35\textwidth}{(b)}
\fig{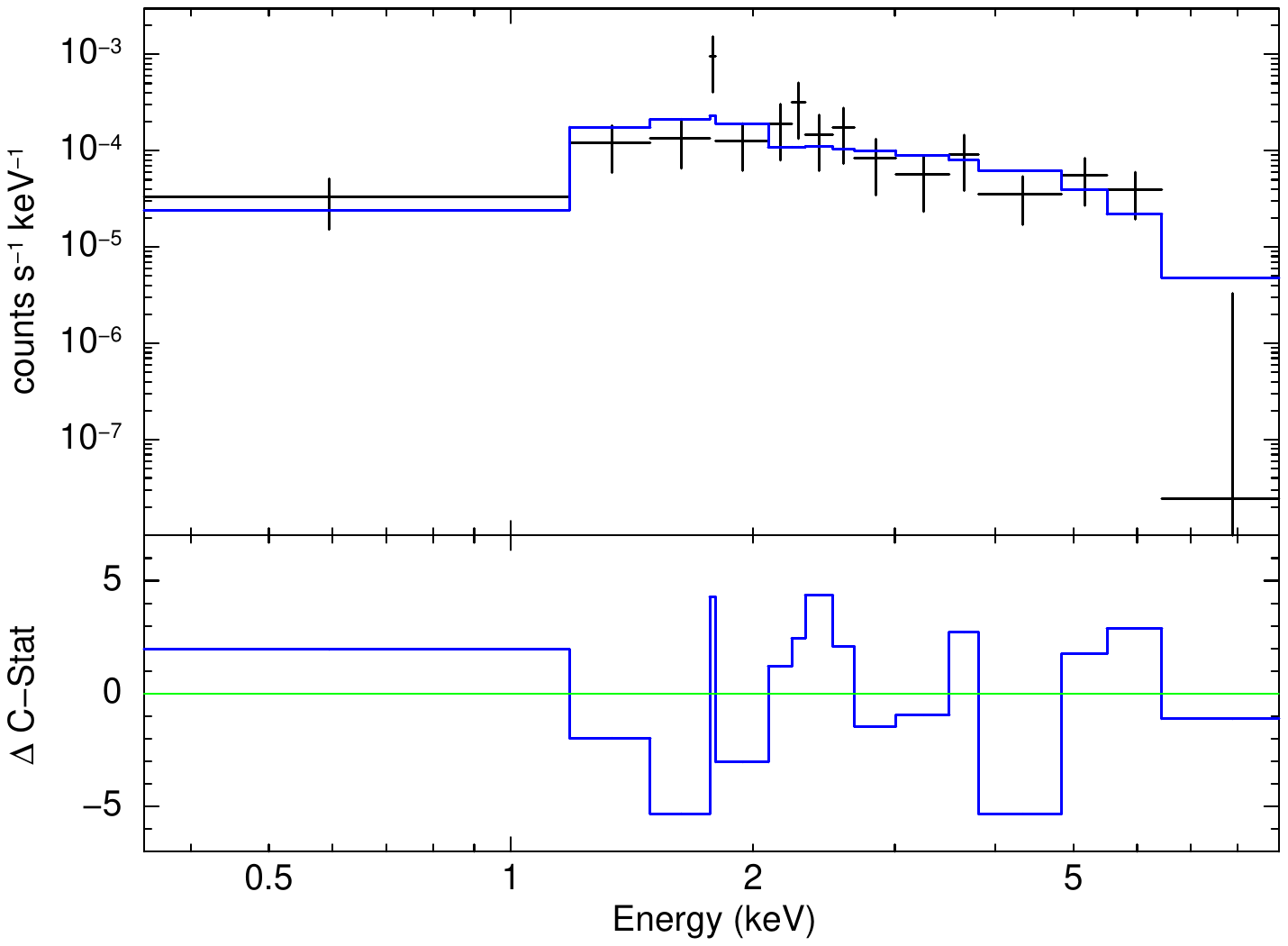}{0.35\textwidth}{(c)}
}
\caption{
Top panels:
the 
Chandra spectra for the diffuse X-ray emission
in X-ray source \#1 (left), \#2 (middle), and \#3 (right).
The best-fit model is overlaid.
The model is plotted in blue when the best-fit model is a power law;
the model is in red when the best model is an APEC model.
The bottom panel provides $\Delta$ C-stat.
The data were fitted with C-stat, but were binned to a S/N $>$ 1.7
for plotting purposes.
\label{fig:xspec_gas1_3}}
\end{figure}

\begin{figure}[ht!]
\gridline{\fig{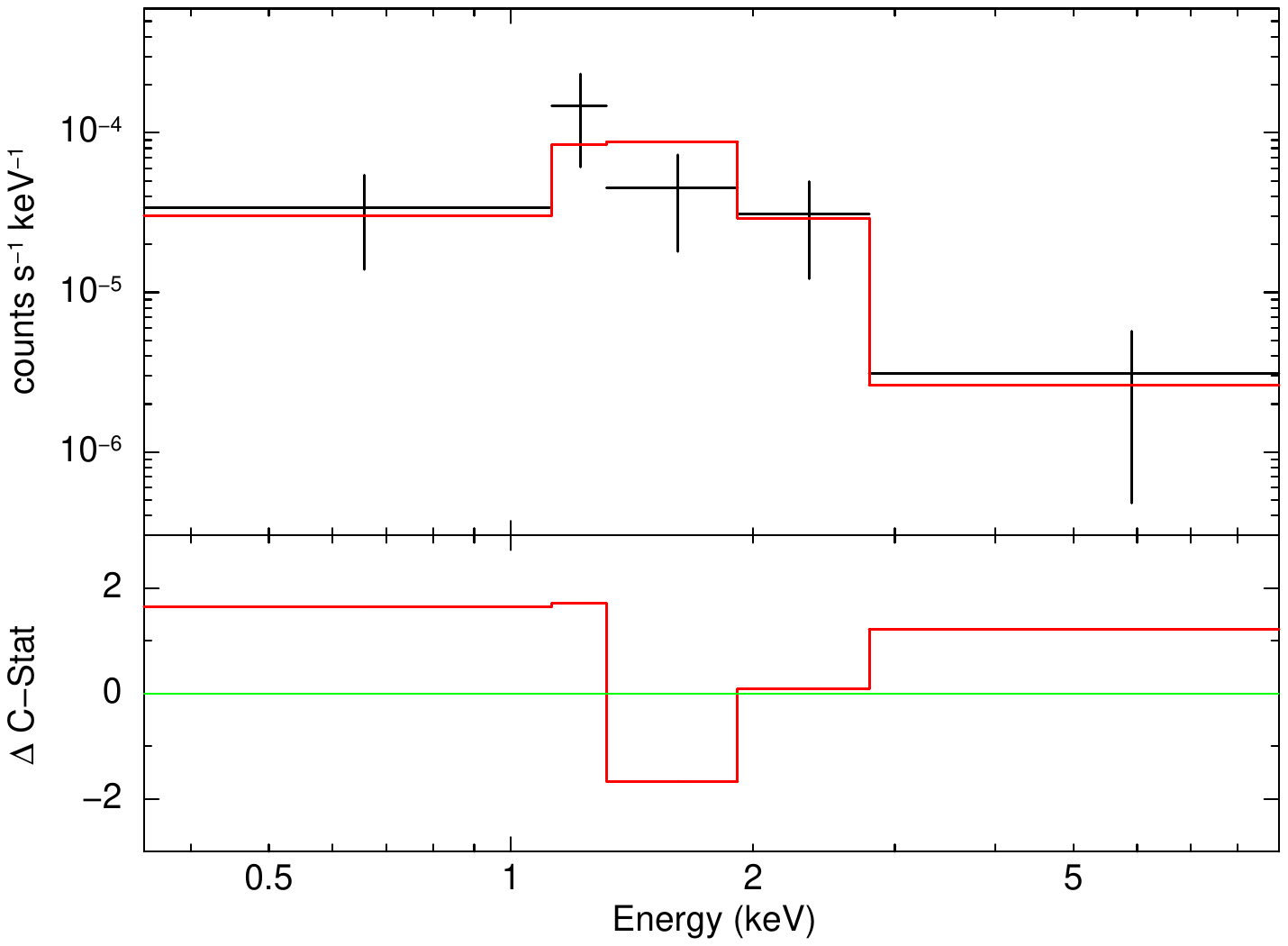}{0.35\textwidth}{(a)}
\fig{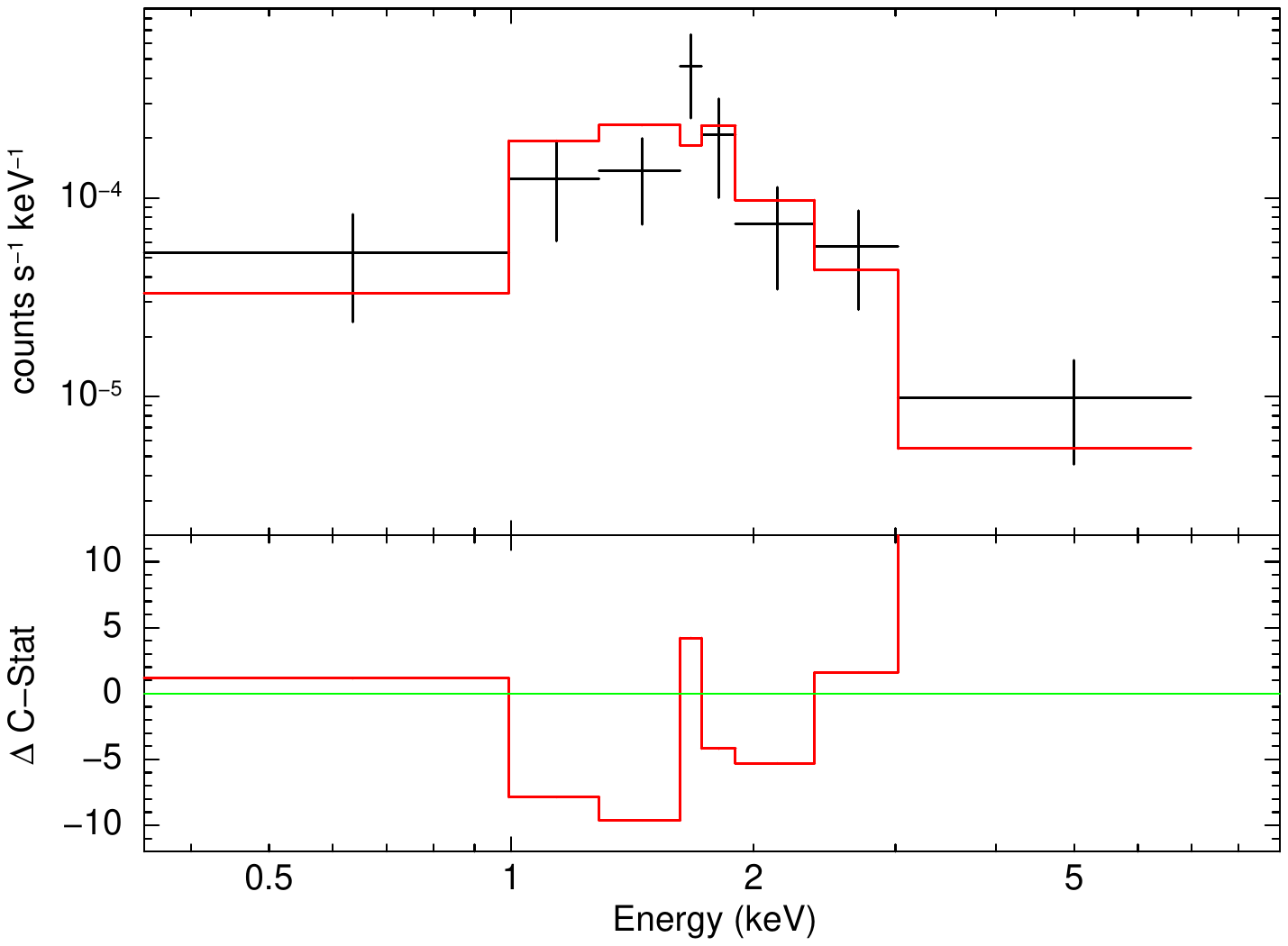}{0.35\textwidth}{(b)}
\fig{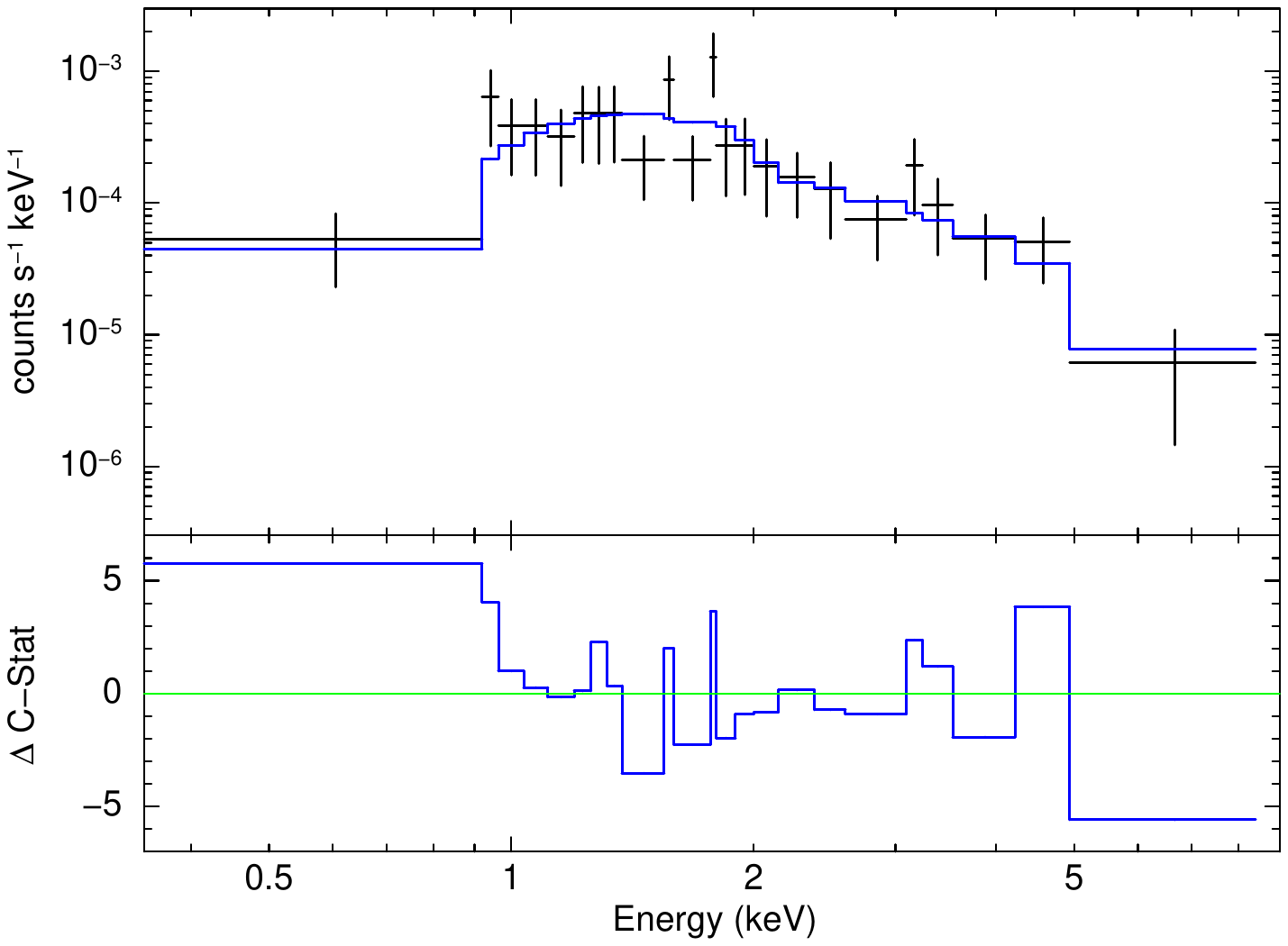}{0.35\textwidth}{(c)}
}
\caption{
Top panels:
The 
Chandra spectra for the diffuse X-ray emission
in X-ray source \#4 (left), \#5 (middle), and \#6 (right).
The best-fit model is overlaid.
The model is plotted in blue when the best-fit model is a power law;
the model is in red when the best model is an APEC model.
The bottom panel provides $\Delta$ C-stat.
The data were fitted with C-stat, but were binned to a S/N $>$ 1.7
for plotting purposes.
\label{fig:xspec_gas4_6}}
\end{figure}

\begin{figure}[ht!]
\gridline{\fig{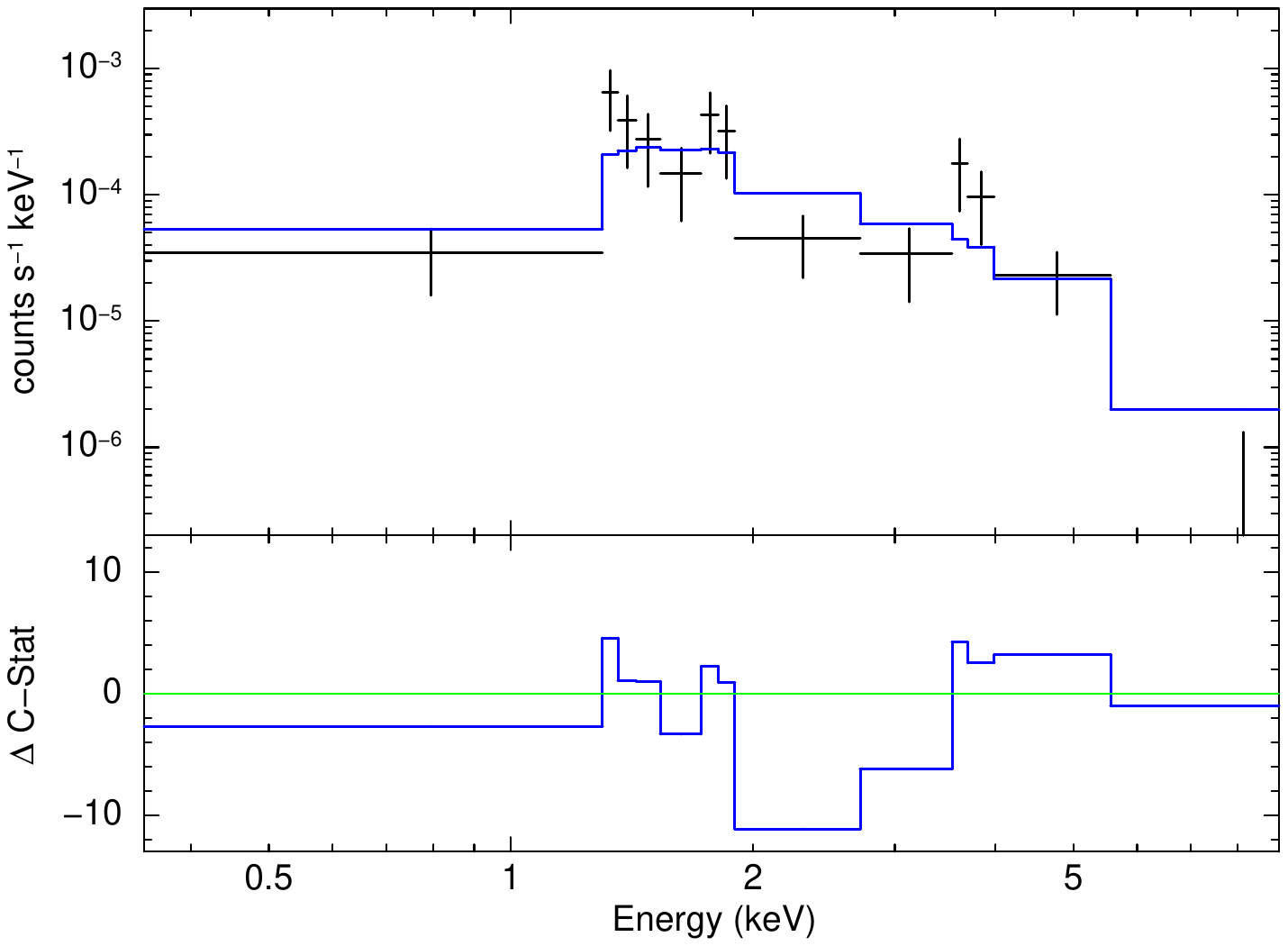}{0.35\textwidth}{(a)}
\fig{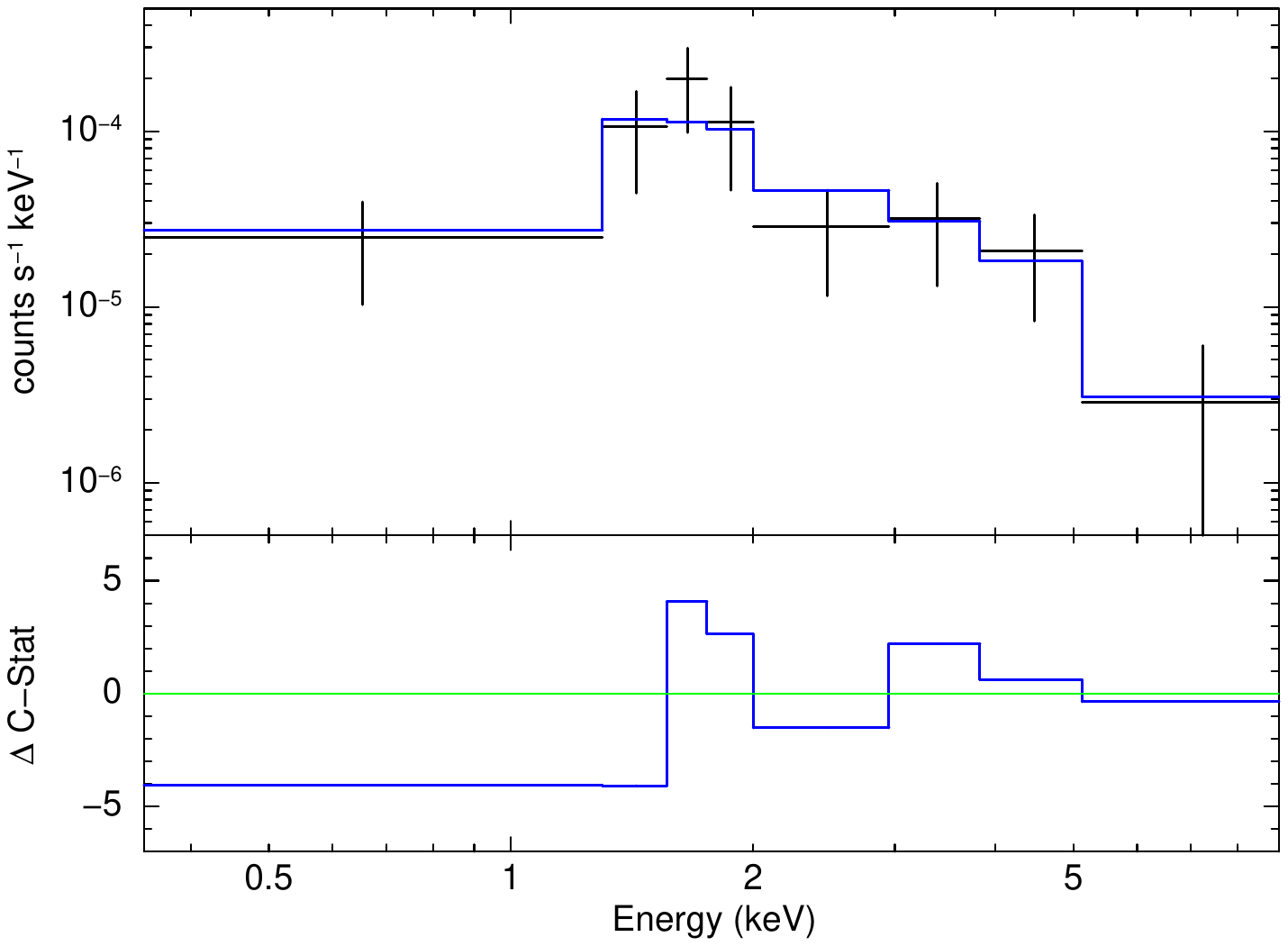}{0.35\textwidth}{(b)}
\fig{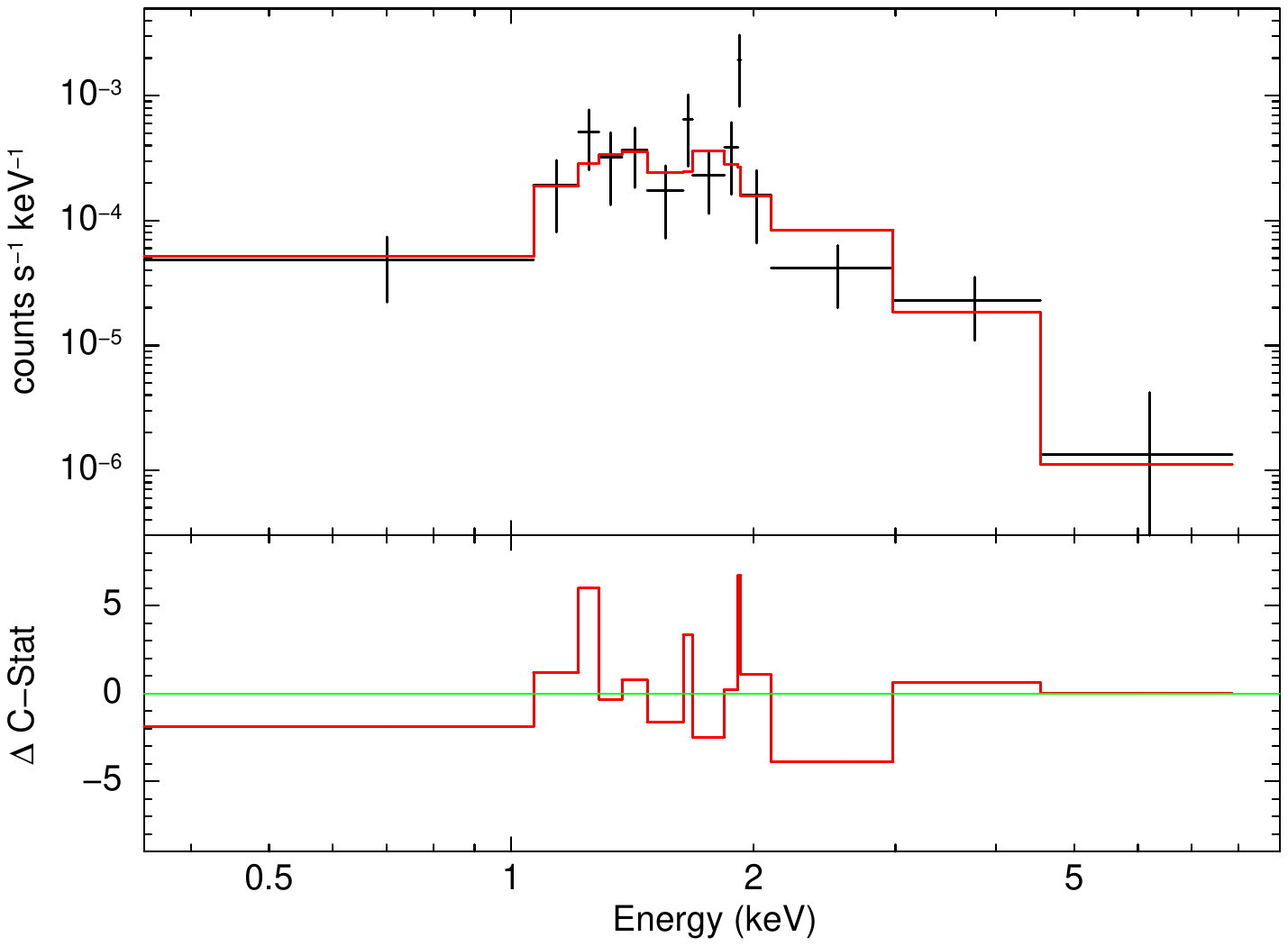}{0.35\textwidth}{(c)}
}
\caption{
Top panels:
The 
Chandra spectra for the diffuse X-ray emission
in X-ray source \#7 (left), \#8 (middle), and \#9 (right).
The best-fit model is overlaid.
The model is plotted in blue when the best-fit model is a power law;
the model is in red when the best model is an APEC model.
The bottom panel provides $\Delta$ C-stat.
The data were fitted with C-stat, but were binned to a S/N $>$ 1.7
for plotting purposes.
\label{fig:xspec_gas7_9}}
\end{figure}

\begin{figure}[ht!]
\plotone{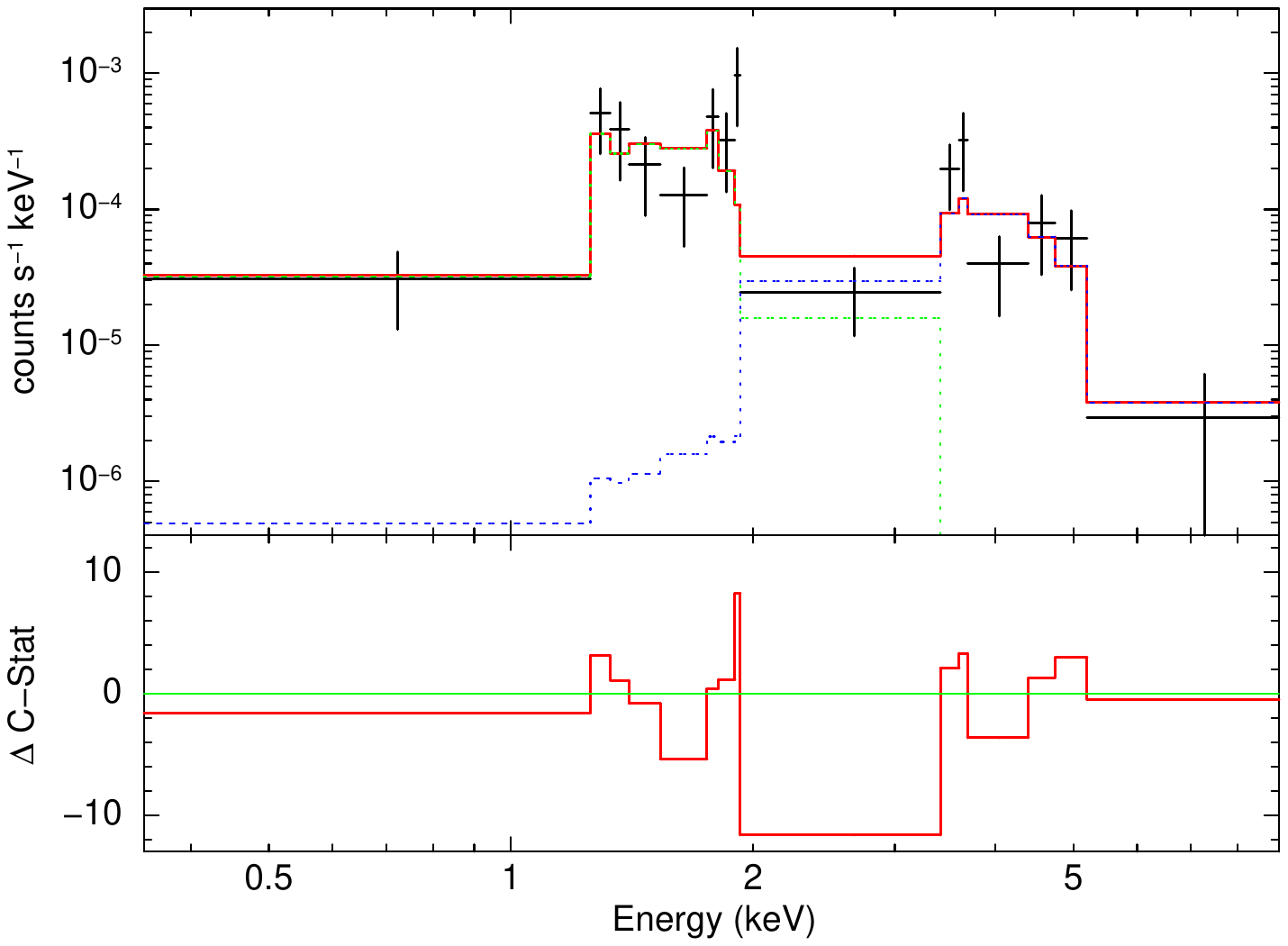}
\caption{
Top panel:
The 
Chandra spectra for the diffuse X-ray emission
in X-ray source \#10.
The best-fit model, a model with two APEC components, is overlaid.
The two APEC components are shown dotted in blue and green.
The bottom panel provides $\Delta$ C-stat.
The data were fitted with C-stat, but were binned to a S/N $>$ 1.7
for plotting purposes.
\label{fig:xspec_gas10_nuc}}
\end{figure}

We also investigated the X-ray spectra 
of the ten sources identified
in the X-ray maps. 
Table \ref{xspectab_gas} provides coordinates and estimated radii 
for these sources.
For each region, we fit for both an APEC-only model
and a power-law-only model,
and compared the two fits to determine which model is most
appropriate.  
In Table \ref{xspectab_gas}, we provide
our fits, including the C-statistic and DOF.
Table \ref{xspectab_gas} also gives 
the unabsorbed luminosity of the APEC or power law spectrum.
The fit listed first for each source
(in boldface) is our favored fit.
The spectra and the best-fit model are displayed in
Figures 
\ref{fig:xspec_gas1_3}, 
\ref{fig:xspec_gas4_6}.
\ref{fig:xspec_gas7_9},
and \ref{fig:xspec_gas10_nuc}.

For five sources (\#1, \#2, \#4, \#5, and \#9), we 
favor the APEC model,
because the power-law model 
gives a photon index larger than typical for AGN or HMXB 
(i.e., $\Gamma$ $>$ 2.1).
For these APEC fits, the temperatures
range between kT = 0.96 $\pm$ $^{2.72}_{0.38}$ keV for source \#2 to 1.95 $\pm$ $^{5.85}_{0.77}$ keV for source \#1.
Source \#5 has a relatively large C-stat/DOF ratio and 
may be variable (see 
Section
\ref{sec:chandra})
thus it may have a power law
contribution as well.
The best-fit values for 
N$_{\rm H}$ for the APEC fits for regions \#1, \#2, and \#9
are significantly larger than the values
found from the global fits, and are very poorly constrained. 
The APEC N$_{\rm H}$ values are a factor of 2 $-$ 4 times
larger than estimates for their surrounding regions (regions A1, A2, and B3,
respectively) from CIGALE.
These sources may be highly
obscured star-forming regions.

\input roberto_gas_without_count_rates.tex

For sources \#3 and \#8, 
we prefer the power law model since $\Gamma$ $<$ 2 and the APEC model
gives a temperature higher than typical for interstellar gas associated
with star formation (i.e., best-fit kT $>$ 5 keV).  
Source \#3 shows some evidence for variability
(Figure \ref{fig:rates}), supporting this interpretation.
Sources \#6 and \#7
have best-fit values
of $\Gamma$ higher than expected for HMXBs or AGN, and the 
temperatures obtained with an APEC fit are also high.
We suggest that the spectra of sources \#6 and \#7 
have contributions from both 
hot gas and HMXBs.  
Source \#6 shows evidence of variability
(Figure \ref{fig:rates}), consistent with this idea.

For source \#10, the apparent nucleus of Galaxy B, 
a power law model gives a reasonable $\Gamma$ but a large C-stat/DOF,
suggesting a second component.
We therefore experimented with
more complex models involving two components with different absorptions.
We select as our best model one with two APEC components, both
with high absorption.    A model with a single APEC component gives an 
unexpectedly large
temperature, while an APEC+power law model produces an unphysically-large $\Gamma$. 
The nature of source \#10 is discussed further in Section 
\ref{sec:nuclB}.

Adding up 
the APEC luminosities
of the best-fit models for the sources in Galaxy A, 
sources \#1, \#2,  and \#4 and half of the luminosity from the bridge
source \#5, we find 
L$_{\rm X}$(gas) = 1.8 $\times$ 10$^{41}$~erg~s$^{-1}$, about 
54\% of the APEC luminosity we derived for Galaxy A as a whole.   
Summing the APEC luminosities of the sources in Galaxy B, 
sources \#6, \#7, and \#9,  assuming source \#6 and \#7 are 
50\% APEC, 50\% power law,
and adding half of L$_{\rm X}$(gas) from source \#5
gives 
L$_{\rm X}$(gas) = 
1.9 $\times$ 10$^{41}$~erg~s$^{-1}$, 
about 19\% of the APEC luminosity we found for Galaxy B as a whole.  
Since the region for source \#10 overlaps with those of sources \#7 and \#8, we
don't include source \#10 in this calculation.
The difference between the global values and the sum of the individual
regions may be due to diffuse emission outside of our small circular apertures.
Summing over all individual sources excluding source \#10, we find 
a total
L$_{\rm X}$(gas) = 3.7 $\times$ 10$^{41}$~erg~s$^{-1}$,
about 32\% of the APEC luminosity within the 12$''$ radius.
In 
the following analysis,
when we compare with other galaxies
(Section \ref{sec:comparison})
we will use the global estimates
of L$_{\rm X}$(gas) and L$_{\rm power~law}$ as more representative of the 
system as a whole.

We estimate the amount of hot gas in each region in IC 2431 using the procedure
outlined in \citet{2019AJ....158..169S}
and the cooling functions of \citet{1977ApJ...215..213M} 
and \citet{1987soap.conf..255M}. 
In Table \ref{hot_gas_mass}, 
we provide estimated radii of the concentrations of hot gas
and their implied volumes, assuming spherical distributions.
We combined these volumes
with the absorption-corrected L$_{\rm X}$(gas) and the
gas temperature from Table 
\ref{xspectab_gas} to calculate 
n$_{\rm e}$$\sqrt{f}$ for each region, 
where 
n$_{\rm e}$ is the electron density and f the filling factor.
We then use these electron densities and volumes to calculate the mass of 
the 
hot X-ray-emitting gas in the region, along with the inferred 
cooling time
(Table \ref{hot_gas_mass}).
Over the entire system, 
the total mass of hot gas is about 5 $\times$ 10$^7$ M$_{\sun}$, approximately 1\% the mass
of neutral hydrogen gas.   
This excludes possible thermal contributions from regions \#6 and \#7.



\input table_hot_gas_mass_full.tex

\section{Comparison with Other Systems} \label{sec:comparison}

\subsection{Comparison Samples and Available Data}

In this section, we compare the Spitzer colors, X-ray properties, star formation rate, stellar
mass, morphological types, and atomic
gas content of the IC 2431 group to 
those of other galaxy systems.  Our comparison systems include
the compact groups studied by 
\citet{2013ApJ...763..121D, 2014ApJ...790..132D}, which includes 
Stephan's Quintet, and the radio-loud group HCG 62, which
hosts a radio galaxy \citep{2010ApJ...714..758G, 2011ApJ...735...11O}. 
We also compare to
a set of 49 equal-mass merging pairs of galaxies in various merger stages 
\citep{2018AJ....155...81S, 2019AJ....158..169S},
and 
the Spitzer Infrared Nearby Galaxies Survey (SINGS) 
\citep{2003PASP..115..928K}
sample of nearby spirals, irregulars, and
ellipticals.   
In addition, 
we compare with the Taffy galaxy pair, the radio-loud galaxy group
NGC 4410, the Cartwheel
collisional ring galaxy, and the Taffy-like system HCG 57.

For the compact groups, we utilize hot gas luminosities
from \citet{2013ApJ...763..121D, 2014ApJ...790..132D}, after converting to
the 0.3 $-$ 8 keV energy range using the 
PIMMS software\footnote{Portable, Interactive Multi-Mission Simulator; https://cxc.harvard.edu/toolkit/pimms.jsp}.  We use
the summed 0.5 $-$ 8 keV luminosities 
of 
the point sources in these groups, from
\citet{2016ApJ...817...95T}.  
We use UV/IR-based SFRs for the compact groups
from 
\citet{2016MNRAS.459.2948L}, after 
adjusting to a 
Kroupa IMF as in 
\citet{2012ARA&A..50..531K}.
For the compact groups, 
we use 
stellar masses from 
\citet{2016MNRAS.459.2948L}, 
atomic hydrogen masses 
from \citet{2013ApJ...763..121D, 2014ApJ...790..132D},
and Spitzer IR colors from \citet{2013ApJ...765...93C}.
For the galaxies in Stephan's Quintet and for two intergalactic
regions in SQ with strong shock emission,
we measured Spitzer IRAC fluxes from 
archival
data.
Morphological types for the compact group
galaxies are available from 
\citet{2013ApJ...763..121D} and
\citet{2023AA...670A..21J}.
Two of these groups host known X-ray AGN
\citep{2016ApJ...817...95T}.  

For the mergers, 
we use the 
\citet{2018AJ....155...81S}
absorption-corrected
L$_{\rm X}$(gas) and 
L$_{\rm X}$(power law) values, 
obtained using
spectral decomposition 
after point-source subtraction. 
We use version 2.1 of 
the Chandra Source Catalog (CSC)
\citep{2024arXiv240710799E}
to make a list of all of the X-ray point sources within
the D25 radius of the mergers, and sum their X-ray luminosities.
We use
UV/IR-based SFRs for the mergers from 
\citet{2018AJ....155...81S}, 
HI and H$_2$ masses as compiled by 
\citet{2019AJ....158..169S},
and Spitzer infrared fluxes extracted as 
in 
\citet{2018AJ....155...81S}. 
We derived sSFRs for the mergers using
stellar masses calculated from the 
Spitzer 3.6 $\mu$m and 4.5 $\mu$m fluxes
using the method of 
\citet{2022A&A...660A..69W}.
Ten of these mergers have a known Seyfert nucleus,
according to classifications compiled by 
\citet{2018AJ....155...81S}.  
Note that the merger sample was specifically selected to 
exclude radio AGN \citep{2018AJ....155...81S}.

For the SINGS galaxies, we use  
the X-ray measurements of 
\citet{2019ApJS..243....3L} and the Spitzer fluxes of
\citet{2017ApJ...837...90D}.
For the Taffy pair, we 
use the X-ray properties, SFR,
and stellar mass 
from
\citet{2015ApJ...812..118A}, the HI mass 
from
\citet{1993AJ....105.1730C}, and Spitzer fluxes extracted from archival
images.
For NGC 4410, we use 
the Chandra results of \citet{2003AJ....126.1763S}, the HI data from
\citet{2000ApJ...541..624S},
the star formation rate from 
\citet{2002AJ....123.1922D},
the morphological types tabulated
in 
\citet{2000ApJ...541..624S}, and derive stellar masses from the WISE
data using the relation from
\citet{2022A&A...660A..69W}.
For the 
Cartwheel galaxy, 
we use the SFR from 
\citet{2005ApJ...620L..35M}, 
the
stellar mass from \citet{2022MNRAS.514.1689Z},
the
HI mass from \citet{1996ApJ...467..241H},
L$_{\rm X}$(gas)
from \citet{2004A&A...426..787W} and \citet{2009A&A...501..445C},
and L$_{\rm X}$ from the point sources plus the diffuse power law component
from 
\citet{2004A&A...426..787W}.
For the compact group HCG 57, we obtain morphological types
and Chandra L$_{\rm X}$(gas) measurements
from 
\citet{2025ApJ...979..240O}, and the HI mass from
\citet{2001AA...377..812V}.

\subsection{Spitzer IR Colors} \label{sec:spitzer_colors}

\begin{figure}[ht!]
\plotone{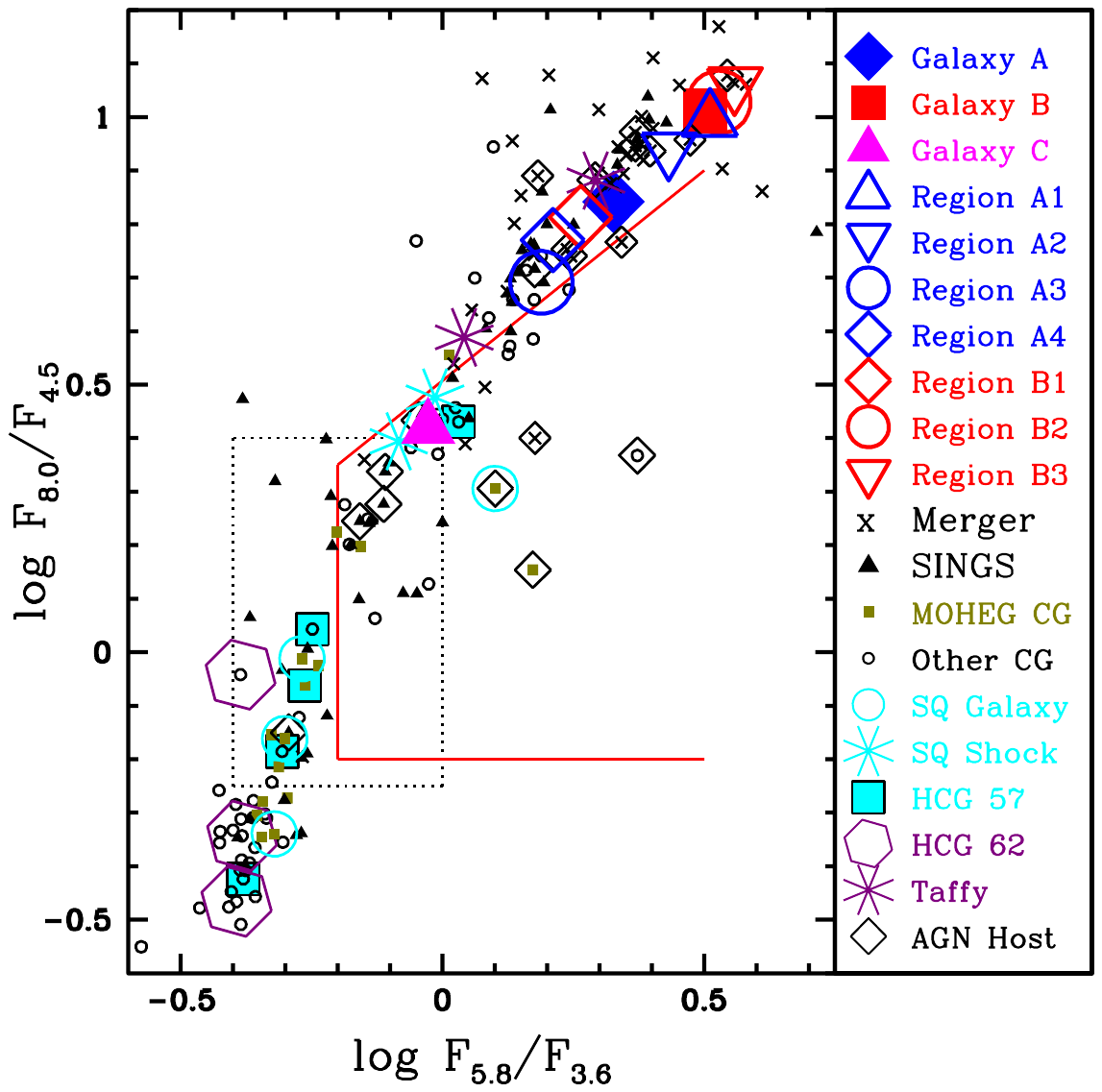}
\vskip -15mm
\caption{
Log (F$_{8.0}$/F$_{4.5}$) vs.\ log (F$_{5.8}$/F$_{3.6}$)
for the regions in IC 2431, and the comparison galaxies.
The colors have been K-corrected as described in the text.
As indicated by the legend on the right,
Galaxy A is a filled blue diamond,
Galaxy B a filled red square,
and 
Galaxy C a filled magenta triangle.
Subregions within each galaxy are plotted as open symbols with the
same color as the galaxy as a whole.
Merging pairs are 
plotted as 
black crosses and 
SINGS galaxies 
as black filled triangles. 
The small olive squares are compact group
galaxies classified 
as MOHEG, while black open circles are other compact groups.
Galaxies in Stephan's Quintet (HCG 92) are
marked with cyan circles,
and two locations in the intergalactic shock in SQ
are marked by cyan asterisks.
Galaxies in 
the Taffy-like system 
HCG 57 are marked by filled cyan squares outlined in black.
The large open purple hexagons mark 
galaxies in the compact group HCG 62, which
hosts a radio galaxy.
The purple asterisks show the two galaxies in
the Taffy pair, using archival data (UGC 12915 is 
the redder galaxy).
Galaxies known to host an optical AGN are marked with open black
diamonds.
The region marked in red is where AGNs are expected to
lie, according to 
\citet{2004ApJS..154..166L}; 
the known optical AGNs are mostly in or near this region.
The dotted box shows the 
`gap' region defined by 
\citet{2013ApJ...765...93C} (see text for details).
\label{fig:spitzer_colors}}
\end{figure}

In Figure \ref{fig:spitzer_colors}, 
we plot 
log (F$_{8.0}$/F$_{4.5}$) vs.\ log (F$_{5.8}$/F$_{3.6}$)
for the various regions in IC 2431 and the
comparison samples,
where 
F$_{3.6}$,
F$_{4.5}$,
F$_{5.8}$,
and 
F$_{8.0}$ are the 
Spitzer flux densities at 3.6, 4.5, 5.8, and 8.0 $\mu$m,
respectively.
These flux densities have been K-corrected to
account for the shifting of PAH features relative to the
Spitzer broadband filters.  The K-corrections were
calculated using the \citet{2007AJ....133..734B} kcorrect\footnote{https://kcorrect.readthedocs.io/}
software
using a large set of infrared galaxy templates from \citet{2024ApJ...977..102P}.
At the redshift of IC 2431 with a starburst spectrum, the K-correction in the 8 $\mu$m filter is small, so our
8 $\mu$m-based SFR calculations are not strongly affected; the largest K-correction is in the 5.8 $\mu$m filter,
causing a shift of about 0.45 dex to the right in 
Figure \ref{fig:spitzer_colors}.

Galaxies in the upper right of 
Figure \ref{fig:spitzer_colors}
tend
to be star-forming, while galaxies in the lower
left are quiescent \citep{2005ApJ...621..256S}.
In 
Figure \ref{fig:spitzer_colors}, we have marked
in red the region in which powerful AGN are
expected to fall, according to
\citet{2004ApJS..154..166L}.  
The galaxies in our sample known to host optical AGNs 
(marked with black diamonds) lie in or near this region.
The dotted box marks
the `gap' region, as defined by  
\citet{2013ApJ...765...93C}, which hosts
relatively few compact group galaxies.
The `gap' region 
is thought to be a transition region; galaxies
in compact groups apparently evolve quickly
through this part of the Spitzer color-color plot \citep{2010AJ....140.1254W}.
Galaxies above the `gap' region are actively forming stars, while galaxies
below the `gap' are quenched.

In IC 2431, 
Galaxy B as a whole 
is particularly red in these Spitzer colors, as are the two southern
regions in Galaxy B (regions B2 and B3), and the two northern regions in Galaxy A (regions
A1 and A2).
These regions are 
therefore very 
strongly star-forming.  
However,
the other regions in Galaxy A and Galaxy B (regions A3, A4, and B3) also lie in the star-forming
zone in this plot, as does Galaxy A as a whole.
In contrast, 
Galaxy C sits
on the upper boundary of the gap region.

In 
Figure \ref{fig:spitzer_colors}, 
we have identified the compact group galaxies classified by
\citet{2013ApJ...765...93C}
as `Molecular Hydrogen Emission-Line Galaxies'
(MOHEG), based on the strength of the mid-IR
H$_2$ lines relative to the PAH features.
For these galaxies, the H$_2$ lines appear
excited above that expected by UV heating
\citep{2013ApJ...765...93C}.
As noted by 
\citet{2013ApJ...765...93C},
the MOHEG galaxies tend to lie in or near the `gap'.
In 
Figure \ref{fig:spitzer_colors}, we have marked
locations of shocked IGM in Stephan's Quintet;
these regions lie at the top of the dotted box, while the 
SQ galaxies themselves lie at the bottom of the dotted box
or below, in the quenched region.  The galaxies in the radio-loud group HCG 
62 also sit near the bottom left of the plot.  In contrast,
the two Taffy galaxies have Spitzer IRAC colors
similar to those of the galaxies in IC 2431, thus are star-forming.

\subsection{Absorbing Column Density}

The large values of L$_{\rm X}$(thermal) that we derive  
for Galaxy A and Galaxy B are a consequence of the relatively high 
absorbing columns 
(6 $\times$
10$^{21}$ cm$^{-2}$ and
9 $\times$
10$^{21}$ cm$^{-2}$, respectively) implied by the X-ray spectrum.
For the galaxies in the merger comparison sample,
the highest 
N$_{\rm H}$ values 
inferred from the UV/IR ratios and obtained from fitting
Chandra X-ray spectra
are 4 $-$ 7 $\times$ 10$^{21}$~cm$^{-2}$
\citep{2018AJ....155...81S}.
For the SINGS galaxies, the median inferred
N$_{\rm H}$(internal) from Chandra data is 2 $\times$ 10$^{21}$~cm$^{-2}$,
with a 75th percentile of 
6 $\times$ 10$^{21}$~cm$^{-2}$
\citet{2019ApJS..243....3L}. 
Thus the absorbing columns for IC 2431 are quite
high compared to most other galaxies.

\subsection{Hot Gas vs.\ Star Formation Rate in IC 2431 vs.\ Other Galaxies} \label{hotgas_discussion}

\begin{figure}[ht!]
\includegraphics[width=7.5cm]{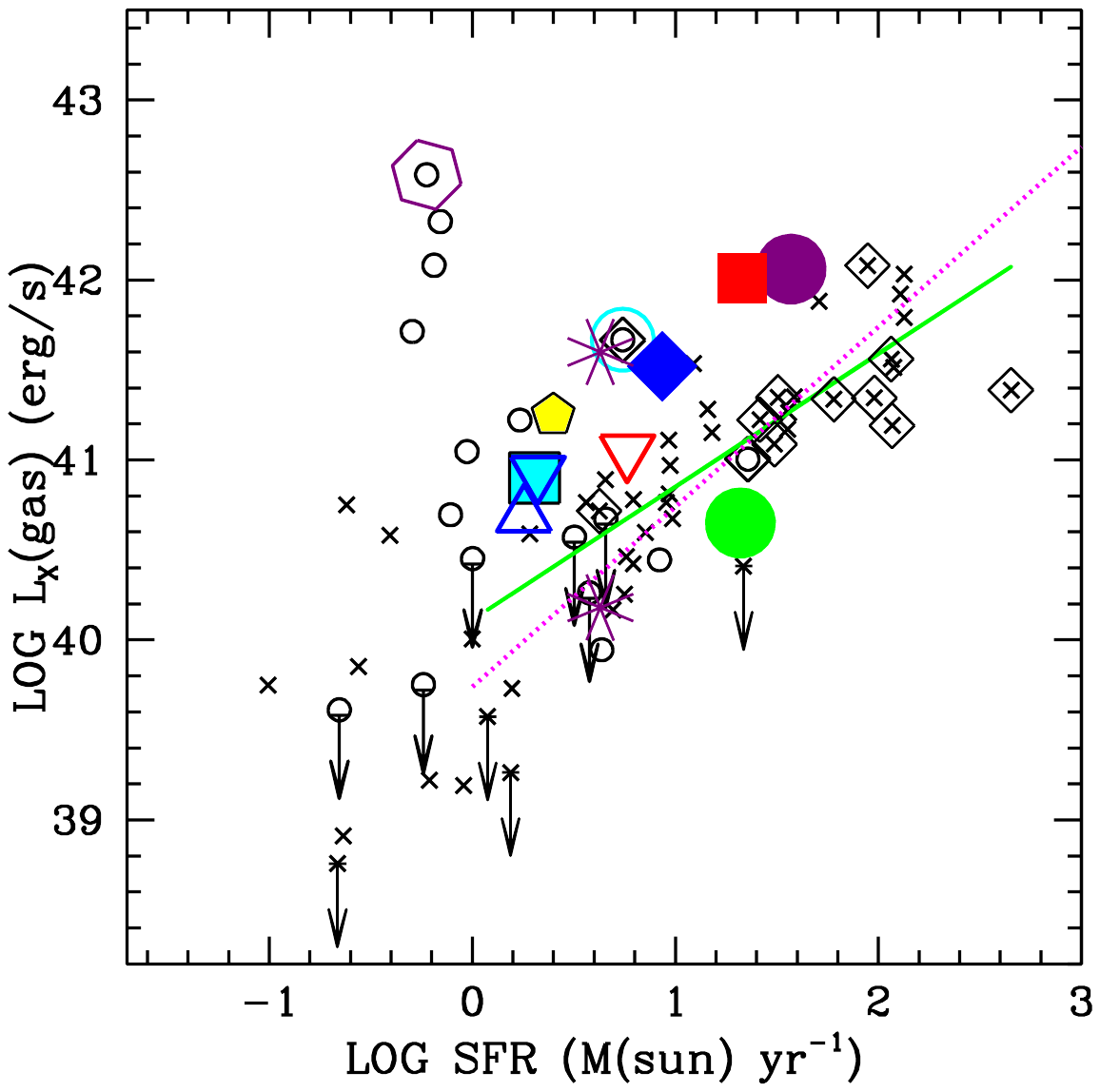}%
\includegraphics[width=7.5cm]{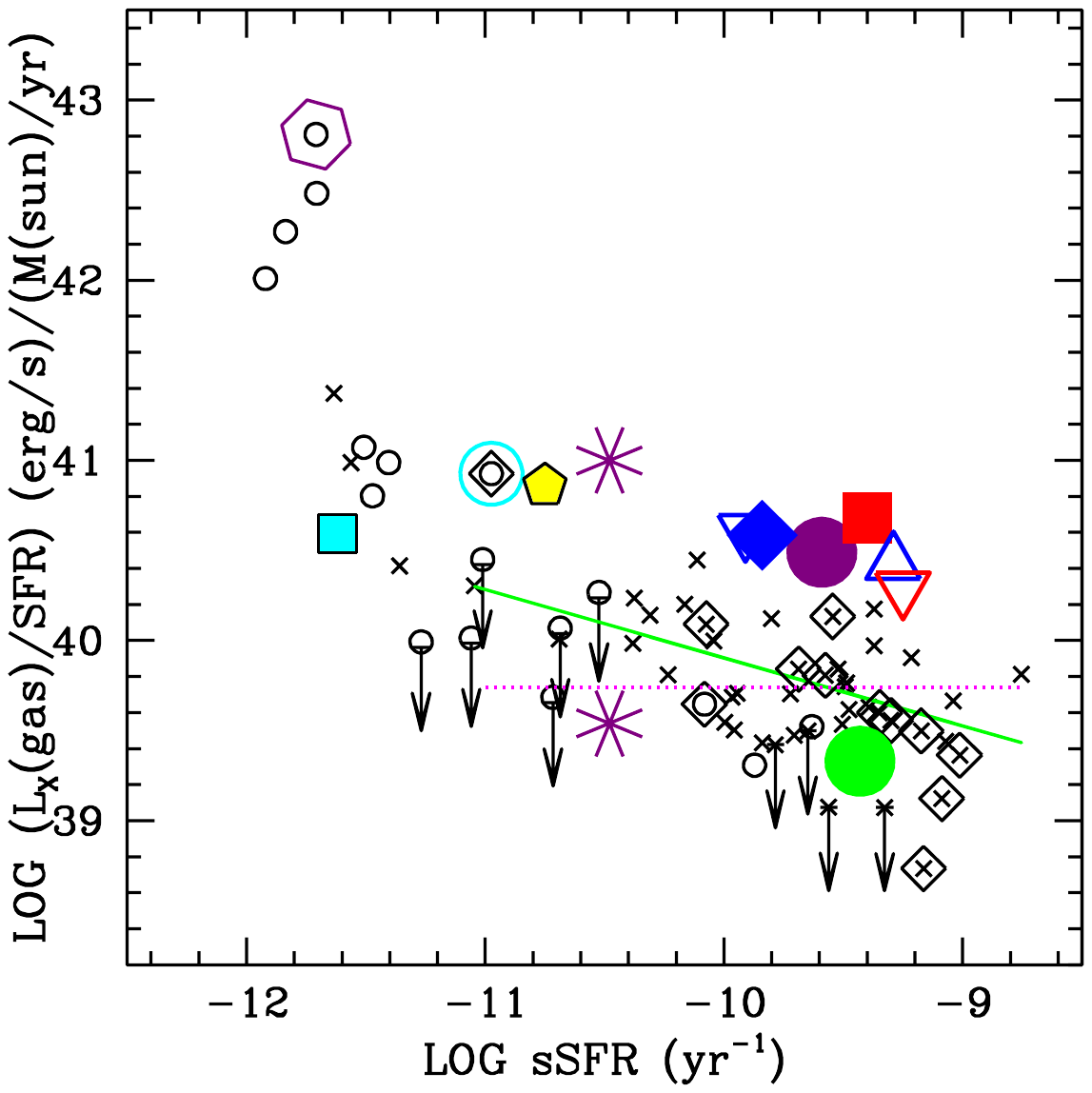}%
\includegraphics[width=6.5cm,trim={2.8cm 8cm 6cm 0}, clip=true]{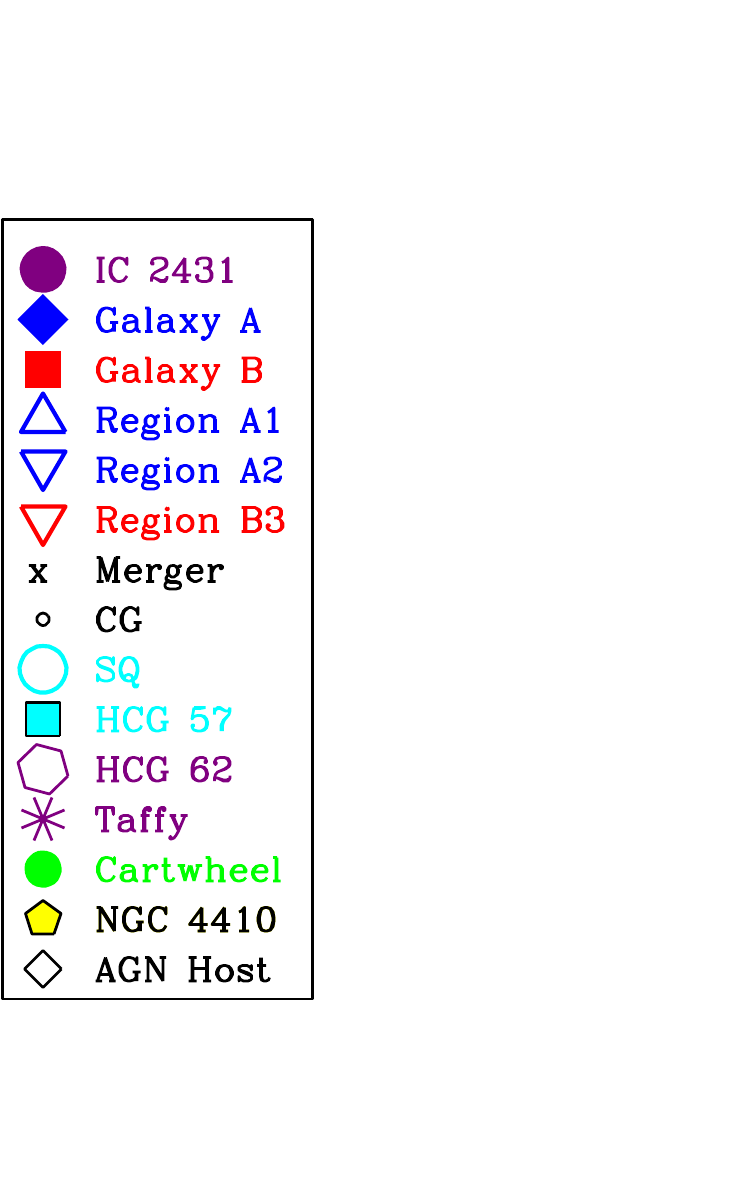}
\vskip -15mm
\caption{
Left: The X-ray luminosity from hot gas plotted against the SFR.
Right: The ratio L$_{\rm X}$(gas)/SFR vs.\ sSFR.
{
The purple filled circle marks the location of IC 2431 as a whole. 
The blue filled
diamond is Galaxy A, and the red filled square is Galaxy B.
The three APEC-dominated sub-regions in IC 2431 are open symbols
as given in the legend on the right.
The black crosses are the merging pairs.
The black open circles are the compact groups.
Stephan's Quintet (HCG 92) 
is marked as an open cyan circle, and the possible Taffy system
HCG 57 is identified by a filled cyan square outlined in black.
The purple open hexagon is HCG 62, which
hosts a radio galaxy.
The purple asterisks show limiting cases for
the Taffy pair as a whole, using data from  
\citet{2015ApJ...812..118A};
the lower value was obtained using a X-ray spectral model with
both a power law and thermal component, with no internal
absorption, while the upper value comes from a model with only an 
absorbed thermal
component.
The green filled circle marks the Cartwheel galaxy.
The yellow pentagon outlined in black is NGC 4410.
Systems hosting a known optical AGN are marked with open black
diamonds.
The dotted magenta line marks the median value of 
L$_{\rm X}$(gas)/SFR
= 5.5 $\times$ 10$^{39}$
(erg~s$^{-1}$)/(M$_{\sun}$~yr$^{-1}$)
for the mergers with
SFR $\ge$ 1~M$_{\sun}$~yr$^{-1}$.
The 
green solid line is the best-fit
line for the mergers with SFR $\ge$ 1~M$_{\sun}$~yr$^{-1}$.
}
\label{fig:desjardins2014_fig7}}
\end{figure}

In the left panel of Figure \ref{fig:desjardins2014_fig7},
we plot L$_{\rm X}$(gas) 
vs SFR for our comparison samples, and 
IC 2431 as a whole and Galaxy A and B separately.
The right panel 
of Figure \ref{fig:desjardins2014_fig7}
shows 
L$_{\rm X}$(gas)/SFR vs.\ sSFR.
In Figure 
\ref{fig:desjardins2014_fig7}, we also mark the three sub-regions
in IC 2431 whose X-ray spectra are dominated by APEC emission:
region A1 (X-ray source \#1), region A2 (X-ray source \#2),
and region B3 (X-ray source \#9).
On these plots, 
we overlay two lines, one marking the median value of L$_{\rm X}$(gas)/SFR 
for the mergers, 
and the other showing the best-fit relation for the mergers.  For these lines,
only
mergers
with SFR $\ge$ 1 M$_{\sun}$~yr$^{-1}$ are included.
IC 2431 as a whole lies above these lines, as do Galaxy A and Galaxy B individually,
particularly Galaxy B.
For the SFR $>$ 1~M$_{\sun}$~yr$^{-1}$ mergers, 
the 
median
absorption-corrected
L$_{\rm X}$(gas)/SFR is 
5.5 $\times$ 10$^{39}$
(erg~s$^{-1}$)/(M$_{\sun}$~yr$^{-1}$)
\citep{2018AJ....155...81S}.
The observed 
L$_{\rm X}$(gas)/SFR for IC 2431 as a whole from Chandra is 
approximately 2 $\times$ 10$^{40}$
(erg~s$^{-1}$)/(M$_{\sun}$~yr$^{-1}$), about
four times higher than the median value for the merging galaxies.
The scatter for the mergers is
large, however (about a factor of 2.3 in Lx(gas)/SFR),
 so IC 2431 as a whole does not
stand out in
Figure \ref{fig:desjardins2014_fig7}.
Galaxy B individually, however, is exceptional in its amount of hot
gas.  Depending upon whether the CIGALE or FUV+8$\mu$m SFR is used, the 
L$_{\rm X}$(gas)/SFR for Galaxy B is 4 $-$ 5 $\times$ 10$^{40}$
(erg~s$^{-1}$)/(M$_{\sun}$~yr$^{-1}$), a factor of 9 times larger 
than the median value
of the mergers.
The three sub-regions in IC 2431 whose X-ray spectra are well-fit 
by a APEC function (A1, A2, and B3) all lie above the best-fit line for the
mergers, particularly region A2, which is associated with the dust lane.

In the L$_{\rm X}$(gas) vs.\ SFR plot, 
the IC 2431 group as a whole
lies close to Arp 236 (VV 114; IC 1623), a pair of disk
galaxies in close contact.  
James Webb Space Telescope
(JWST) observations of Arp 236 reveal a very dust-enshrouded core,
H$_2$ and [Fe~II] shock lines,
but no high ionization lines indicative of an AGN
\citep{2023MNRAS.519.3691D}.
Four other merging systems lie $\sim$ 0.5 dex to the right of
IC 2431 
with similar L$_{\rm X}$(gas) values but SFRs three times higher.
In order from highest L$_{\rm X}$, these four galaxies
are NGC 6240, IRAS 17208-0014, AM 2055-425 (IRAS 20551-4250), 
and AM 2312-591 (IRAS 23128-5919).
Of these four systems, only NGC 6240 has a known Seyfert nucleus,
although star formation appears to dominate its infrared luminosity
\citep{2006ApJ...640..204A}.
All of these four galaxies except AM 2312-591 have strong 
mid-IR H$_2$ lines \citep{2018AJ....156..295P}.

The optically-identified AGNs among the mergers tend to have high SFRs, and 
lower  
L$_{\rm X}$(gas) for their SFRs, compared to mergers not
identified as AGN
(see Figure \ref{fig:desjardins2014_fig7}).
One possible explanation 
for the lower L$_{\rm X}$(gas)/SFR for the AGNs
is that
the SFRs in these galaxies may be over-estimated because of
contributions to the UV and IR from the AGN
\citep{2018AJ....155...81S}. 
In any case, the optical AGNs are not over-luminous in hot gas,
and, except for NGC 6240, do not lie close to IC 2431 in 
Figure \ref{fig:desjardins2014_fig7}.

In Figure \ref{fig:desjardins2014_fig7}, we also compare
with the 
\citet{2013ApJ...763..121D, 2014ApJ...790..132D}
compact groups.
For SFRs $\ge$ 1~M$_{\sun}$~yr$^{-1}$, the compact groups
agree reasonably well with the mergers.
At SFRs $<$ 1~M$_{\sun}~$yr$^{-1}$, some of the compact groups show large excesses
relative to the L$_{\rm X}$(gas)/SFR ratio for the high SFR mergers. 
The group with the highest L$_{\rm X}$(gas)
is HCG 62, which hosts a radio galaxy \citep{2010ApJ...714..758G, 2011ApJ...735...11O}. 
Stephan's Quintet (HCG 92) 
and NGC 4410 
also
show elevated L$_{\rm X}$(gas)/SFR ratios compared to mergers
with similar SFRs.
Figure 
\ref{fig:desjardins2014_fig7} includes two limiting cases for
the Taffy galaxies, using data from 
\citet{2015ApJ...812..118A}.
In the extreme case of an absorbed thermal spectrum with
no power law component, the Taffy lies close to Stephan's
Quintet in 
Figure \ref{fig:desjardins2014_fig7}.
At the other extreme (a composite power law plus thermal
spectrum with no internal absorption), the Taffy system
has a modest L$_{\rm X}$(gas) for its SFR, near the
bottom of the range for its SFR compared to the mergers.

For both 
the 
compact groups 
and the 
mergers,
L$_{\rm X}$(gas)/SFR is anti-correlated with sSFR
(right panel, Figure \ref{fig:desjardins2014_fig7}).
Systems with 
high L$_{\rm X}$(gas)/SFR tend to have low sSFR.  
The compact groups tend to follow the same trend as
the mergers, but extend to lower sSFRs and higher
L$_{\rm X}$(gas)/SFRs.
As the galaxies in a group or merging pair quench and the sSFR drops,
the amount of hot gas relative to the SFR increases.
This trend is visible even for strongly star-forming galaxies
(log sSFR $>$ -11).
Relative to this trend, IC 2431 stands out as having a high L$_{\rm X}$/SFR,
particularly Galaxy B, which is about 1 dex above the trend.
Galaxy A and sub-regions A1, A2, and B3 are also enhanced in hot gas, 
lying 0.8 dex above
the best-fit relation for the mergers. 

Stephan's Quintet and NGC 4410, which have lower sSFRs than
IC 2431, have hot gas excesses of about 0.6 dex relative to this trend.
HCG 62, a group with a radio AGN, has the highest
L$_{\rm X}$(gas)/SFRs in the sample.
It is noteable that the three systems in this sample with 
apparent radio AGNs (IC 2431, NGC 4410, and HCG 62) all show excess hot
gas relative to their SFRs.
We note, however, 
that the excess hot gas in IC 2431 is not only in the vicinity of the 
radio jet, but also in the far north of Galaxy A and within Galaxy B.

\subsection{Morphology and HI Content}

\begin{figure}[ht!]
\includegraphics[width=7.5cm]{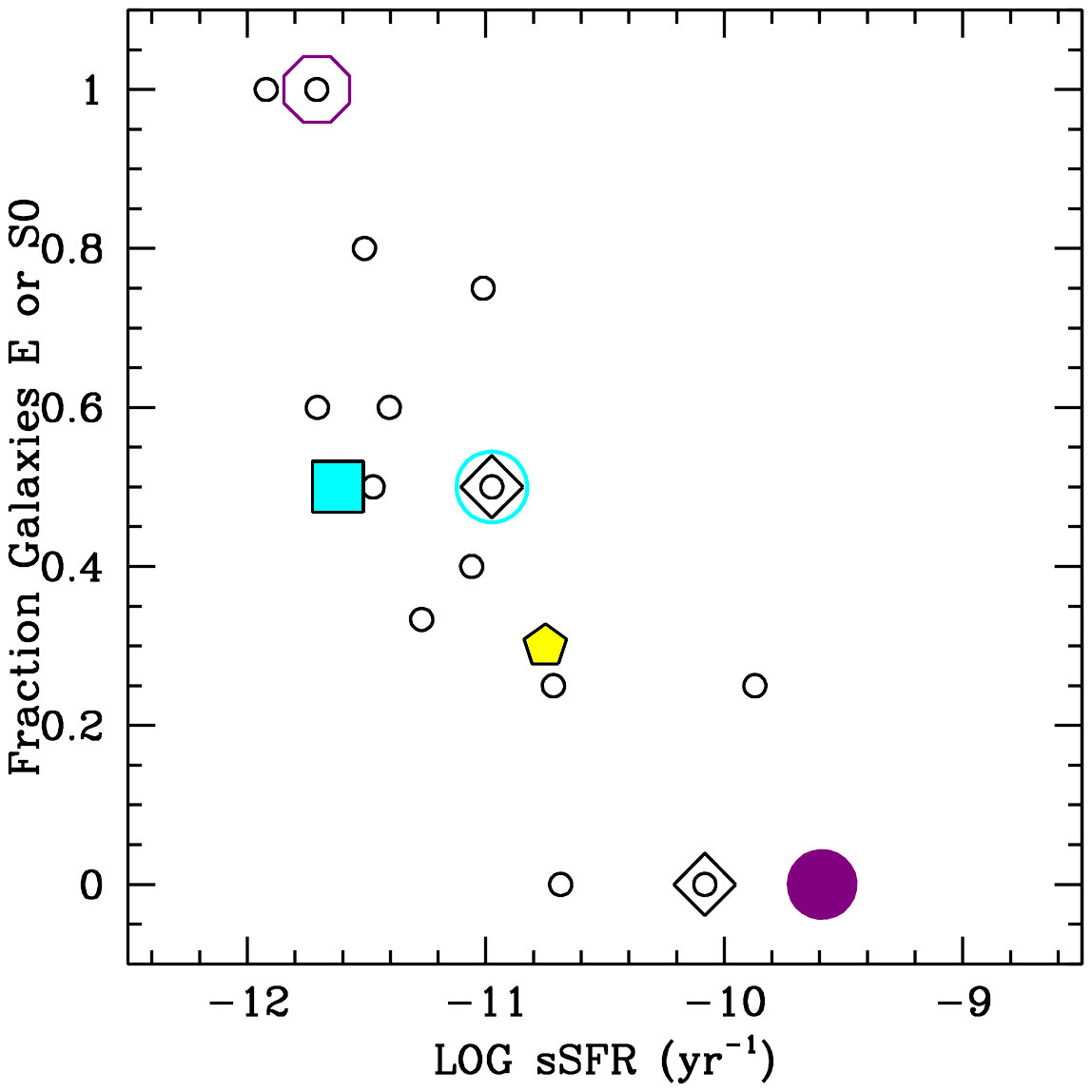}%
\includegraphics[width=7.5cm]{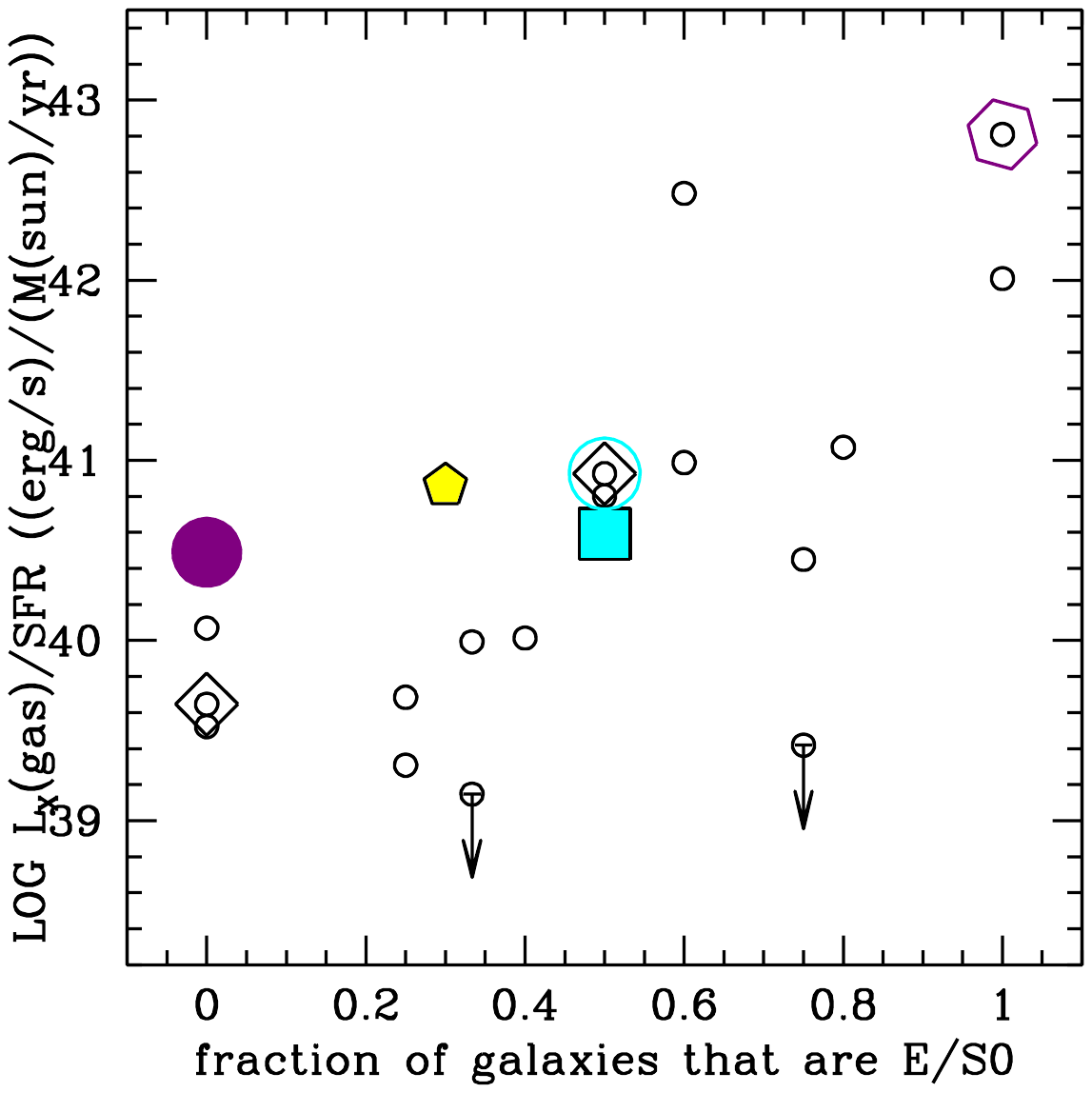}%
\includegraphics[width=6.5cm,trim={2.8cm 8cm 6cm 0}, clip=true]{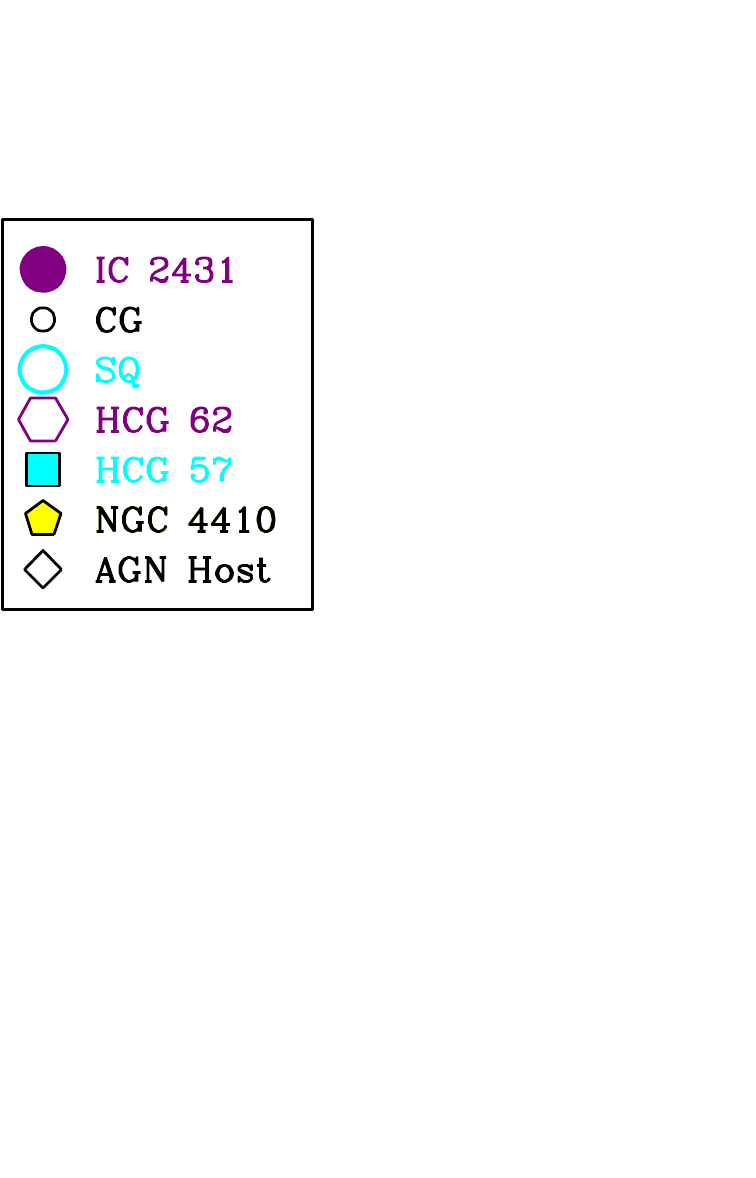}
\vskip -15mm
\caption{
Left: The fraction of non-dwarf galaxies that
are elliptical or S0, plotted against the sSFR.
Right: The L$_{\rm X}$(gas)/SFR ratio plotted against
the fraction of non-dwarf galaxies that
are elliptical or S0.
{ 
The purple filled circle marks the location of IC 2431.
The black open circles are the compact groups.
Stephan's Quintet (HCG 92) 
is marked by a cyan circle.
The large open purple hexagon marks HCG 62, which hosts a radio galaxy.
The Taffy-like system 
HCG 57 is marked by a filled cyan square outlined in black.
The yellow pentagon outlined in black is NGC 4410.
Known optical AGN are identified by black open diamonds.
}
\label{fig:fS0}}
\end{figure}

Another tracer of the evolutionary stage of a compact group
is the fraction of non-dwarf galaxies which are elliptical or S0.
This is plotted against sSFR in the left panel of Figure \ref{fig:fS0}
for the compact groups and IC 2431.
In making this plot, we assumed that all
three of the galaxies in IC 2431 are spirals.  
The compact groups show an anti-correlation; 
groups with high fractions of E/S0
galaxies have low sSFRs.  
IC 2431 is in
the bottom right of the plot, consistent with the 
trend seen in the other compact groups, but with a higher sSFR.
The radio-loud group HCG 62 is at the other extreme, with very low
sSFR and all E/S0 galaxies.
Stephan's Quintet is not unusual in this plot; it lies in the middle of the 
range for the 
compact groups,
with about half of the galaxies identified as E/S0,
and a moderate sSFR.  

In the right panel of Figure \ref{fig:fS0}, 
we plot 
L$_{\rm X}$(gas)/SFR 
for the compact groups and IC 2431 
against the fraction of galaxies that 
are E/S0.
As the fraction of early-type galaxies in a compact group increases, the 
L$_{\rm X}$(gas)/SFR of the group tends to increase.
HCG 62 is at the upper right of the plot, with large 
L$_{\rm X}$(gas)/SFR and all E/S0 morphologies.
IC 2431 has 
a relatively high L$_{\rm X}$(gas)/SFR 
ratio, compared to other compact groups containing no E or S0 galaxies.
One possibility is that 
Galaxy A in the process of transforming from a spiral to an S0
galaxy via ram pressure stripping.  The lack of star formation
south of the nucleus of Galaxy A may support this idea.
It is also possible that Galaxy C, which is more quenched 
than the other two galaxies, is an S0 rather than spiral galaxy.

\begin{figure}[ht!]
\includegraphics[width=7.5cm]{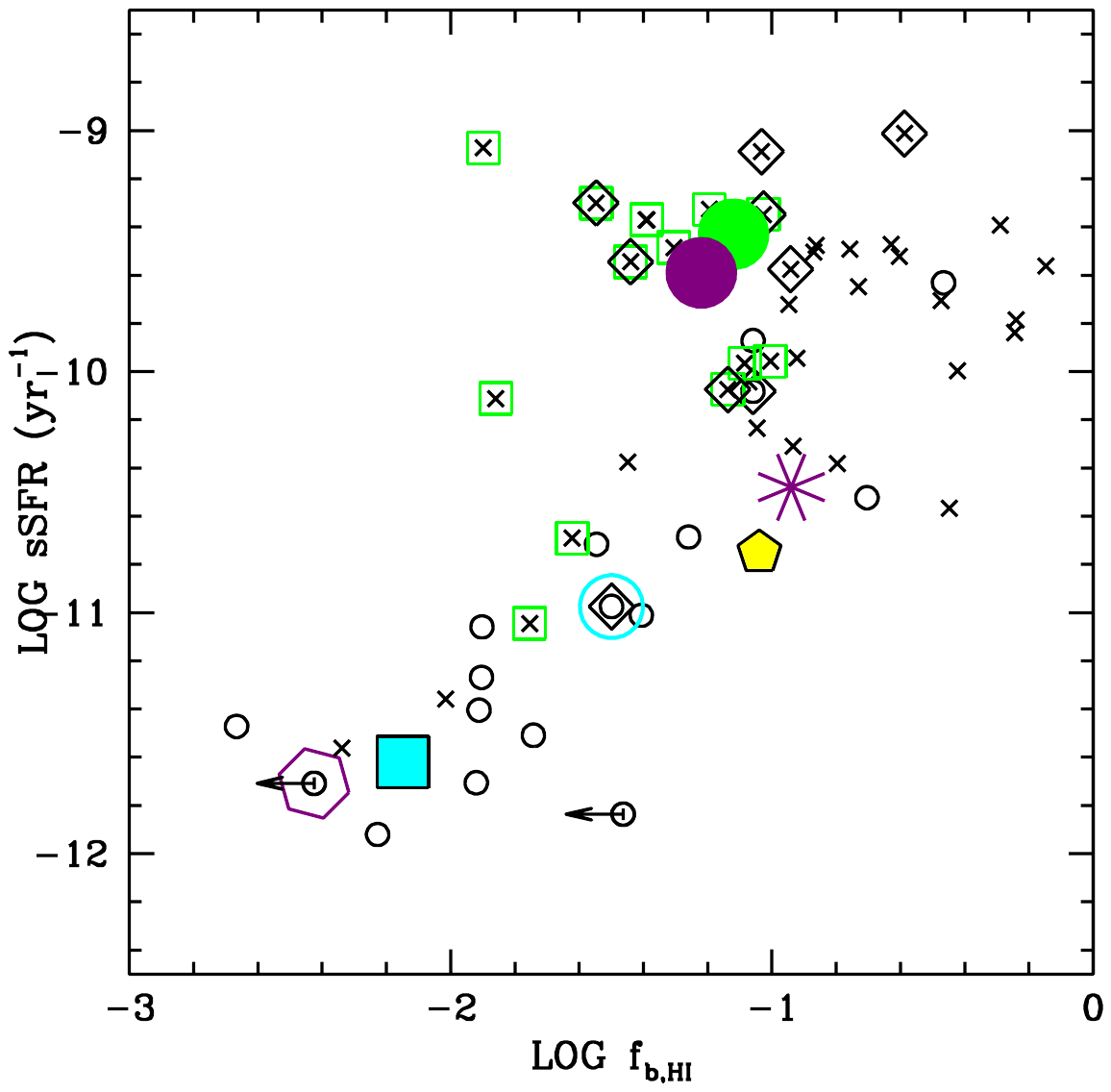}%
\includegraphics[width=7.5cm]{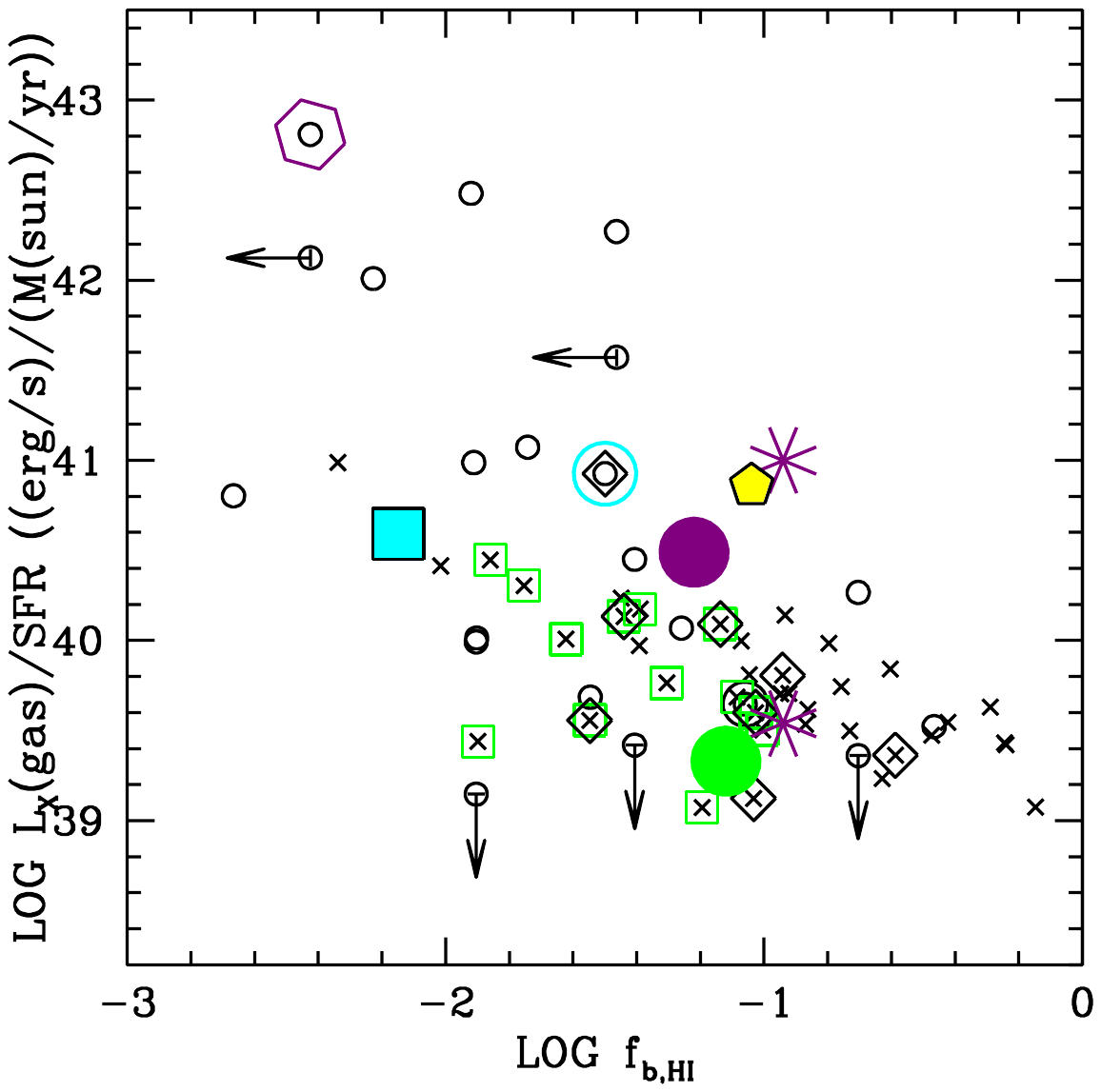}%
\includegraphics[width=6.5cm,trim={2.8cm 8cm 6cm 0}, clip=true]{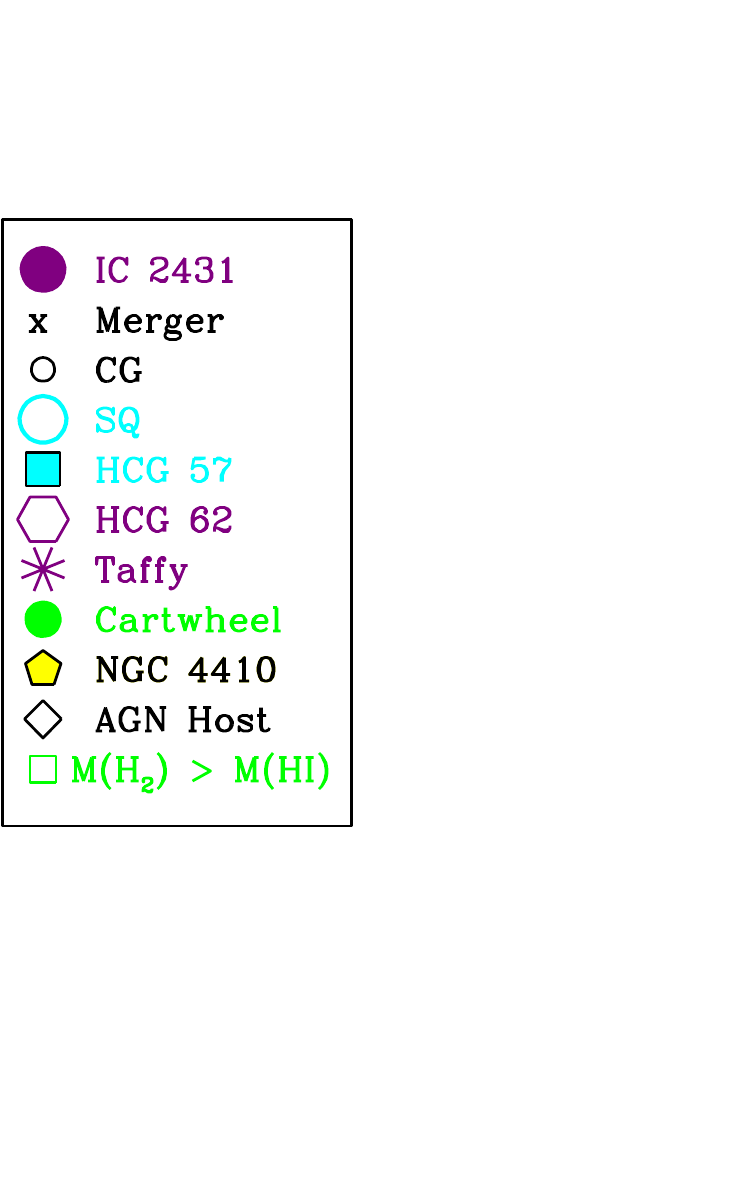}
\vskip -15mm
\caption{
Left: The sSFR plotted against f$_{\rm b,HI}$, the
fraction of
baryons that are HI, defined as the ratio 
of the HI mass to the sum of the mass of stars plus 
HI gas (ignoring molecular gas).
Right: L$_{\rm X}$(gas)/SFR vs.\ log f$_{\rm b,HI}$.
Symbols are defined in the legend on the right.  
{ 
The purple filled circle
marks the location of IC 2431.
The black crosses are the merging pairs.
The black open circles are the compact groups.
Stephan's Quintet (HCG 92) 
is marked as an open cyan circle.
HCG 57 is marked by a filled cyan square outlined in black.
The open purple hexagon marks HCG 62, which hosts a radio galaxy.
The purple asterisks give
the limiting cases for 
the Taffy galaxies.
The green filled circle marks the Cartwheel galaxy.
The yellow pentagon outlined in black is NGC 4410.
Known optical AGNs are identified by black open diamonds.
Open green
squares mark mergers
with M(H$_2$) $>$ M(HI).
}
\label{fig:f_b}}
\end{figure}

Another proxy of the evolutionary stage of a compact group
is the fraction of the baryons that are in the form of atomic interstellar
hydrogen gas, f$_{\rm b,HI}$, defined by 
\citet{2014ApJ...790..132D} to be the ratio of the HI mass to
the sum of the stellar mass plus the HI mass (ignoring molecular
gas).   This quantity is plotted with sSFR in the left panel of 
Figure \ref{fig:f_b}.  This plot includes the mergers
as well as the compact groups and IC 2431.  
A strong correlation is
seen for the compact groups, in that galaxies with proportionally
more HI gas have higher sSFRs.  
The radio loud system HCG 62 is at the lower left in this plot,
with low sSFR and low HI content.
Stephan's Quintet is in the middle of the range
for the compact groups and is consistent with the trend seen
in the other compact groups.
The Taffy galaxies and NGC 4410 also follow the same relation as the compact groups.

Most of the mergers follow this same
trend of increasing 
f$_{\rm b,HI}$ with increasing sSFR, however, there are some with high sSFRs but
only moderate values of f$_{\rm b,HI}$.  
We have confirmed that 
these discrepant mergers 
tend to have large quantities of interstellar molecular gas.
In Figure \ref{fig:f_b},
we identify mergers with H$_2$ masses 
greater than the HI mass, 
using published 
2.6 mm CO observations compiled by 
\citet{2019AJ....158..169S}.
IC 2431 lies above the trend for the compact groups,
but is not as discrepant as some of the mergers.
IC 2431 has proportionally much more HI than Stephan's Quintet as well
as a much higher sSFR.   The location of IC 2431 on this plot suggests that,
like some of the mergers, it may be rich in molecular gas.
It has not yet been observed in CO.

In the right panel of Figure \ref{fig:f_b},
L$_{\rm X}$(gas)/SFR 
is plotted with the fraction of baryons in HI, including
both the compact groups and the mergers.
As the fraction of the baryons in atomic hydrogen gas decreases, the 
L$_{\rm X}$(gas)/SFR ratio tends to increase.
The 
IC 2431 group has 
a L$_{\rm X}$(gas)/SFR ratio and a sSFR comparable
to mergers and other compact groups with similar fractional amounts
of atomic gas.

Based on its high HI mass relative to its
stellar mass, its high sSFR, and its low fraction of E/S0
galaxies, the IC 2431 group would be classified 
as being in an early stage of evolution.
According to 
\citet{2013ApJ...763..121D}, as the fraction of E/S0 galaxies
in a compact group
increases and both the HI content and star formation rate decrease, 
more of the hot gas is found outside of the main bodies
of the galaxies.   
In IC 2431, with the exception of the bridge
gas we find little extended diffuse X-ray emission
outside of the disks of the galaxies.  This is consistent with
the status of IC 2431 as a relatively early-stage group merger.   

\subsection{The Power Law Component of the X-Ray Spectrum}

From spectral analysis 
of the Chandra data we find
an absorption-corrected
L$_{\rm X}$(power law) = 3 $\times$ 10$^{41}$ erg~s$^{-1}$
for the IC 2431 group as a whole.
Outside of the galactic nuclei, most of this radiation 
likely originates from an unresolved population of HMXBs
associated with the starburst.
In non-AGN star forming galaxies, HMXB emission dominates the power law
component in the X-ray \citep{2003MNRAS.339..793G,
2012MNRAS.419.2095M}.

\begin{figure}[ht!]
\includegraphics[width=7.5cm]{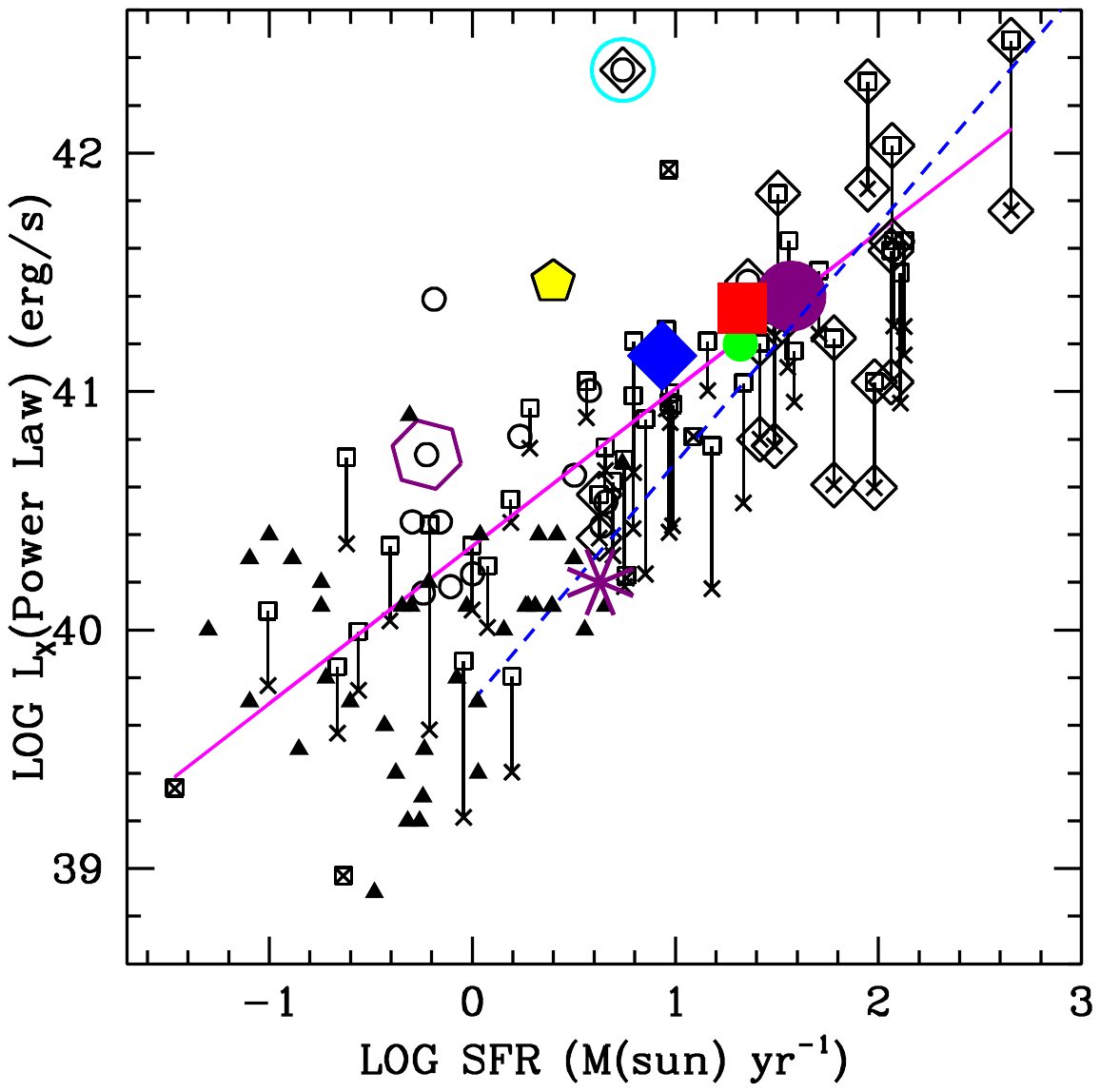}%
\includegraphics[width=7.5cm]{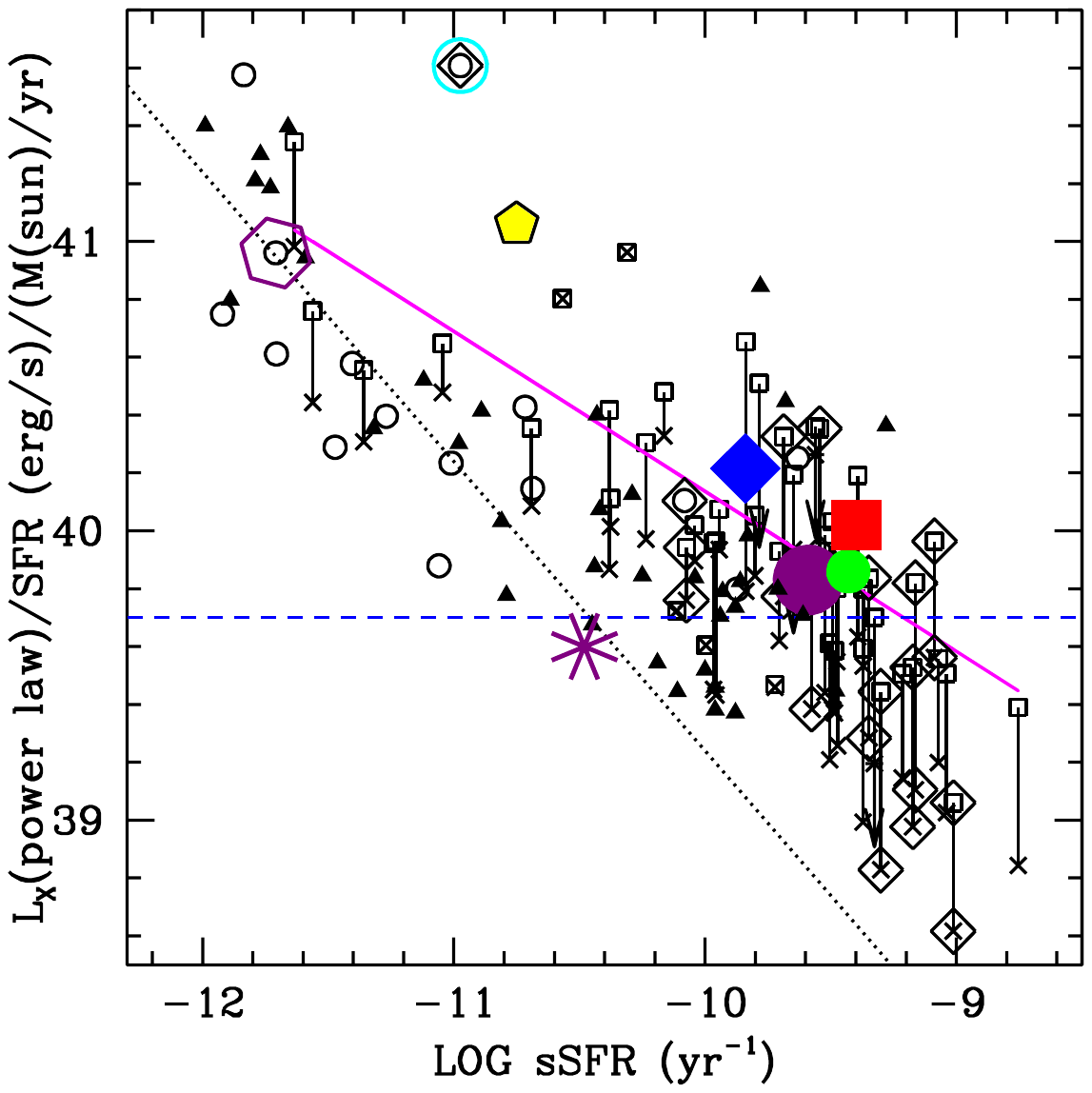}%
\includegraphics[width=6.5cm,trim={2.8cm 8cm 6cm 0}, clip=true]{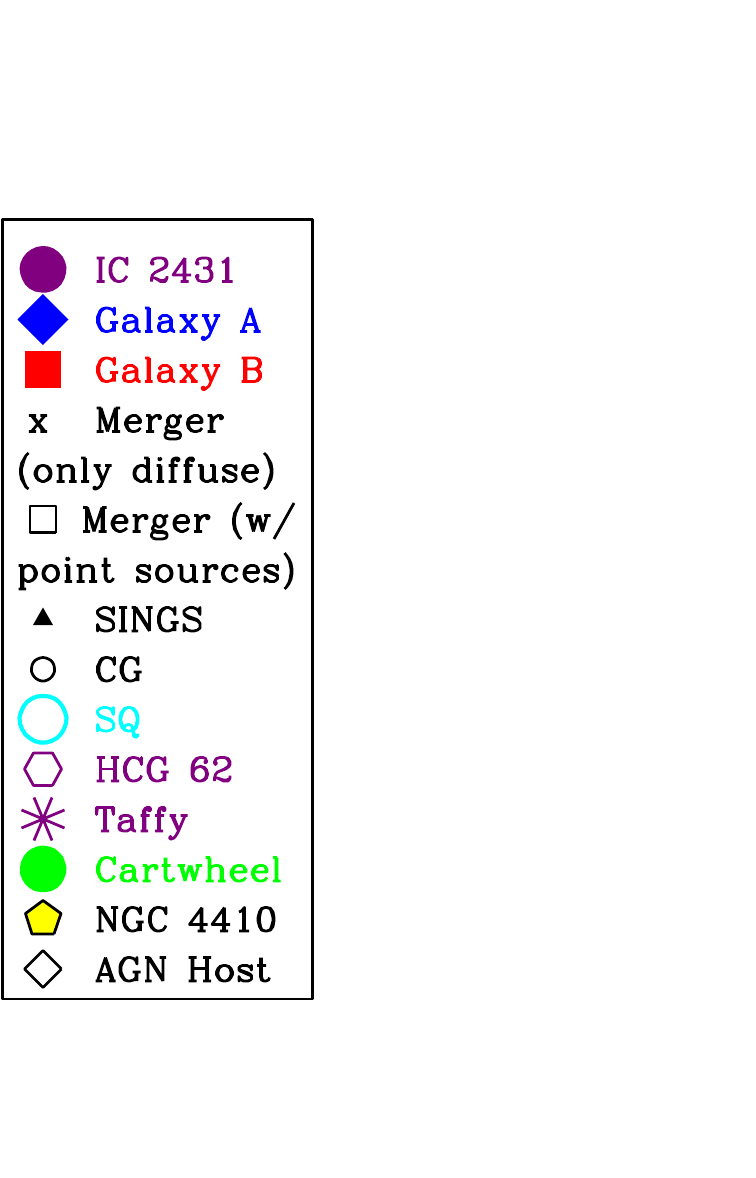}
\vskip -15mm
\caption{
Left: The power law component
of the X-ray luminosity plotted against the SFR.
Right: The ratio L$_{\rm X}$(power law)/SFR vs.\ sSFR.
{ 
The symbols are explained in the legend on the right.
The purple filled circle marks the location of IC 2431 as a whole.
The filled blue diamond is Galaxy A and the filled red square
is Galaxy B.
Each pair in the merging galaxy sample is marked by two datapoints:
the black crosses give the power law component of only the diffuse
emission for the mergers, 
while the black open squares
give 
the diffuse power law component added to the sum of the resolved
point sources.
The black filled triangles mark the sum of the 
X-ray luminosities of the point
sources in the SINGS galaxies.
The black open circles are the compact groups.
Stephan's Quintet 
is marked as a cyan circle.
HCG 62 is plotted as an open magenta hexagon.
The Taffy pair 
is marked by a purple asterisk.
The green filled circle marks the Cartwheel galaxy.
The yellow pentagon outlined in black is NGC 4410.
The black open diamonds mark known optical AGN.
The magenta
solid line is the best-fit
line for the mergers after adding the resolved point
sources to the power law component of the diffuse emission:
log (L$_{\rm X}$(PL)/SFR) = ($-$0.55 $\pm$ 0.07) log sSFR + (34.6 $\pm$ 0.7).
The blue dashed line marks the 
\citet{2019ApJS..243....3L} 
best-fit value
for the HMXB component in SINGS galaxies of 
L$_{\rm X}$(PL)/SFR = 
5.0 $\times$ 10$^{39}$ erg~s$^{-1}$. 
In the right panel, the black dashed line is the 
\citet{2022ApJ...930..135L} best-fit value for
LMXBs in the SINGS galaxies, log L$_{\rm X}$(LMXB)/M*
= 29.25.
}
\label{fig:LxPL}}
\end{figure}

In the 
left panel of Figure \ref{fig:LxPL}, we plot 
L$_{\rm X}$(power~law)
vs.\ the SFR for IC 2431 and the comparison samples.  
The right panel of Figure  
\ref{fig:LxPL} shows
L$_{\rm X}$(power~law)/SFR vs.\ sSFR.
For the mergers, we have included both the unresolved power law component as
seen by Chandra
\citep{2018AJ....155...81S}, 
as well 
as the sum of the luminosities of the X-ray point sources in the CSC,
added to the unresolved power law component.
For the compact groups and the SINGS galaxies,
we only plot the sum of the luminosities of the X-ray point sources
as tallied by 
\citet{2016ApJ...817...95T}  and 
\citet{2019ApJS..243....3L}, respectively.

For the full set of galaxies, a clear anti-correlation is seen
between 
L$_{\rm X}$(power~law)/SFR and sSFR.  Galaxies with low sSFR
have higher L$_{\rm X}$(power~law)/SFR, probably because of
contributions from LMXBs to the power law emission.
High sSFR mergers tend to have low
L$_{\rm X}$(power~law)/SFR ratios, compared to galaxies
with moderate sSFRs, although in both cases HMXBs likely
dominate the power law emission.  Three
factors may contribute to the low ratios in the high sSFR mergers: 1) obscured
AGN may contribute to the UV/IR fluxes, causing the SFRs to be
over-estimated, 
2) the extinction has been under-estimated.
or 
3) these systems may have young stellar ages,
thus there may not have been enough time for a large population of HMXBs
to form.  

In contrast to the extreme starbursts in the merger sample,
IC 2431 as a whole and Galaxy B individually have 
L$_{\rm X}$(power~law)/SFR ratios close to the best-fit value for the 
HMXB component for the 
SINGS galaxies and to the relation for the mergers.  
The power-law emission from Galaxy A is enhanced by about a factor of four
relative to the SINGs median value, but is only slightly high
(a factor of about 1.6) relative to the best-fit relation for the mergers.
This excess may be due to contributions from an AGN, although most
of the mergers in 
Figure \ref{fig:LxPL}
known to have AGNs don't show excesses in their power-law
X-ray emission.

In contrast to IC 2431, 
Stephan's Quintet stands out 
in Figure 
\ref{fig:LxPL}
as having a high
L$_{\rm X}$(power~law)/SFR for its sSFR.  
The vast majority of this excess emission is from the Seyfert 2 galaxy 
NGC 7319; without that galaxy, the L$_{\rm X}$(power law) for
Stephan's Quintet would be only 10$^{41}$ erg~s$^{-1}$, 
closer to the other systems in Figure 
\ref{fig:LxPL} with similar SFRs. 
The radio galaxy HCG 62 does not show a strong excess in power law emission,
when compared to its SFR.

\section{Discussion} \label{sec:discussion}

\subsection{Overview}

IC 2431 is a triple galaxy system, with
two galaxies of 
approximately equal stellar mass
of $\sim$5 $\times$ 10$^{10}$~M$_{\sun}$,
and a third galaxy 
with a lower stellar mass of $\sim$1.5 $\times$ 10$^{10}$ M$_{\sun}$.
IC 2431 is undergoing a powerful starburst, with an
overall SFR of about 37 M$_{\sun}$~yr$^{-1}$. About
65$\%$ of this star formation is associated with Galaxy B,
30$\%$ with Galaxy A, and 5$\%$ with 
Galaxy C.
Galaxy C has a moderate SFR of 0.5 $-$ 1 M$_{\sun}$~yr$^{-1}$
(Table \ref{SFR_table}),
which
places it near 
the star-forming main sequence
(e.g., 
\citealp{2016MNRAS.462.1749S, 
2020MNRAS.491.5406T}),
thus it has a typical star formation rate for its stellar
mass compared to normal star-forming galaxies, similar to that
of normal spiral galaxies with similar stellar mass
\citep{2022AJ....164..146S, 
2024AJ....168...12S}.
In contrast, 
Galaxies A and B have SFRs that are 
enhanced relative to the main sequence by 
factors of about 6 and 15, respectively.

The IC 2431 system is tidally disturbed, with at least five tidal tails
visible in optical images.
IC 2431 has several other morphological peculiarities, including a prominent
dust lane crossing Galaxy A, an apparent mid-infrared bridge between the two
main 
galaxies, a radio ridge/jet extending perpendicularly out of the disk of Galaxy A,
and spatial offsets between the optical, infrared, and radio in the outskirts of 
the two larger galaxies. 
IC 2431 may also have an excess of hot gas relative to its SFR, compared to
other star-forming systems.

All three galaxies in IC 2431 have possible AGNs.
Galaxy C is listed as an optical AGN by 
\citet{2017ApJ...835..280L}, although it is not detected in
either the radio or X-ray maps.
The putative AGNs in Galaxy A and Galaxy B are discussed in detail
in Sections \ref{sec:radioAGN} and 
\ref{sec:nuclB}, respectively.

The interaction history of IC 2431 is uncertain; 
a direct head-on collision between two of the galaxies
may have happened in the recent past, as in the 
Taffy galaxy pair; alternatively, 
a grazing tidal encounter may have occurred.
These possibilities are discussed below.

\subsection{A Possible Radio AGN in Galaxy A} \label{sec:radioAGN}

Galaxy A has a radio continuum excess relative to its SFR,
and the X-ray spectrum of the apparent nucleus 
is best-fit by a power law model. 
This nucleus may 
be variable in both the radio and the X-ray.
A radio jet/ridge extends about 4$''$ 
(4 kpc projected size)
from this nucleus.  
This feature is anti-coincident with the hot 
gas as seen in the 1.0 $-$ 2.5 keV X-ray map,
but approximately aligned with the mid-infrared bridge, 
and parallel to, but offset from, the
dust lane seen in the optical images.
Its apparent connection to the nucleus is strong evidence
that this structure is indeed a radio jet, although
it does not meet the formal definition of a jet given 
by \citet{1984ARA&A..22..319B}:
at least four times as long as it is wide.
The core power of Galaxy A in IC 2431 at 4.86 GHz is 
about 
6 $\times$ 10$^{21}$ W~Hz$^{-1}$, placing it near the bottom of the range
for the radio galaxies with jets
in the \citet{1984ARA&A..22..319B} sample. 
The total 1.49 GHz power for Galaxy A is
2 $\times$ 10$^{23}$ W~Hz$^{-1}$
\citep{1992ApJS...81...49S}.
Although
this luminosity is below the luminosity 
cutoff 
of 10$^{24}$~W~Hz$^{-1}$ 
used by \citet{2016A&ARv..24...10T}
to define a radio AGN,
IC 2431
has a high radio/FIR 
ratio relative to that expected from star formation alone
\citep{1992ApJS...81...49S,
1996ApJ...460..225C}. 

The jet/ridge in Galaxy A has a steeper radio spectral 
index ($\alpha$ = $-$0.7 to $-$1.1;
S$_{\nu}$ $\propto$ $\nu$$^{\alpha}$) 
than the nucleus ($\alpha$ $\sim$ $-$0.6), with $\alpha$ steepening
with increasing distance from the nucleus.
The Centaurus A radio jet has $\alpha$
ranging from $-$0.39 to $-$0.72, also decreasing 
along the length of the jet
\citep{1983ApJ...273..128B}.
Such a variation along a jet 
suggests aging of relativistic electrons
\citep{1983ApJ...273..128B}.
Such steepening of the spectral index along radio
lobes is seen in other galaxies as well
\citep{1985ApJ...291...52M}.
Only one apparent jet is visible in IC 2431.
One-sided jets are common in radio
galaxies \citep{1988gera.book..563K, 1984ARA&A..22..319B}.  
Relativistic beaming and Doppler-boosting \citep{1979ApJ...232...34B}
may contribute to one-sidedness in radio jets.

In IC 2431, the radio jet may have been 
disturbed by the on-going interaction between the galaxies. 
Its morphology, with
enhanced flux at the end of the ridge/jet, is consistent
with an interaction with the surrounding interstellar medium.
The diffuse 1.49 GHz emission to the northeast of the jet/ridge 
may
be a consequence of an interaction of the jet with interstellar 
matter during a close encounter between galaxies.
Jets within
groups or clusters
can be distorted by ram pressure due to motion of 
a galaxy through a cluster/group
intracluster/intragroup medium 
\citep{1980AJ.....85..204B,
2001AJ....121.2915B,
2011ApJ...738..145F,
2018AJ....155...14S}.
In colliding galaxies, ram pressure due to the interstellar
matter in a companion galaxy may also distort a jet
\citep{1986ApJ...311...58V, 
1987ApJ...319..671S,
1993ApJ...416..157B, 1996MNRAS.283..673S}.
For example, the very asymmetric radio jets/lobes in NGC 4410
may have been caused by interactions in a group
\citep{1987ApJ...319..671S, 2000ApJ...541..624S}.

The Hubble type of Galaxy A in IC 2431 is uncertain,
but it may be a disk galaxy.
Most radio galaxies have elliptical morphologies \citep{1995ApJ...438...62W, 
1993ARA&A..31..473A,
2020ApJ...901..159C},
but spiral galaxies have also been known to host 
active nuclei with radio jets
\citep{1987ApJ...319..671S,
1983ApJ...275....8W,
2021A&A...648A..17V,
2023A&A...675A..58P} and/or 
large radio lobes
\citep{1998ApJ...495..227L, 
2011MNRAS.417L..36H, 
2015MNRAS.454.1556S,
2015MNRAS.446.4176M,
2023RAA....23c5005G}.

The anti-coincidence of the radio jet/ridge in 
IC 2431 with the medium-energy X-ray emission is suggestive of 
an interaction of a jet with the surrounding medium.
In some compact groups with radio jets/lobes, 
X-ray 
cavities are observed to surround the radio features
\citep{2004ApJ...607..800B,
2013MNRAS.428...58R,
2015ApJ...805..112R,
2020MNRAS.496.1471K}.
In IC 2431, 
the 
X-ray source that lies between the two galaxies,
source \#5, is adjacent
to the radio structure and has a thermal spectrum,
thus may be shocked gas associated with
jet impact.  
This source does not have strong mid-IR emission
associated with it,
unlike sources \#1 and \#2, thus it may not be a dusty star-forming
region.

AGN feedback may be responsible for 
the apparent excess of hot X-ray emitting gas in IC 2431 
relative to the SFR, compared to the other galaxy systems
in our comparison set
(Figure 
\ref{fig:desjardins2014_fig7}).
According to 
\citet{2005MNRAS.357..279C}, groups that host radio AGNs
have hotter intragroup gas than radio-quiet groups.
AGN feedback is a popular explanation for the hot gas seen
in elliptical galaxies \citep{2012ARA&A..50..455F, 2014MNRAS.442..440C,  2015MNRAS.449.4105C}
and in galaxy groups \citep{2015MNRAS.449.4105C, 2020MNRAS.496.1471K}.
In numerical simulations of elliptical formation via major mergers,
AGN feedback contributes significantly to heating the interstellar gas
\citep{2006ApJ...643..692C, 
2015MNRAS.449.4105C}.
It is suggestive that the two groups in our comparison sample
that host radio galaxies, HCG 62 and NGC 4410, also show
excess hot gas relative to their SFR and sSFR, compared to other 
star-forming galaxies.
However, the hot gas excess in IC 2431 is associated with
both Galaxy A and Galaxy B, not Galaxy A alone, so this explanation
is uncertain.

Although a jet/interstellar medium interaction can account for
many of the observed properties of IC 2431, 
the spectacular
dust lane crossing Galaxy A that is visible in optical images
cannot easily be explained with this scenario alone; it also
requires a strong tidal effect or a collision between galaxies.
The dust lane is likely either: a) a disturbed tidal arm/tail
of either Galaxy B or Galaxy C projected in front of Galaxy A,
or b) interstellar matter ram pressure-stripped from Galaxy B
or Galaxy C during a head-on collision between two 
galaxies.  If a radio AGN is present,
the jet may impinge on stripped or tidal gas pulled
out during an interaction/collision, 
so its morphology
may be affected by the motions in the group.

There is, however, 
an alternative scenario that can account for both the
radio jet/ridge and the dust lane simultaneously:
the radio structure may have been caused by 
gas 
stripped from a galaxy
during a head-on collision with another galaxy,
as in 
the radio continuum bridge in the Taffy galaxies
\citep{1993AJ....105.1730C}.  
Perhaps a similar process has happened in IC 2431.
This possibility is discussed further in Section
\ref{sec:hot_gas_collision}.

\subsection{The Galactic Nucleus of Galaxy B: 
A Possible Obscured AGN?} \label{sec:nuclB}

The possible nucleus of Galaxy B, X-ray source \#10, may be variable,
and is best-fit
by two APEC components, both highly obscured.
Possible explanations include:

1) It may be a young supernova with a relatively high temperature,
but still thermal and declining exponentially.  
However, 
the higher temperature APEC component 
of source \#10
has an unabsorbed L$_{\rm X}$ $\sim$ 10$^{43}$ erg~s$^{-1}$.
This is 
high for a supernova, 
which typically peak near or below 0.3 $-$ 1 $\times$ 10$^{42}$ erg~s$^{-1}$
\citep{2015ApJ...810...32C, 
2016MNRAS.462.1101D,
2017ApJ...850..111N,
2018ApJ...864...45M, 2019MNRAS.490.4536Q}.
Only very rare super-luminous supernovae
sometimes reach higher X-ray luminosities
\citep{2013ApJ...771..136L, 2018ApJ...864...45M, 2019ARA&A..57..305G}.

2) It may be a super-Eddington stellar mass black hole,
in which the X-ray spectrum typically 
has a sharp downturn above a few keV;
the exact value of the downturn depends on viewing angle, outflow density, and
accretion rate \citep{2007MNRAS.377.1187P, 2017MNRAS.468.2865P}.
For example, it may be a ULX like NGC 247 X-1 
\citep{2016ApJ...831..117F, 2021MNRAS.505.5058P}
or NGC 55 X-1 
\citep{2017MNRAS.468.2865P, 2022MNRAS.516.3972B}, which  
have similar slope/temperature above 2 keV as our nuclear source.

3) It may be an IMBH, with a mass between 10$^3$ $-$ 10$^5$ M$_{\sun}$. 
Likely IMBH candidates have soft and variable X-ray counterparts with
L$_{\rm X}$ $\sim$ 10$^{42}$ L$_{\sun}$ 
\citep{2009Natur.460...73F,
2017MNRAS.469..886S,
2025arXiv250300904C}.
IMBHs may exist in the nuclei of dwarf galaxies and late-type
spiral galaxies, if the correlation between central
black hole mass and the spheroidal component of the host galaxy
holds to low masses 
\citep{2018ApJ...863....1C, 2021ApJ...923..246G}.
Because the morphology of IC 2431 is peculiar, a disk/bulge decomposition
for Galaxy B has not been done.
However,
given the relatively large stellar mass of 5 $\times$ 10$^{10}$ M$_{\sun}$
for Galaxy B, the nuclear black hole might be expected to be 
more massive than an IMBH.

4) It may be a thermal tidal disruption
event (TDE), which happens when a star gets too close
to a massive black hole and is destroyed by 
tidal forces (e.g., \citealp{1988Natur.333..523R}).
Such events
would produce a transient 
thermal X-ray spectrum \citep{1996A&A...309L..35B,
2020SSRv..216...85S,
2023A&A...671A..33S}.
Observationally, TDEs may appear as sources
with very steep X-ray spectra ($\Gamma$ $>$ 3) and high
L$_{\rm X}$ $>$ 10$^{41}$ erg~s$^{-1}$
\citep{2023A&A...671A..33S}.
These events would be 
highly absorbed if inside a 
starburst nucleus, and therefore may be missed in the 
optical because of the extinction.
TDEs are very rare, however, with rates of approximately
1 $-$ 2 $\times$ 10$^{-5}$~galaxy$^{-1}$~yr$^{-1}$
\citep{2002AJ....124.1308D,
2021MNRAS.508.3820S,
2023A&A...671A..33S}, thus the probability that one
would occur in IC 2431 during our observation is small.

5) Alternatively, the nucleus
of Galaxy B may be a changing-look AGN, which vary dramatically
on short timescales ($<$1 day to years)
\citep{2023NatAs...7.1282R}.
These sometimes have soft X-ray spectra
\citep{2012ApJ...752..154T,
2023A&A...671A..33S}.
The variability in the X-ray may be caused in part by 
changes 
in obscuration with time, with N$_{\rm H}$ varying from
$<$10$^{22}$~cm$^{-2}$ to $>$2 $\times$ 10$^{24}$~cm$^{-2}$ 
\citep{2002ApJ...571..234R, 2014MNRAS.439.1403M, 2015ApJ...815...55R,
2022ApJ...935..114M,
2023NatAs...7.1282R}.
For an intrinsic power law spectrum,
the ratio of the flux 
in the 7 $-$ 10 keV band vs.\ that in the 2 $-$ 4 keV band
(i.e., the observed hardness ratio 
F$_{7-10 keV}$/F$_{2-4 keV}$)
increases with N$_{\rm H}$
to N$_{\rm H}$ $\sim$ 5 $\times$ 10$^{23}$~cm$^{-2}$, then 
drops precipitously 
\citep{2014MNRAS.439.1403M}.  When in a high absorption state,
such sources are expected to appear very soft. 
In our 2-component fits for the nucleus of Galaxy B,
the harder component has an unabsorbed luminosity 
$\sim$10$^{43}$ erg~s$^{-1}$, consistent with 
a 
soft changing-look AGN.

\subsection{A Head-On Collision 
Between Gas Rich Galaxies?} \label{sec:hot_gas_collision}

There are some morphological signs that a head-on collision 
between two galaxies may have occurred in IC 2431 in the recent past.
These include
the dust lane visible in the optical,
the dust bridge seen
in the mid-infrared, 
the hot gas between the galaxies in the 
X-ray maps (i.e., X-ray source \#5),
and 
the mid-IR/optical offsets in Galaxy A.
These features may have been produced by ram pressure stripping 
of interstellar matter during
such a collision.  Which of the three galaxies were involved in this
collision is uncertain, however.
Galaxy C is faint in the
mid-IR, in spite of its flattened disk-like shape, and the dust lane crossing
Galaxy A is aligned with the disk of Galaxy C.
This suggests that Galaxy C might have traveled through Galaxy B
and/or Galaxy A and got stripped in the process.
Alternatively, Galaxy A and Galaxy B might have suffered a 
face-on collision,
like the Taffy galaxy pair. 
This idea is supported by the observation that 
the southern portion of Galaxy A lacks mid-IR emission,
and the mid-IR is offset from the optical in the north of Galaxy A.

The radio continuum ridge in Galaxy A is another peculiar feature that 
could be accounted for by 
a head-on collision.
This structure may be similar to the
radio continuum bridge in the Taffy galaxies 
\citep{1993AJ....105.1730C}, perhaps at an earlier stage.
In the Taffy system,
cosmic rays, magnetic fields, and HI gas were removed
from the disk, and the radio continuum in the gas enhanced by
synchrotron emission from relativistic
electrons trapped in the bridge
\citep{1993AJ....105.1730C,
2020MNRAS.492.4892Y,
2024arXiv240911707Y}.
Such features are called `splash bridges'
\citep{2020MNRAS.492.4892Y,
2024arXiv240911707Y}.
In the Taffy radio continuum bridge,
the radio spectral index $\alpha$ = -1.4 is
considerably steeper than 
than the values of -0.7 to -0.9 
in the galactic disks of the Taffy galaxies 
\citep{1993AJ....105.1730C}.
The latter is similar to the median spectral index for spiral 
galaxies.   
According to 
\citet{1993AJ....105.1730C},
the steep spectral index in the Taffy bridge is 
due to synchrotron and inverse Compton losses, with a time scale of 
$\sim$ 2 $\times$ 10$^7$~yr$^{-1}$.
The $\alpha$ for the Taffy bridge is steeper than the $\alpha$ = -1.1
at the tip of the IC 2431 radio ridge, however, given the uncertainty
on $\alpha$ of $\sim$0.2 for the IC 2431 feature, these spectral indices
agree within 1$-$2$\sigma$.

Our detection of high L$_{\rm X}$(gas)/SFR in IC 2431
compared to other star-forming galaxies is another observation that
might be explained by a head-on encounter.
The production of hot gas during a head-on collision 
of equal-mass galaxies was modeled by 
\citet{2006ApJ...643..692C}, who found that 
L$_{\rm X}$(gas) can increase by more than an order of magnitude
at the time of the first disk impact due to shock heating.
In contrast to these models, however, 
in the 
\citet{2018AJ....155...81S} study of 
L$_{\rm X}$(gas)/SFR ratios of
49 major mergers in different
merger stages, 
no trend with stage is evident.
The lack of an observed trend with merger stage may be due to variations
in the interaction parameters from system to system, causing the
timing and intensity of the X-ray enhancement from shocks
to vary from system to 
system.  Perhaps IC 2431 is viewed at an optimal time and 
has favorable orbital parameters for shock heating of the gas.

The Taffy galaxy pair is generally considered the prototype of this
kind of collision, as it has a bridge of hot gas, a radio continuum bridge,
and spectral signatures of strong interstellar shocks
\citep{1993AJ....105.1730C,
2012ApJ...751...11P,
2015ApJ...812..118A,
2019ApJ...878..161J}.
As discussed in Section 
\ref{hotgas_discussion}, however, 
it is uncertain whether the 
Taffy system as a whole 
has a global excess of hot gas relative to the SFR.

We can estimate the amount of energy that would be available for gas 
heating from such a head-on collision.  
This energy would be approximately
(f/2)Mv$^2$, where f is an efficiency factor,
M the total gas mass, and v the relative velocity.  
This gives
f $\times$ 2.5 $\times$ 10$^{56}$ ergs of energy, 
assuming a total mass of hot gas of 
10$^{8}$~M$_{\sun}$ and v = 500 km~s$^{-1}$.
If this is mostly radiated away in X-rays,
the implied X-ray luminosity would be f $\times$ 
8 $\times$ 10$^{40}$ 
erg~s$^{-1}$ $\times$ (100 Myrs/$\tau$), where
$\tau$ is the cooling time for the hot gas in units of Myrs.  
Assuming an efficiency of close to 100\% and a cooling time of $\sim$30 Myrs,
this estimate is consistent with the observed
excess 
X-ray luminosity relative to the SFR.

\section{Summary} \label{sec:summary}

We have conducted a multi-wavelength analysis of the compact galaxy group IC 2431,
using new Chandra X-ray images and archival UV, optical, IR, and radio maps.
IC 2431 has some very unusual morphological features
indicative of a recent collision and/or strong gravitational 
interactions between
the galaxies, and possible AGN activity.
These peculiar structures 
include:
a prominent dust lane visible in optical images, 
a mid-IR bridge connecting two galaxies in Spitzer images, 
a massive concentration (2 $\times$ 10$^7$ M$_{\sun}$)
of X-ray-emitting hot gas between
the galaxies, and a 4~kpc-long radio continuum jet/ridge 
extending from a galactic nucleus.  
The Chandra data show an excess of hot gas in the system, 
relative to the
star formation rate, compared to other star-forming systems.

One possible explanation of these peculiarities 
is a recent head-on collision between two gas-rich disk 
galaxies.  Ram pressure stripping of interstellar matter
during such a collision may have pulled out
gas and dust from the disks, creating a bridge
between the galaxies.  
Shock heating during the impact
may be responsible for the hot gas observed between the two
disks, as well as the excess of hot gas relative to 
other star-forming galaxies.
The radio continuum ridge may have been
created by this process, powered by relativistic electrons
trapped by interstellar magnetic fields.

Alternatively, the observed radio continuum ridge in IC 2431
may be a one-sided radio jet powered by an AGN.  In this
picture, the global excess of hot gas and the 
large concentration of hot gas between the galaxies may have been
caused by this jet impinging on interstellar matter.
In this second scenario, 
tidal interactions have perturbed the interstellar matter
in these galaxies, which in turn has disturbed and distorted the radio jet.
A third possibility is a combination of these two: a head-on collision
causing ram pressure stripping, plus a radio jet disturbed by an interaction
with interstellar gas.

More observations are needed to better understand this
system, and to distinguish between these possibilities.
For example, 
optical and/or
infrared imaging spectroscopy
could be used to search for signatures of shock heating, and to 
measure the kinematics
of the system. 
Interferometric 21 cm radio and 2.6 mm millimeter maps
of the cold interstellar gas and its kinematics
would help constrain the evolutionary
state of the system.
More sensitive and/or higher resolution
radio continuum maps would be useful to distinguish
between a radio AGN and a splash bridge as the cause of the
radio ridge/jet in Galaxy A.


This paper employs a list of Chandra datasets, obtained by
the Chandra X-ray Observatory, contained in the 
Chandra Data Collection (CDC) 416 https://doi.org/10.25574/cdc.416.
Support for this research was provided by the National Aeronautics and Space Administration through Chandra Award Number GO3-24108X issued by the Chandra X-ray Center, which is operated by the Smithsonian Astrophysical Observatory for and on behalf of the National Aeronautics Space Administration under contract NAS8-03060.  
RS acknowledges the INAF grant No. 1.05.23.04.04.
We thank the anonymous referee for carefully reading the paper and providing
helpful suggestions.
We also thank Michael Blanton for his work in developing the kcorrect software,
and for his help in using the code.

This work is also based [in part] on observations made with the Spitzer Space Telescope, which was operated by the Jet Propulsion Laboratory, California Institute of Technology under a contract with NASA.
This research is also
based in part on observations made with the NASA/ESA Hubble Space Telescope, and obtained from the Hubble Legacy Archive, which is a collaboration between the Space Telescope Science Institute (STScI/NASA), the Space Telescope European Coordinating Facility (ST-ECF/ESA) and the Canadian Astronomy Data Centre (CADC/NRC/CSA).
This research also uses data from the Very Large Array, operated by
the National Radio Astronomy Observatory.
The National Radio Astronomy Observatory is a facility of the National Science Foundation operated under cooperative agreement by Associated Universities, Inc.
This research used data from the Galaxy Evolution Explorer (GALEX), a NASA mission.
This research has made use of the NASA/IPAC Extragalactic Database, which is funded by the National Aeronautics and Space Administration and operated by the California Institute of Technology.

This research also made use of data from the Legacy Surveys.
The Legacy Surveys consist of three individual and complementary projects: 
the Dark Energy Camera Legacy Survey (DECaLS; Proposal ID \#2014B-0404; PIs: David Schlegel and Arjun Dey), the Beijing-Arizona Sky Survey (BASS; NOAO Prop. ID \#2015A-0801; PIs: Zhou Xu and Xiaohui Fan), and the Mayall z-band Legacy Survey (MzLS; Prop. ID \#2016A-0453; PI: Arjun Dey). DECaLS, BASS and MzLS together include data obtained, respectively, at the Blanco telescope, Cerro Tololo Inter-American Observatory, NSF’s NOIRLab; the Bok telescope, Steward Observatory, University of Arizona; and the Mayall telescope, Kitt Peak National Observatory, NOIRLab. Pipeline processing and analyses of the data were supported by NOIRLab and the Lawrence Berkeley National Laboratory (LBNL). The Legacy Surveys project is honored to be permitted to conduct astronomical research on Iolkam Du’ag (Kitt Peak), a mountain with particular significance to the Tohono O’odham Nation.
NOIRLab is operated by the Association of Universities for Research in Astronomy (AURA) under a cooperative agreement with the National Science Foundation. LBNL is managed by the Regents of the University of California under contract to the U.S. Department of Energy.
This project used data obtained with the Dark Energy Camera (DECam), which was constructed by the Dark Energy Survey (DES) collaboration. Funding for the DES Projects has been provided by the U.S. Department of Energy, the U.S. National Science Foundation, the Ministry of Science and Education of Spain, the Science and Technology Facilities Council of the United Kingdom, the Higher Education Funding Council for England, the National Center for Supercomputing Applications at the University of Illinois at Urbana-Champaign, the Kavli Institute of Cosmological Physics at the University of Chicago, Center for Cosmology and Astro-Particle Physics at the Ohio State University, the Mitchell Institute for Fundamental Physics and Astronomy at Texas A\&M University, Financiadora de Estudos e Projetos, Fundacao Carlos Chagas Filho de Amparo, Financiadora de Estudos e Projetos, Fundacao Carlos Chagas Filho de Amparo a Pesquisa do Estado do Rio de Janeiro, Conselho Nacional de Desenvolvimento Cientifico e Tecnologico and the Ministerio da Ciencia, Tecnologia e Inovacao, the Deutsche Forschungsgemeinschaft and the Collaborating Institutions in the Dark Energy Survey. The Collaborating Institutions are Argonne National Laboratory, the University of California at Santa Cruz, the University of Cambridge, Centro de Investigaciones Energeticas, Medioambientales y Tecnologicas-Madrid, the University of Chicago, University College London, the DES-Brazil Consortium, the University of Edinburgh, the Eidgenossische Technische Hochschule (ETH) Zurich, Fermi National Accelerator Laboratory, the University of Illinois at Urbana-Champaign, the Institut de Ciencies de l’Espai (IEEC/CSIC), the Institut de Fisica d’Altes Energies, Lawrence Berkeley National Laboratory, the Ludwig Maximilians Universitat Munchen and the associated Excellence Cluster Universe, the University of Michigan, NSF’s NOIRLab, the University of Nottingham, the Ohio State University, the University of Pennsylvania, the University of Portsmouth, SLAC National Accelerator Laboratory, Stanford University, the University of Sussex, and Texas A\&M University.
BASS is a key project of the Telescope Access Program (TAP), which has been funded by the National Astronomical Observatories of China, the Chinese Academy of Sciences (the Strategic Priority Research Program “The Emergence of Cosmological Structures” Grant \# XDB09000000), and the Special Fund for Astronomy from the Ministry of Finance. The BASS is also supported by the External Cooperation Program of Chinese Academy of Sciences (Grant \# 114A11KYSB20160057), and Chinese National Natural Science Foundation (Grant \# 12120101003, \# 11433005).  
The Legacy Surveys imaging of the DESI footprint is supported by the Director, Office of Science, Office of High Energy Physics of the U.S. Department of Energy under Contract No. DE-AC02-05CH1123, by the National Energy Research Scientific Computing Center, a DOE Office of Science User Facility under the same contract; and by the U.S. National Science Foundation, Division of Astronomical Sciences under Contract No. AST-0950945 to NOAO.


\appendix

\section{Photometry} \label{sec:appendixB}

In 
Table \ref{Photometrytab}, we provide the UV/optical/IR fluxes for the three galaxies in IC 2431,
plus the global values within 12$''$ and 24$''$ radii.
In Table \ref{Photometry3tab}, we provide UV/optical/IR fluxes in 18 bands for the 
seven smaller rectangular regions in IC 2431: regions A1, A2, A3,
A4, B1, B2, and B3 (see Figures \ref{fig:boxes} and
\ref{fig:SEDs}). 

\input full_photometry.tex

\input third_photometry.tex

\section{Details of the Population Synthesis Modeling} \label{appendix:pop_syn}

In this section, we provide additional information
about the method we 
used to model the stellar populations in IC 2431.
In addition to the parameters mentioned in Section 
\ref{sec:pop_syn},
CIGALE also requires a number of other parameters, 
including the e-folding time of the old stellar population,
the onset age for the main stellar population,
the ratio E(B$-$V)(stars)/E(B$-$V)(lines), 
the slope of the power law of the 
\citet{2000ApJ...533..682C}
attenuation curve,
the \citet{2014ApJ...780..172D},
dust emission parameter
q$_{\rm PAHs}$ 
(the fraction of the dust mass in PAHs), 
the minimum UV radiation field U$_{\rm min}$,
the power law slope $\alpha$ of the dust emission curve, as defined by
dM$_{dust}$/dU $\propto$ U$^{-\alpha}$, 
and the illumination fraction from U$_{\rm min}$
to U$_{\rm max}$.
In 
Table \ref{CIGALE_input}, we provide 
the choices we permitted for the parameters that were allowed to vary.
The onset age for the main stellar population is fixed to 13 Gyrs.
In modeling
the 
nebular emission, we fixed the ionization parameter
to log U = $-$3.0, the electron density to 100 cm$^{-2}$,
the Lyman continuum escape fraction to zero, the Lyman continuum dust absorption
fraction to zero, and the emission line width to 300 km~s$^{-1}$.
When the best-fit value of an input parameter
is an extremum of the set of input parameters 
listed in Table \ref{CIGALE_input},
we extended 
the input parameter range and reran CIGALE, repeating until no best-fit
parameters are extrema.
Many of the CIGALE output parameters were already presented in
Table \ref{SFR_table};
the best-fit values of the remaining parameters are given 
Table \ref{CIGALE_table} (for the galaxies as a whole), 
and 
Table \ref{CIGALE_table_second} (for the seven smaller rectangular regions).

\input input_to_cigale_table.tex

\input cigale_latex_table.tex

\input cigale_latex_table_second.tex


%

\vspace{5mm}
\facilities{CXO}

\bibliography{IC2431_paper.bib}{}
\bibliographystyle{aasjournal}

\end{document}

%% file: SFR_table.tex
\begin{deluxetable}{c|ccc|ccccccccc}
\rotate
\tablecolumns{13}
\tablewidth{0pc}
\caption{Star Formation Rates and Absorption in Various Regions in IC 2431\label{SFR_table}}
\tablehead{   
\colhead{{\bf Region}}    
$|$
&\multicolumn{3}{c}{{\bf From FUV plus 8 $\mu$m}}   
$|$
&\multicolumn{9}{c}{{\bf From CIGALE}}   
\\ 
\hline
\colhead{}   
& \colhead{SFR}    
& \colhead{E(B-V)} 
& \colhead{N(H)}
& \colhead{SFR}    
& \colhead{log M*}    
& \colhead{log sSFR}    
& \colhead{E(B-V)} 
& \colhead{E(B-V)} 
& \colhead{N(H)}
& \colhead{N(H)}
& \colhead{Age}
& \colhead{r$_{\rm SFR}$}
\\ 
\colhead{} 
& \colhead{(M$_{\sun}$/yr)}
& \colhead{}    
& \colhead{($\times$ 10$^{22}$}
& \colhead{(M$_{\sun}$/yr)}
& \colhead{(M$_{\sun}$)}    
& \colhead{(yr$^{-1}$)}
& \colhead{(stars)} 
& \colhead{(lines)} 
& \colhead{($\times$ 10$^{22}$}
& \colhead{($\times$ 10$^{22}$}
& \colhead{(Myrs)} 
&\colhead{} 
\\ 
\colhead{} 
& \colhead{}
&\colhead{} 
& \colhead{cm$^{-2}$)}
& \colhead{}
& \colhead{} 
& \colhead{} 
& \colhead{} 
& \colhead{} 
& \colhead{cm$^{-2}$)}
& \colhead{cm$^{-2}$)}
& \colhead{} 
&\colhead{} 
\\ 
\colhead{} 
& \colhead{}
&\colhead{} 
& \colhead{}
& \colhead{}
& \colhead{} 
& \colhead{} 
& \colhead{} 
& \colhead{} 
& \colhead{(stars)}
& \colhead{(lines)}
& \colhead{} 
&\colhead{} 
}
\startdata
12$''$ circle & 34.8 $\pm$ 0.01 & 0.31 $\pm$ 0.01 & 0.18 $\pm$ 0.01 & \\
24$''$ circle & 38.76 $\pm$ 0.02 & 0.3 $\pm$ 0.01 & 0.18 $\pm$ 0.01 & \\
Galaxy A & 8.97 $\pm$ 0.01 & 0.45 $\pm$ 0.01 & 0.26 $\pm$ 0.01 & 8.5 $\pm$  2.06 & 10.75 $\pm$  0.04 & -9.81 $\pm$  0.11 & 0.28 $\pm$  0.04 & 0.81 $\pm$  0.28 & 0.16 $\pm$  0.02 & 0.47 $\pm$  0.16 & 34 $\pm$  35 & 1.3 $\pm$  1.16 \\
Galaxy B & 23.76 $\pm$ 0.01 & 0.28 $\pm$ 0.01 & 0.16 $\pm$ 0.01 & 18.32 $\pm$  9.37 & 10.74 $\pm$  0.07 & -9.47 $\pm$  0.23 & 0.19 $\pm$  0.04 & 0.62 $\pm$  0.25 & 0.11 $\pm$  0.02 & 0.36 $\pm$  0.15 & 44 $\pm$  38 & 5.3 $\pm$  2.54 \\
Galaxy C & 0.5 $\pm$ 0.01 & 0.18 $\pm$ 0.02 & 0.11 $\pm$ 0.01 & 1.04 $\pm$  0.45 & 10.18 $\pm$  0.07 & -10.15 $\pm$  0.2 & 0.12 $\pm$  0.04 & 0.35 $\pm$  0.22 & 0.07 $\pm$  0.02 & 0.2 $\pm$  0.13 & 48 $\pm$  34 & 0.49 $\pm$  0.46 \\
Region A1 & 2.2 $\pm$ 0.01 & 0.42 $\pm$ 0.01 & 0.24 $\pm$ 0.01 & 1.82 $\pm$  0.85 & 9.56 $\pm$  0.08 & -9.29 $\pm$  0.22 & 0.25 $\pm$  0.04 & 0.64 $\pm$  0.24 & 0.15 $\pm$  0.02 & 0.37 $\pm$  0.14 & 53 $\pm$  37 & 6.32 $\pm$  2.42 \\
Region A2 & 3.35 $\pm$ 0.01 & 0.53 $\pm$ 0.02 & 0.31 $\pm$ 0.01 & 2.12 $\pm$  1.28 & 10.24 $\pm$  0.06 & -9.91 $\pm$  0.27 & 0.39 $\pm$  0.03 & 0.91 $\pm$  0.33 & 0.22 $\pm$  0.01 & 0.53 $\pm$  0.19 & 30 $\pm$  29 & 8.07 $\pm$  2.68 \\
Region A3 & 2.89 $\pm$ 0.01 & 0.48 $\pm$ 0.02 & 0.28 $\pm$ 0.01 & 2.25 $\pm$  1.23 & 10.56 $\pm$  0.05 & -10.2 $\pm$  0.24 & 0.23 $\pm$  0.05 & 0.73 $\pm$  0.27 & 0.14 $\pm$  0.03 & 0.42 $\pm$  0.16 & 62 $\pm$  33 & 0.28 $\pm$  0.24 \\
Region A4 & 0.53 $\pm$ 0.01 & 0.26 $\pm$ 0.02 & 0.15 $\pm$ 0.01 & 0.61 $\pm$  0.11 & 9.68 $\pm$  0.03 & -9.89 $\pm$  0.09 & 0.12 $\pm$  0.02 & 0.33 $\pm$  0.17 & 0.07 $\pm$  0.01 & 0.19 $\pm$  0.1 & 39 $\pm$  35 & 0.6 $\pm$  0.39 \\
Region B1 & 2.7 $\pm$ 0.01 & 0.17 $\pm$ 0.01 & 0.1 $\pm$ 0.01 & 2.61 $\pm$  0.39 & 10.2 $\pm$  0.03 & -9.77 $\pm$  0.07 & 0.07 $\pm$  0.02 & 0.13 $\pm$  0.07 & 0.04 $\pm$  0.01 & 0.07 $\pm$  0.04 & 28 $\pm$  34 & 1.22 $\pm$  0.94 \\
Region B2 & 13.58 $\pm$ 0.01 & 0.32 $\pm$ 0.01 & 0.19 $\pm$ 0.01 & 8.76 $\pm$  3.86 & 10.47 $\pm$  0.08 & -9.52 $\pm$  0.21 & 0.23 $\pm$  0.03 & 0.82 $\pm$  0.19 & 0.14 $\pm$  0.01 & 0.48 $\pm$  0.11 & 39 $\pm$  34 & 6.18 $\pm$  2.45 \\
Region B3 & 7.49 $\pm$ 0.01 & 0.29 $\pm$ 0.01 & 0.17 $\pm$ 0.01 & 5.8 $\pm$  2.66 & 10.02 $\pm$  0.09 & -9.25 $\pm$  0.22 & 0.22 $\pm$  0.03 & 0.69 $\pm$  0.24 & 0.13 $\pm$  0.02 & 0.4 $\pm$  0.14 & 48 $\pm$  34 & 8.84 $\pm$  2.18 \\
\enddata
\end{deluxetable}

%% file: roberto_global.tex
\begin{deluxetable}{ccccccc}
\tablecolumns{7}
\tablewidth{0pc}
\setlength{\tabcolsep}{5pt}
\caption{Chandra Spectral Results for Galaxy A, Galaxy B, and IC 2431 as a Whole \label{xspectab}}
\tablehead{   
\colhead{ID}   
& \colhead{N$_{\rm H}$}
& \colhead{kT}    
& \colhead{$\Gamma$} 
& \colhead{C-stat/DOF}
&\colhead{APEC}
&\colhead{Power Law}\\ 
\colhead{} 
&\colhead{($\times$10$^{22}$}
& \colhead{(keV)}    
&\colhead{} 
& \colhead{}
& \colhead{Log L$_{\rm X}$}
& \colhead{Log L$_{\rm X}$}
\\ 
\colhead{} 
& \colhead{cm$^{-2}$)}
& \colhead{}    
& \colhead{}
& \colhead{}
& \colhead{(erg~s$^{-1}$)}
& \colhead{(erg~s$^{-1}$)}
\\ 
\colhead{} 
& \colhead{}    
& \colhead{}
& \colhead{}
& \colhead{}
& \colhead{0.3-8 keV}
& \colhead{0.3-8 keV}
}
\startdata
A & 0.58 $\pm$ $^{0.36}_{0.46}$ & 0.85 $\pm$ $^{0.18}_{0.48}$ & 1.05 $\pm$ $^{0.64}_{0.85}$ & 121.27/178 & 41.51 $\pm$ $^{0.25}_{0.25}$& 41.16 $\pm$ $^{0.18}_{0.14}$\\
B & 0.90 $\pm$ $^{0.38}_{0.5}$ & 0.58 $\pm$ $^{0.35}_{0.21}$ & 1.80 $\pm$ $^{0.71}_{0.6}$ & 183.60/184 & 41.99 $\pm$ $^{0.3}_{0.51}$& 41.34 $\pm$ $^{0.31}_{0.3}$\\
12\farcs radius & 0.76 $\pm$ $^{0.21}_{0.26}$ & 0.92 $\pm$ $^{0.13}_{0.58}$ & 1.06 $\pm$ $^{0.78}_{0.91}$ & 282.91/274 & 42.06 $\pm$ $^{0.14}_{0.21}$& 41.41 $\pm$ $^{0.37}_{0.13}$\\
\enddata
\tablenotetext{ }{Quoted uncertainties are 90\% confidence.  The luminosities are unabsorbed.}
\end{deluxetable}

%% file: roberto_gas_without_count_rates.tex
\begin{deluxetable}{cccccccccccc}
\tablecolumns{12}
\tablewidth{0pc}
\setlength{\tabcolsep}{1pt}
\caption{Chandra Spectral Results for the X-Ray Sources within IC 2431\label{xspectab_gas}}
\tablehead{   
\colhead{X-Ray}   
& \colhead{RA}
& \colhead{DEC}
& \colhead{Radius}
& \colhead{N$_{\rm H}$}
& \colhead{kT}    
& \colhead{$\Gamma$} 
& \colhead{C-stat/DOF}
&\colhead{APEC}
&\colhead{Power Law}\\ 
\colhead{Source} 
&\colhead{(deg)} 
&\colhead{(deg)} 
&\colhead{($''$)} 
&\colhead{($\times$10$^{22}$}
& \colhead{(keV)}    
&\colhead{} 
& \colhead{}
& \colhead{Log L$_{\rm X}$}
& \colhead{Log L$_{\rm X}$}
\\ 
\colhead{ID} 
&
&
&
& \colhead{cm$^{-2}$)}
& \colhead{}    
& \colhead{}
& \colhead{}
& \colhead{(erg~s$^{-1}$)}
& \colhead{(erg~s$^{-1}$)}
\\ 
\colhead{} 
&
&
&
& \colhead{}    
& \colhead{}
& \colhead{}
& \colhead{}
& \colhead{0.3-8 keV}
& \colhead{0.3-8 keV}
}
\startdata
{\bf \#1  }& 136.14752 & 14.59499 & 1.7 
& 1.78 $\pm$ $^{2.39}_{1.75}$ & 1.95 $\pm$ $^{5.85}_{0.77}$ & ....   & 28.90/25 & 40.69 $\pm$ $^{0.49}_{0.32}$& ....  \\
\#1 &  &    
   &  
& 1.09 $\pm$ $^{3.45}_{1.09}$ & ....   & 2.29 $\pm$ $^{1.78}_{1.13}$ & 31.48/25 & ....  & 40.7 $\pm$ $^{1.32}_{0.38}$\\
{\bf \#2  }& 136.14657 & 14.59378 & 1.7 
& 1.82 $\pm$ $^{1.47}_{1.82}$ & 0.96 $\pm$ $^{2.72}_{0.38}$ & ....   & 6.39/20 & 40.93 $\pm$ $^{0.58}_{0.78}$& ....  \\
\#2 &  &    
   &  
& 1.13 $\pm$ $^{2.88}_{1.13}$ & ....   & 3.48 $\pm$ $^{2.62}_{1.58}$ & 7.32/20 & ....  & 41.08 $\pm$ $^{1.61}_{0.84}$\\
{\bf \#3  }& 136.14574 & 14.59329 & 1.5 
& 0.29 $\pm$ $^{1.50}_{0.29}$ & ....   & 1.39 $\pm$ $^{0.91}_{0.56}$ & 42.86/50 & ....  & 40.72 $\pm$ $^{0.3}_{0.13}$\\
\#3 &  &    
   &  
& 0.85 $\pm$ $^{1.29}_{0.85}$ & 5.11 $\pm$ $^{999}_{2.74}$ & ....   & 42.41/50 & 40.77 $\pm$ $^{0.19}_{0.18}$& ....  \\
{\bf \#4  }& 136.14509 & 14.59290 & 1.3 
& 0.0 $\pm$ $^{1.27}_{0}$ & 1.64 $\pm$ $^{1.56}_{0.77}$ & ....   & 6.33/12 & 40.1 $\pm$ $^{0.58}_{0.21}$& ....  \\
\#4 &  &    
   &  
& 0.035 $\pm$ $^{1.63}_{0.035}$ & ....   & 3.06 $\pm$ $^{2.40}_{0.99}$ & 7.85/12 & ....  & 40.46 $\pm$ $^{1.61}_{0.37}$\\
{\bf \#5  }& 136.14488 & 14.59385 & 2.0 
& 0.79 $\pm$ $^{0.84}_{0.79}$ & 1.17 $\pm$ $^{0.79}_{0.38}$ & ....   & 54.21/24 & 40.76 $\pm$ $^{0.44}_{0.4}$& ....  \\
\#5 &  &    
   &  
& 0.0 $\pm$ $^{0.95}_{0}$ & ....   & 2.76 $\pm$ $^{1.24}_{0.65}$ & 60.42/24 & ....  & 40.7 $\pm$ $^{0.77}_{0.21}$\\
{\bf \#6  }& 136.14466 & 14.59569 & 1.6 
& 0.0 $\pm$ $^{0.22}_{0}$ & ....   & 2.39 $\pm$ $^{0.44}_{0.41}$ & 45.83/60 & ....  & 40.95 $\pm$ $^{0.18}_{0.12}$\\
\#6 &  &    
   &  
& 0.0 $\pm$ $^{0.27}_{0}$ & 2.80 $\pm$ $^{1.76}_{0.81}$ & ....   & 49.16/60 & 40.8 $\pm$ $^{0.06}_{0.09}$& ....  \\
{\bf \#7  }& 136.14412 & 14.59513 & 1.2 
& 0.61 $\pm$ $^{1.64}_{0.61}$ & ....   & 2.42 $\pm$ $^{1.25}_{0.87}$ & 44.13/34 & ....  & 40.81 $\pm$ $^{0.81}_{0.32}$\\
\#7 &  &    
   &  
& 0.013 $\pm$ $^{1.35}_{0.013}$ & 5.11 $\pm$ $^{23.6}_{3.34}$ & ....   & 45.78/34 & 40.55 $\pm$ $^{0.25}_{0.13}$& ....  \\
{\bf \#8  }& 136.14361 & 14.59463 & 1.2 
& 0.0 $\pm$ $^{1.81}_{0}$ & ....   & 1.80 $\pm$ $^{1.42}_{0.7}$ & 19.56/21 & ....  & 40.35 $\pm$ $^{0.7}_{0.16}$\\
\#8 &  &    
   &  
& 0.0 $\pm$ $^{1.06}_{0}$ & 8.68 $\pm$ $^{999}_{6.37}$ & ....   & 20.24/21 & 40.31 $\pm$ $^{0.18}_{0.16}$& ....  \\
{\bf \#9  }& 136.14307 & 14.59408 & 1.5 
& 1.24 $\pm$ $^{0.74}_{1.24}$ & 1.08 $\pm$ $^{0.96}_{0.31}$ & ....   & 30.20/33 & 41.05 $\pm$ $^{0.37}_{0.49}$& ....  \\
\#9 &  &    
   &  
& 1.14 $\pm$ $^{1.63}_{1.14}$ & ....   & 4.14 $\pm$ $^{1.80}_{1.4}$ & 29.18/33 & ....  & 40.38 $\pm$ $^{0.44}_{0.14}$\\
{\bf \#10  }& 136.14390 & 14.59509 & 1.2 
 &   
4.08 $\pm$ $^{3.46}_{2.18}$/51.1 $\pm$ $^{56.3}_{24.9}$ & 0.23 $\pm$ $^{0.36}_{0.09}$/0.73$\pm$ $^{0.55}_{0.37}$ & ....   & 47.24/36 & 43.08 $\pm$ $^{1.41}_{1.53}$& 43.54 $\pm$ $^{0.84}_{1.67}$\\
\#10 &     &
   &  
& 0.17 $\pm$ $^{1.33}_{0.17}$ & ....   & 1.68 $\pm$ $^{0.95}_{0.75}$ & 62.91/39 & ....  &  40.67 $\pm$ $^{0.38}_{0.13}$\\
\#10 &  &   
   &  
& 0.0 $\pm$ $^{0.82}_{0.0}$ & 18.24 $\pm$ $^{99.9}_{14.52}$ & ....   & 63.4/39 & 40.62 $\pm$ $^{0.13}_{0.11}$& ....  \\
\#10 & & &  & 4.10 $\pm$ $^{3.46}_{2.17}$/60.35 $\pm$ $^{27.85}_{34.4}$ & 0.23 $\pm$ $^{0.30}_{0.09}$ & 8.37 $\pm$ $^{99.9}_{5.17}$ & 48.2/36 & 43.1 $\pm$ $^{2.09}_{1.57}$& 99.9 $\pm$ $^{99.9}_{99.9}$\\
\enddata
\tablenotetext{ }{Quoted uncertainties are 90\% confidence.  The luminosities are unabsorbed.
Source \#3 is the presumed nucleus of Galaxy A, while Source \#10
is assumed to be the nucleus of Galaxy B.
}
\end{deluxetable}

%% file: table_hot_gas_mass_full.tex
\begin{deluxetable}{cccccccc}
\tablecolumns{6}
\tablewidth{0pc}
\caption{Derived Properties of the Hot Gas in IC 2431\label{hot_gas_mass}}
\tablehead{   
\colhead{X-Ray}   
& \colhead{Assumed}
& \colhead{Volume}    
& \colhead{n$_{\rm e}$$\sqrt{f}$} 
& \colhead{Log}
& \colhead{Cooling}\\
\colhead{Source} 
&\colhead{Radius}
& \colhead{(kpc$^3$)}    
&\colhead{(cm$^{-3}$)} 
& \colhead{Hot}
& \colhead{Time}\\
\colhead{ID} 
&\colhead{($''$)}
& \colhead{}
& \colhead{}
& \colhead{Gas}    
& \colhead{(Myrs)}\\
\colhead{} 
&\colhead{}
&\colhead{}
& \colhead{}
& \colhead{Mass}    
& \colhead{}\\
\colhead{} 
&\colhead{}
& \colhead{}
& \colhead{}
& \colhead{(M$_{\sun}$)}    
& \colhead{}\\
}
\startdata
\#1 & 0.7 &  1.44 & 0.22 & 6.88 & 29.4 \\
\#2 & 0.7 &  1.44 & 0.25 & 6.93 & 9.4 \\
\#4 & 0.75 &  1.77 & 0.1 & 6.62 & 51.4 \\
\#5 & 1.25 &  8.18 & 0.09 & 7.25 & 34.7 \\
\#9 & 0.7 &  1.44 & 0.29 & 7 & 9.6 \\
Total & &  14.27 &  & 7.68& \
\enddata
\end{deluxetable}

%% file: full_photometry.tex
\begin{deluxetable}{cccccc}
\tablecolumns{6}
\tablewidth{0pc}
\caption{Photometry of the Individual Galaxies in IC 2431$^a$\label{Photometrytab}}
\tablehead{   
\colhead{Filter}   
&\colhead{24$''$}   
& \colhead{12$''$}
& \colhead{Galaxy}    
& \colhead{Galaxy} 
& \colhead{Galaxy}
\\ 
\colhead{} 
&\colhead{Radius} 
& \colhead{Radius}
& \colhead{A}    
& \colhead{B} 
& \colhead{C}
\\ 
}
\startdata
shape &  circle & circle & rectangle & rectangle & rectangle \\
RA (2000) &  136.145030 & 136.145030 & 136.146100 & 136.144340 & 136.142800 \\
Dec (2000) &  14.594220 & 14.594220 & 14.593750 & 14.595020 & 14.599040 \\
Box &    &    & 6.9" $\times$ 22.9" & 8" $\times$ 21.5" & 19.2" $\times$ 7.4" \\
P.A. & & &  43$^{\circ}$ & 43$^{\circ}$ & 43$^{\circ}$ \\
\hline
FUV & 0.959 $\pm$ 0.005 & 0.847 $\pm$ 0.002 & 0.076 $\pm$ 0.001 & 0.681 $\pm$ 0.002 & 0.029 $\pm$ 0.001 \\
NUV & 1.676 $\pm$ 0.005 & 1.447 $\pm$ 0.002 & 0.199 $\pm$ 0.001 & 1.047 $\pm$ 0.001 & 0.08 $\pm$ 0.001 \\
SDSS u & 3.091 $\pm$ 0.027 & 2.711 $\pm$ 0.013 & 0.687 $\pm$ 0.008 & 1.936 $\pm$ 0.008 & 0.155 $\pm$ 0.007 \\
SDSS g & 7.761 $\pm$ 0.008 & 6.549 $\pm$ 0.004 & 2.095 $\pm$ 0.002 & 4.213 $\pm$ 0.002 & 0.674 $\pm$ 0.002 \\
SDSS r & 11.405 $\pm$ 0.013 & 9.451 $\pm$ 0.007 & 3.603 $\pm$ 0.004 & 5.448 $\pm$ 0.004 & 1.202 $\pm$ 0.004 \\
SDSS i & 15.773 $\pm$ 0.021 & 13.098 $\pm$ 0.01 & 5.208 $\pm$ 0.006 & 7.3 $\pm$ 0.006 & 1.555 $\pm$ 0.006 \\
SDSS z & 18.831 $\pm$ 0.068 & 15.769 $\pm$ 0.034 & 6.742 $\pm$ 0.02 & 8.427 $\pm$ 0.021 & 1.932 $\pm$ 0.019 \\
HST & 10.909 $\pm$ 0.003 & 9.236 $\pm$ 0.001 & 3.39 $\pm$ 0.001 & 5.523 $\pm$ 0.001 & 1.064 $\pm$ 0.001 \\
DES g & 7.938 $\pm$ 0.002 & 6.708 $\pm$ 0.001 & 2.194 $\pm$ 0.001 & 4.267 $\pm$ 0.001 & 0.704 $\pm$ 0.001 \\
DES r & 13.127 $\pm$ 0.003 & 11.017 $\pm$ 0.002 & 4.143 $\pm$ 0.001 & 6.403 $\pm$ 0.001 & 1.265 $\pm$ 0.001 \\
DES z & 18.787 $\pm$ 0.008 & 15.777 $\pm$ 0.004 & 6.725 $\pm$ 0.002 & 8.446 $\pm$ 0.002 & 1.908 $\pm$ 0.002 \\
J & 28.87 $\pm$ 0.146 & 24.022 $\pm$ 0.073 & 9.879 $\pm$ 0.043 & 12.27 $\pm$ 0.045 & 2.558 $\pm$ 0.041 \\
H & 34.164 $\pm$ 0.252 & 28.449 $\pm$ 0.126 & 12.442 $\pm$ 0.075 & 14.445 $\pm$ 0.078 & 3.098 $\pm$ 0.07 \\
K & 33.832 $\pm$ 0.212 & 28.875 $\pm$ 0.106 & 12.637 $\pm$ 0.063 & 14.345 $\pm$ 0.065 & 2.685 $\pm$ 0.059 \\
W1 & 20.509 $\pm$ 0.006 & 16.178 $\pm$ 0.003 \\
W2 & 15.925 $\pm$ 0.005 & 12.345 $\pm$ 0.003 \\
W3 & 104.2 $\pm$ 0.036 & 69.815 $\pm$ 0.018 \\
W4 & 270.427 $\pm$ 0.083 & 127.739 $\pm$ 0.042 \\
IRAC ch1 & 23.837 $\pm$ 0.009 & 20.731 $\pm$ 0.004 & 8.271 $\pm$ 0.003 & 11.535 $\pm$ 0.003 & 1.28 $\pm$ 0.002 \\
IRAC ch2 & 18.164 $\pm$ 0.011 & 15.935 $\pm$ 0.006 & 6.107 $\pm$ 0.003 & 9.183 $\pm$ 0.003 & 0.847 $\pm$ 0.003 \\
IRAC ch3 & 40.187 $\pm$ 0.047 & 35.187 $\pm$ 0.024 & 11.288 $\pm$ 0.014 & 22.338 $\pm$ 0.015 & 0.919 $\pm$ 0.013 \\
IRAC ch4 & 177.351 $\pm$ 0.051 & 159.378 $\pm$ 0.025 & 44.947 $\pm$ 0.015 & 105.91 $\pm$ 0.016 & 2.151 $\pm$ 0.014 \\
\enddata
\tablenotetext{a}{Units are mJy.   The quoted uncertainties are statistical, and do not include uncertainties in the flux calibration.}
\end{deluxetable}

%% file: third_photometry.tex
\begin{deluxetable}{cccccccc}
\tablecolumns{8}
\tablewidth{0pc}
\caption{Photometry of Smaller Rectangular Regions in IC 2431$^a$\label{Photometry3tab}}
\tablehead{   
\colhead{Filter}   
& \colhead{Galaxy}    
& \colhead{Galaxy} 
& \colhead{Galaxy}
& \colhead{Galaxy}
& \colhead{Galaxy}
& \colhead{Galaxy}
& \colhead{Galaxy}
\\ 
\colhead{} 
& \colhead{A1}    
& \colhead{A2} 
& \colhead{A3}
& \colhead{A4}
& \colhead{B1} 
&\colhead{B2} 
&\colhead{B3} 
\\ 
}
\startdata
shape &  rectangle & rectangle & rectangle & rectangle & rectangle & rectangle & rectangle \\
RA (2000) &  136.147770 & 136.146630 & 136.145510 & 136.144380 & 136.145750 & 136.144340 & 136.142920 \\
Dec (2000) &  14.595530 & 14.594350 & 14.593190 & 14.592020 & 14.596480 & 14.595020 & 14.593560 \\
Box & 6.9" $\times$ 5.7" & 6.9" $\times$ 5.7" & 6.9" $\times$ 5.7" & 6.9" $\times$ 5.7" & 8" $\times$ 7.2" & 8"$\times$ 7.2" & 8"$\times$ 7.2"\\
P.A. & 43$^{\circ}$ & 43 $^{\circ}$ & 43$^{\circ}$ & 43$^{\circ}$ & 43$^{\circ}$ & 43$^{\circ}$ & 43$^{\circ}$ \\
\hline
FUV & 0.024 $\pm$ 0.001 & 0.016 $\pm$ 0.001 & 0.02 $\pm$ 0.001 & 0.174 $\pm$ 0.001 & 0.298 $\pm$ 0.001 & 0.209 $\pm$ 0.001 & 0.037 $\pm$ 0.001 \\
NUV & 0.058 $\pm$ 0.001 & 0.053 $\pm$ 0.001 & 0.057 $\pm$ 0.001 & 0.272 $\pm$ 0.001 & 0.471 $\pm$ 0.001 & 0.305 $\pm$ 0.001 & 0.103 $\pm$ 0.001 \\
SDSS u & 0.152 $\pm$ 0.004 & 0.107 $\pm$ 0.004 & 0.32 $\pm$ 0.004 & 0.628 $\pm$ 0.005 & 0.737 $\pm$ 0.005 & 0.563 $\pm$ 0.005 & 0.256 $\pm$ 0.006 \\
SDSS g & 0.336 $\pm$ 0.001 & 0.289 $\pm$ 0.001 & 1.158 $\pm$ 0.001 & 1.506 $\pm$ 0.001 & 1.638 $\pm$ 0.001 & 1.04 $\pm$ 0.001 & 0.631 $\pm$ 0.002 \\
SDSS r & 0.448 $\pm$ 0.002 & 0.543 $\pm$ 0.002 & 2.135 $\pm$ 0.002 & 1.896 $\pm$ 0.002 & 2.296 $\pm$ 0.002 & 1.226 $\pm$ 0.002 & 1.037 $\pm$ 0.003 \\
SDSS i & 0.647 $\pm$ 0.003 & 0.903 $\pm$ 0.003 & 3.027 $\pm$ 0.003 & 2.336 $\pm$ 0.004 & 3.213 $\pm$ 0.004 & 1.728 $\pm$ 0.004 & 1.661 $\pm$ 0.004 \\
SDSS z & 0.797 $\pm$ 0.01 & 1.233 $\pm$ 0.01 & 3.997 $\pm$ 0.01 & 2.665 $\pm$ 0.012 & 3.852 $\pm$ 0.012 & 1.844 $\pm$ 0.012 & 2.277 $\pm$ 0.014 \\
HST & 0.476 $\pm$ 0.001 & 0.53 $\pm$ 0.001 & 1.937 $\pm$ 0.001 & 1.869 $\pm$ 0.001 & 2.269 $\pm$ 0.001 & 1.361 $\pm$ 0.001 & 1.021 $\pm$ 0.001 \\
DES g & 0.334 $\pm$ 0.001 & 0.312 $\pm$ 0.001 & 1.213 $\pm$ 0.001 & 1.496 $\pm$ 0.001 & 1.701 $\pm$ 0.001 & 1.063 $\pm$ 0.001 & 0.648 $\pm$ 0.001 \\
DES r & 0.574 $\pm$ 0.001 & 0.686 $\pm$ 0.001 & 2.348 $\pm$ 0.001 & 2.054 $\pm$ 0.001 & 2.745 $\pm$ 0.001 & 1.592 $\pm$ 0.001 & 1.316 $\pm$ 0.001 \\
DES z & 0.713 $\pm$ 0.001 & 1.233 $\pm$ 0.001 & 4.055 $\pm$ 0.001 & 2.667 $\pm$ 0.001 & 3.926 $\pm$ 0.001 & 1.835 $\pm$ 0.001 & 2.186 $\pm$ 0.002 \\
J & 0.79 $\pm$ 0.021 & 2.498 $\pm$ 0.022 & 5.594 $\pm$ 0.021 & 3.371 $\pm$ 0.026 & 5.953 $\pm$ 0.025 & 2.901 $\pm$ 0.027 & 3.786 $\pm$ 0.031 \\
H & 1.022 $\pm$ 0.037 & 3.022 $\pm$ 0.038 & 7.401 $\pm$ 0.037 & 3.616 $\pm$ 0.044 & 7.456 $\pm$ 0.044 & 3.335 $\pm$ 0.046 & 4.892 $\pm$ 0.053 \\
K & 1.12 $\pm$ 0.031 & 3.325 $\pm$ 0.032 & 7.149 $\pm$ 0.031 & 3.354 $\pm$ 0.037 & 7.605 $\pm$ 0.037 & 3.342 $\pm$ 0.039 & 5.315 $\pm$ 0.044 \\
IRAC ch1 & 1.139 $\pm$ 0.001 & 2.283 $\pm$ 0.001 & 4.236 $\pm$ 0.001 & 2.076 $\pm$ 0.002 & 6.343 $\pm$ 0.002 & 3.123 $\pm$ 0.002 & 4.35 $\pm$ 0.002 \\
IRAC ch2 & 0.965 $\pm$ 0.002 & 1.768 $\pm$ 0.002 & 2.983 $\pm$ 0.002 & 1.498 $\pm$ 0.002 & 5.17 $\pm$ 0.002 & 2.52 $\pm$ 0.002 & 3.439 $\pm$ 0.002 \\
IRAC ch3 & 2.284 $\pm$ 0.007 & 3.883 $\pm$ 0.007 & 4.471 $\pm$ 0.007 & 2.355 $\pm$ 0.008 & 13.12 $\pm$ 0.008 & 6.877 $\pm$ 0.008 & 7.373 $\pm$ 0.01 \\
IRAC ch4 & 10.712 $\pm$ 0.008 & 16.884 $\pm$ 0.007 & 14.941 $\pm$ 0.008 & 10.072 $\pm$ 0.009 & 62.523 $\pm$ 0.009 & 33.38 $\pm$ 0.009 & 31.799 $\pm$ 0.011 \\
\enddata
\tablenotetext{a}{Units are mJy.   The quoted uncertainties are statistical, and do not include uncertainties in the flux calibration.}
\end{deluxetable}

%% file: input_to_cigale_table.tex
\begin{deluxetable}{cc}
\tablecolumns{2}
\tablewidth{0pc}
\caption{CIGALE Input Parameters\label{CIGALE_input}}
\tablehead{
\colhead{Parameter}   
&\colhead{Choices}   
\\ 
}
\startdata
e-folding time of the old population ($\tau$$_{\rm main}$) & 1, 3, 5, 7.5, 10 Gyr \\
Age of the late burst/quench & 2, 5, 10, 25, 50, 75, 100 Myrs \\
Ratio of the SFR after/before the burst (r$_{\rm SFR}$) & 0.2, 0.6, 1.0, 1.25, 2.5, 5.0, 10.0 \\
E(B$-$V)(line emission)&0.1, 0.2, 0.5, 0.6, 0.7, 0.8, 0.9, 1.0, 1.2, 1.5 \\
Reduction factor E(B$-$V)(stars)/E(B$-$V)(lines)& 0.25, 0.50, 0.75\\
Slope of power law of attenuation curve & -1.5, -1.2, -1., -0.7, -0.5,-0.3,  0.0 \\
PAH dust mass fraction (q$_{\rm PAH}$) & 1.12, 2.50, 5.26, 6.63 \\
Miminum radiation field (U$_{\rm min}$) & 0.1, 0.25, 0.50, 1.0, 2.5, 5.0, 10.0, 25.0, 50.0 \\
Power-law slope ($\alpha$, where dM$_{dust}$/dU $\propto$ U$^{-\alpha}$) & 1.5, 2.0, 2.5 \\
Illumination fraction from U$_{\rm min}$ to U$_{\rm max}$ & 0.001, 0.003, 0.01, 0.03, 0.1, 0.3 \\
\enddata
\end{deluxetable}

%% file: cigale_latex_table.tex
\begin{deluxetable}{cccc}
\tablecolumns{4}
\tablewidth{0pc}
\caption{Additional CIGALE Parameters for SED Fits for the Galaxies in IC 2431\label{CIGALE_table}}
\tablehead{
}
\startdata
Region &  A &  B &  C \\
\hline
best ${\tau}_{\rm main}$ (Gyrs)& 10 & 10 & 5 \\
best E(B$-$V) factor &  0.5  &  0.25  &  0.25 \\
best attenuation powerlaw slope &  -0.5  &  -0.5  &  -1.0 \\
best q$_{\rm pah}$ &  2.5  &  2.5  &  2.5 \\
best dust U$_{\rm mean}$ &  11.75  &  0.66  &  0.25 \\
best dust U$_{\rm min}$ & 5.0  & 0.25  & 0.25 \\
\enddata
\end{deluxetable}

%% file: cigale_latex_table_second.tex
\begin{deluxetable}{cccccccc}
\tablecolumns{8}
\tablewidth{0pc}
\caption{Additional CIGALE Parameters for SED Fits for the Smaller Rectangular Regions\label{CIGALE_table_second}}
\tablehead{
}
\startdata
Region &  A1 &  A2 &  A3 &  A4 &  B1 &  B2 &  B3 \\
\hline
best ${\tau}_{\rm main}$ (Gyrs)& 10 & 3 & 5 & 10 & 10 & 10 & 7 \\
best E(B$-$V) factor &  0.5  &  0.5  &  0.25  &  0.5  &  0.5  &  0.25 & 0.5 \\
best attenuation powerlaw slope &  -0.7  &  -0.3  &  -0.7  &  -1.0  &  -1.5  &  -0.3 & -0.3 \\
best q$_{\rm pah}$ &  2.5  &  2.5  &  6.63  &  6.63  &  6.63  &  2.5 & 2.5 \\
best dust U$_{\rm mean}$ &  13.15  &  66.81  &  40.79  &  48.45  &  5.54  &  0.62 & 33.92 \\
best dust U$_{\rm min}$ & 2.5  & 50.0  & 25.0  & 10.0  & 1.0  & 0.1 & 25.0 \\
\enddata
\end{deluxetable}